\documentclass[submission,copyright,creativecommons]{eptcs}
\providecommand{\event}{Quantum Physics and Logic (QPL) 2022} 

\usepackage{tikz}
\usepackage{qcircuit}
\usepackage{amsmath}
\usepackage{blkarray, bigstrut}
\usepackage{xparse}
\usepackage{graphicx}
\usepackage{amssymb}

\usepackage{amsthm}
\usepackage{caption}
\usepackage{subcaption}
\usepackage{braket}
\usepackage{multirow}
\usepackage{varioref}
\usepackage[noabbrev]{cleveref}
\renewcommand{\cref}[1]{\Cref{#1}}
\usepackage{todonotes}
\usepackage{algorithm2e}
\usepackage{csvsimple}
\usepackage{siunitx}
\newcommand{\CNOT}{\text{CNOT}}

\newcommand{\vect}[1]{\textbf{#1}}

\DeclareMathOperator{\real}{\mathbb{R}}

\theoremstyle{plain}

\theoremstyle{plain}

\theoremstyle{plain}

\theoremstyle{plain}

\theoremstyle{plain}

\theoremstyle{plain}

\theoremstyle{definition}
\newtheorem{definition}{Definition}[section]

\theoremstyle{remark}

\theoremstyle{definition}

\definecolor{IMSBlue}{HTML}{4db8ff}
\definecolor{IMSOrange}{HTML}{e67300}
\definecolor{IMSYellow}{HTML}{ffcc80}
\definecolor{IMSRed}{HTML}{cc0000}
\definecolor{IMSGreen}{HTML}{009933}
\definecolor{IMSPink}{HTML}{ff0066}

\usepackage{iftex}

\ifpdf
  \usepackage{underscore}         
  \usepackage[T1]{fontenc}        
\else
  \usepackage{breakurl}           
\fi

\usepackage{placeins}

\title{Dynamic Qubit Routing with CNOT Circuit Synthesis\\for Quantum Compilation}
\author{Arianne Meijer - van de Griend
\institute{Department of Computer Science\\University of Helsinki}
\email{ariannemeijer@gmail.com}
\and
Sarah Meng Li
\institute{Department of Combinatorics \& Optimization\\ Institute for Quantum Computing, University of Waterloo}
\email{sarah.li@uwaterloo.ca}
}
\def\titlerunning{Dynamic Qubit Routing with CNOT Circuit Synthesis for Quantum Compilation}
\def\authorrunning{A. Meijer - van de Griend \& S. M. Li} 
\begin{document}
\maketitle

\begin{abstract}
Many quantum computers have constraints regarding which two-qubit operations are locally allowed. To run a quantum circuit under those constraints, qubits need to be mapped to different quantum registers, and multi-qubit gates need to be routed accordingly. Recent developments have shown that compiling strategies based on Steiner tree provide a competitive tool to route CNOTs. However, these algorithms require the qubit map to be decided before routing. Moreover, the qubit map is fixed throughout the computation, i.e. the logical qubit will not be moved to a different physical qubit register. This is inefficient with respect to the CNOT count of the resulting circuit.
    
In this paper, we propose the algorithm \textit{PermRowCol} for routing CNOTs in a quantum circuit. It dynamically remaps logical qubits during the computation, and thus results in fewer output CNOTs than the algorithms \textit{Steiner-Gauss}~\cite{kissinger2020cnot} and \textit{RowCol}~\cite{wu2023optimization}.

Here we focus on circuits over CNOT only, but this method could be generalized to a routing and mapping strategy on Clifford+T circuits by slicing the quantum circuit into subcircuits composed of CNOTs and single-qubit gates. Additionally, \textit{PermRowCol} can be used in place of \textit{Steiner-Gauss} in the synthesis of phase polynomials as well as the extraction of quantum circuits from ZX-diagrams.
\end{abstract}

\section{Introduction}
Recent strides in quantum computing have made it possible to execute quantum algorithms on real quantum hardware \cite{arute2019quantum,zhu2022quantum}. Contrary to classical computing, efficient quantum circuits are necessary for successful execution due to the decoherence of qubits \cite{nielsen2001quantum}. If a quantum circuit takes too long to execute, it will not produce any usable results. Moreover, due to poor gate fidelities, each additional gate in the quantum circuit adds a small error to the computation. In the absence of fault-tolerant quantum computers, circuits with more gates produce less accurate results. Therefore, we need to reduce the gate complexity of the executed quantum circuits. This requires resource-efficient algorithms and improved quantum compiling procedures.

When mapping a quantum circuit to the physical layer, one has to consider the numerous constraints imposed by the underlying hardware architecture. For example, in a superconducting quantum computer~\cite{stassi2020scalable}, connectivity of the physical qubits restricts multi-qubit operations to adjacent qubits. These restrictions are known as \textit{connectivity constraints} and can be represented by a \textit{connected graph} (also known as a \textit{topology}). Each vertex represents a distinct physical qubit. When two qubits are adjacent, there is an edge between the corresponding vertices. 

Thus, we are interested in improving the routing of a quantum circuit onto a quantum computer. Current routing strategies are dominated by SWAP-based approaches~\cite{li2019tackling,qiskit,sivarajah2020tket,mcts}. These strategies move the logical qubits around on different quantum registers. The drawback of this is that every SWAP-gate adds $3$ CNOTs to the circuit (\cref{fig:gates}.a), adding only more gates to the original circuit. As a result, it will take much longer to execute a routed quantum circuit, and thus introduce more errors to the computation.

Additionally, these SWAP-based strategies can be replaced by a \textit{bridge template} (or \textit{bridge}) that acts like a remote CNOT. As shown in \cref{fig:gates}.b, the bridge template only requires $4$ CNOTs while swapping the qubits would have cost $7$ CNOTs (\cref{fig:gates}.c). The CNOT ladders in the bridge template can be generalized for remote CNOTs with more qubits in between. Note that usually in SWAP-based strategies, the last SWAP (\cref{fig:gates}.c) is omitted, resulting in $4$ CNOTs instead of $7$. Therefore, unlike with SWAP gates, the subsequent parts of the circuit cannot benefit from the new qubit placement because the bridge template does not move the qubits. Thus, we need to make a trade-off between swapping qubits and remote CNOTs when there is a sequence of CNOTs to be routed.

\begin{figure}[!h]
  \renewcommand{\thefigure}{(a)}
  \begin{subfigure}{.5\textwidth}
    \[
\Qcircuit @C=.9em @R=.9em @!R {
          \lstick{\ket{1}} & \qswap & \qw & & & \targ & \ctrl{1} & \targ & \qw & \rstick{\ket{2}} \\
          \lstick{\ket{2}} & \qswap \qwx & \qw  & \push{\rule{.3em}{0em}=\rule{.3em}{0em}} & & \ctrl{-1} & \targ & \ctrl{-1} & \qw & \rstick{\ket{1}}\\
          & & & & & & & & & & & & &\\
  } 
\] 
\caption{A SWAP gate implemented by $3$ CNOTs.}
\label{fig:swapgate}
\end{subfigure}
  \renewcommand{\thefigure}{(b)}
  \begin{subfigure}{.5\textwidth}
    \[
\Qcircuit @C=.9em @R=.9em @!R {
          \lstick{\ket{1}} & \ctrl{2} & \qw & & & \qw & \ctrl{1} & \qw & \ctrl{1} & \qw & \rstick{\ket{1}} \\
          \lstick{\ket{2}} & \qw & \qw  & \push{\rule{.2em}{0em}=\rule{.2em}{0em}} & & \ctrl{1} & \targ & \ctrl{1} & \targ & \qw & \rstick{\ket{2}}\\
          \lstick{\ket{3}} & \targ & \qw & & & \targ & \qw & \targ & \qw & \qw & \rstick{\ket{1 \oplus 3}}\\
  } 
\] 
\caption{A bridge template implemented by $4$ CNOTs.}
\label{fig:bridge}
\end{subfigure}
  \renewcommand{\thefigure}{(c)}
  \begin{subfigure}{1\textwidth}
    \[
\Qcircuit @C=1em @R=1em @!R {
          \lstick{\ket{1}} & \qswap & \qw & \qswap & \qw & \rstick{\ket{1}} & & & & & \targ & \ctrl{1} & \targ & \qw & \targ & \ctrl{1} & \targ & \qw & \rstick{\ket{1}} \\
          \lstick{\ket{2}} & \qswap \qwx & \ctrl{1}  & \qswap \qwx  & \qw & \rstick{\ket{2}} & & &  \push{\rule{.2em}{0em}=\rule{.2em}{0em}} & & \ctrl{-1} & \targ & \ctrl{-1} & \ctrl{1} & \ctrl{-1} & \targ & \ctrl{-1} & \qw & \rstick{\ket{2}}\\
          \lstick{\ket{3}} & \qw & \targ & \qw & \qw & \rstick{\ket{1 \oplus 3}} & & & & & \qw & \qw & \qw & \targ & \qw & \qw & \qw & \qw & \rstick{\ket{1 \oplus 3}}\\
  } 
\] 
\caption{Routing $\CNOT(1,3)$ with SWAP gates results in $7$ CNOTs.}
\label{fig:swapcomparison}
\end{subfigure}
  \renewcommand{\thefigure}{1}
  \caption{Visualize how a SWAP gate and a bridge template is implemented by CNOTs respectively. Note that for a single CNOT constrained by a $3$-qubit line topology, the bridge gate is the same as the output of the \textit{Steiner-Gauss} algorithm .}\label{fig:gates}
\end{figure}
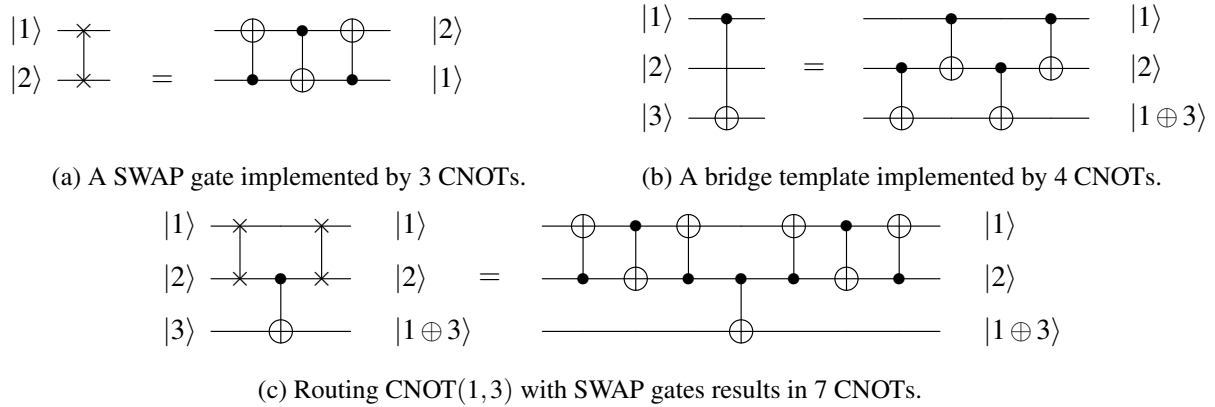
  
\begin{figure}[t!]
\renewcommand{\thefigure}{(a)}
  \begin{subfigure}{1\textwidth}
\centering
\begin{tikzpicture}[scale=0.5]
\begin{scope}[every node/.style={circle,minimum size= .07 cm,draw}]
    \node (1) at (-2,2) {$\mathbf{1}$};
    \node (2) at (0,2) {$2$};
    \node (3) at (2,2) {$3$};
    \node (4) at (-2,0) {$4$};
    \node (5) at (0,0) {$5$};
    \node (6) at (2,0) {$6$};
    \node (7) at (-2,-2) {$7$};
    \node (8) at (0,-2) {$8$};
    \node (9) at (2,-2) {$\mathbf{9}$};
\end{scope}
\draw[thick] (1)--(2);
\draw (2)--(3)--(6)--(9);
\draw[thick] (9)--(8);
\draw (8)--(7)--(4)--(1);
\draw[thick] (2)--(5);
\draw (5)--(6) (4)--(5);
\draw[thick] (5)--(8);
\end{tikzpicture} 
\caption{The $9$-qubit square grid.}
\label{fig2:a}
  \end{subfigure}
  \renewcommand{\thefigure}{(b)}
  \begin{subfigure}{.5\textwidth}
    \[
\Qcircuit @C=.7em @R=.7em @!R {
          \lstick{\ket{1}} & \qswap & \qw & \qw & \qw & \qw & \qw & \qswap & \qw & \rstick{\ket{1}}\\
          \lstick{\ket{2}} & \qswap \qwx & \qw &  \qswap & \qw & \qswap & \qw  & \qswap \qwx & \qw  & \rstick{\ket{2}}\\
          \lstick{\ket{5}} & \qw & \qw & \qswap \qwx & \ctrl{1} & \qswap \qwx & \qw & \qw & \qw & \rstick{\ket{5}}\\
          \lstick{\ket{8}} & \qswap & \qw & \qw & \targ & \qw & \qw & \qswap & \qw & \rstick{\ket{8}}\\
          \lstick{\ket{9}} & \qswap \qwx & \qw & \qw & \qw & \qw & \qw & \qswap \qwx & \qw &  \rstick{\ket{1 \oplus 9}}\\
  }
\]
\caption{SWAP template with a fixed qubit map.}
\label{fig2:b}
\end{subfigure}
  \renewcommand{\thefigure}{(c)}
   \begin{subfigure}{.5\textwidth}
    \[
\Qcircuit @C=.7em @R=.7em @!R {
          \lstick{\ket{1}} & \qswap & \qw & \qw & \qw & \qw & \qw  & \rstick{\ket{2}}\\
          \lstick{\ket{2}} & \qswap \qwx & \qw &  \qswap & \qw & \qw & \qw  & \rstick{\ket{5}}\\
          \lstick{\ket{5}} & \qw & \qw & \qswap \qwx & \ctrl{1} & \qw & \qw & \rstick{\ket{1}}\\
          \lstick{\ket{8}} & \qswap & \qw & \qw & \targ & \qw & \qw  & \rstick{\ket{1 \oplus 9}}\\
          \lstick{\ket{9}} & \qswap \qwx & \qw & \qw & \qw & \qw  & \qw &  \rstick{\ket{8}}\\
  }
\]
\caption{SWAP template with dynamic qubits maps.}
\label{fig2:c}
\end{subfigure}
  \renewcommand{\thefigure}{(d)}
  \begin{subfigure}{1\textwidth}
    \[
\Qcircuit @C=.7em @R=.7em @!R {
          \lstick{\ket{1}} & \qw    & \qw    & \qw      & \ctrl{1}& \qw     & \qw   & \qw   & \qw   & \qw   &\ctrl{1}& \qw  & \qw   & \qw   & \rstick{\ket{1}}\\
          \lstick{\ket{2}} & \qw    & \qw    & \ctrl{1} & \targ  & \ctrl{1} & \qw   &\qw    &  \qw  &\ctrl{1}&\targ &\ctrl{1}& \qw  & \qw   & \rstick{\ket{2}}\\
          \lstick{\ket{5}} & \qw    & \ctrl{1}&\targ    & \qw   & \targ     &\ctrl{1}&\qw   &\ctrl{1}& \targ& \qw   &\targ  &\ctrl{1}& \qw   & \rstick{\ket{5}}\\
          \lstick{\ket{8}} & \ctrl{1}&\targ  & \qw      & \qw   & \qw       & \targ &\ctrl{1}&\targ & \qw   & \qw   & \qw   &\targ  & \qw   & \rstick{\ket{8}}\\
          \lstick{\ket{9}} & \targ  & \qw    & \qw      & \qw   & \qw       & \qw   &\targ  & \qw   & \qw   & \qw   & \qw   &\qw    & \qw   & \rstick{\ket{1 \oplus 9}}\\
  }
\]
\caption{\textit{Steiner-Gauss} with a fixed qubit map~\cite{kissinger2020cnot}.}
\label{fig2:d}
  \end{subfigure}
  \renewcommand{\thefigure}{2}\caption{Given the constrained topology in \cref{fig:swapTemplate}.a, three different routing strategies are compared for implementing $\CNOT(1,9)$. The CNOT count corresponding to each template is $19$, $10$, and $12$ respectively.}
\label{fig:swapTemplate} 
\end{figure}
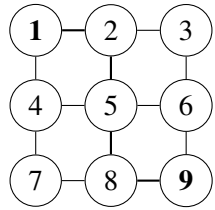
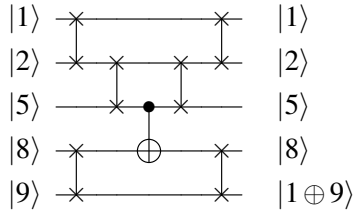
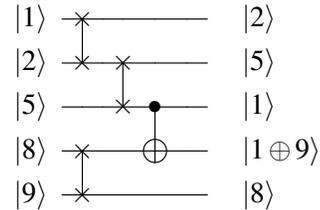
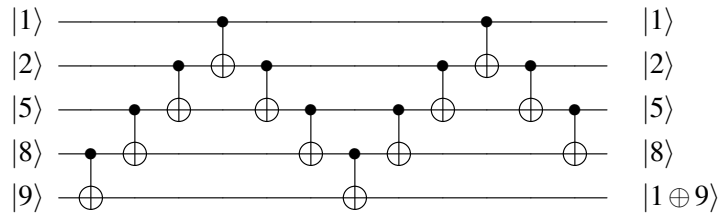

Alternatively, we can use Steiner-tree based synthesis~\cite{amy2018controlled, kissinger2020cnot, nash2020quantum,meijer-vandegriend2020architecture,vandaele2022phase,gheorghiu2022reducing} to find a generalized bridge gate for a CNOT circuit. The new sequence of CNOTs has the same effect on the logical states as the original CNOT circuit, and every gate is permitted by the hardware's connectivity constraint. This is done by changing the rigid representation of the quantum circuit to a more flexible one (\cref{subsec:paritymatrix}), with which we could synthesize a routed circuit (\cref{subsec:synthesis}). This way, we can make global improvements to the CNOT circuit re-synthesis more easily. By re-synthesis, we refer to the process of turning a CNOT circuit into a parity matrix and synthesizing an equivalent CNOT circuit from that parity matrix, up to permutation.

The \textit{Steiner-Gauss} algorithm provides the first Steiner-tree based synthesis approach and it is used to synthesize CNOT circuits~\cite{kissinger2020cnot, nash2020quantum}. Later on, it is used for synthesizing circuits over CNOT and $R_z$ gates~\cite{nash2020quantum, meijer-vandegriend2020architecture,vandaele2022phase}, and further for circuits over CNOT, $R_z$, and NOT gates. Finally, this algorithm is generalized for Clifford$+T$ circuits with the \textit{slice-and-build} procotol based on the locality of Hadamard gates in the circuits~\cite{gheorghiu2022reducing}. 

One major drawback of the existing Steiner-tree based methods is that they are not flexible with respect to the mapping of qubits. Logical qubits of the synthesized circuit will always be stored in the same physical qubit registers where they were originally allocated. However, this is not always optimal. Figure \ref{fig:swapTemplate} shows an example where there exists a smaller synthesized circuit by reallocating logical qubits to the physical registers (\cref{fig:swapTemplate}.c) than using \textit{Steiner-Gauss} (\cref{fig:swapTemplate}.d). Note that with the fixed qubit map, \textit{Steiner-Gauss} produces fewer CNOTs than the SWAP-based method (\cref{fig:swapTemplate}.b). Thus, we conclude that the synthesized circuit in \cref{fig:swapTemplate}.d contains implicit SWAP gates.

Moreover, remapping logical outputs of a quantum circuit to physical registers based on the original qubit mapping can be done by a classical operation. Hence, such operation is considered trivial in quantum computing. In other words, preparing a qubit in a register, routing it to a different register and measuring it does not influence the computation of a quantum circuit. Thus, it is desirable to have a synthesis procedure that permits dynamic qubit maps.

The first CNOT synthesis procedures~\cite{nash2020quantum,kissinger2020cnot} under topological constraints are based on \emph{Gaussian elimination}: the parity matrix representing the synthesized CNOT circuit is eliminated to the identity matrix (\cref{subsec:synthesis}). In \cref{subsec:permutation}, we show that dynamically changing the qubit maps is the same as eliminating the parity matrix into the identity matrix up to permutation. To do this, we need to determine to which permutation of the identity matrix to synthesize a priori, which is not a trivial task. However, by adjusting the algorithm \textit{RowCol}~\cite{wu2023optimization}, we can determine the new qubit map whilst synthesizing a CNOT circuit.

Here, we propose the algorithm \textit{PermRowCol}: a new Steiner-tree based synthesis method for CNOT circuits re-synthesis under topological constraints. It dynamically determines the output qubit maps. This method could be generalized to synthesizing an arbitrary quantum circuit, as articulated in \cref{sec:extension}.

The paper is structured as follows. In \cref{sec:background}, we introduce the Steiner-tree based synthesis approach. In \cref{sec:methods}, we describe our algorithm \textit{PermRowCol}. In \cref{sec:results}, we show how well it performs against \textit{Steiner-Gauss}~\cite{kissinger2020cnot} and \textit{RowCol}~\cite{wu2023optimization}. In \cref{sec:conclusion}, we discuss the results and other open problems. Finally, we explain some ongoing improvements for \textit{PermRowCol} in \cref{sec:future}. Looking forward, our goal is to fairly compare \textit{PermRowCol} with CNOT re-synthesis algorithms in quantum compilers such as SABRE~\cite{li2019tackling}, Qiskit~\cite{qiskit}, and TKET~\cite{sivarajah2020tket}.

\section{Preliminaries}
\label{sec:background}

Here we introduce the core concepts required to understand the proposed algorithm \textit{PermRowCol}. In \cref{subsec:paritymatrix}, we define the matrix representation of a $\CNOT$ circuit. In \cref{sec:steinertree}, we describe the concepts of Steiner tree. In \cref{subsec:synthesis}, we use it for synthesizing a CNOT circuit under a constrained topology. In \cref{subsec:permutation}, we explain the core idea behind algorithm \textit{PermRowCol}: the dynamic mapping of qubits using permutation matrices.

\subsection{The parity matrix of a CNOT circuit}
\label{subsec:paritymatrix}

In this paper, we consider circuits composed of only CNOTs, and call them \textit{CNOT circuits}. $\CNOT$ is short for "controlled not". It acts on two qubits: a control and a target. We write $\CNOT(c,t)$ to denote a CNOT applied between a control qubit $c$ and a target qubit $t$. The control qubit $c$ decides whether a NOT gate is applied to the target qubit $t$. When $\ket{c} = \ket{0}$, $\CNOT(c,t)$ acts trivially on $\ket{t}$, leaving it unchanged. Otherwise, $\ket{t} = \ket{0}$ is changed to $\ket{t} = \ket{1}$ and vice versa. Alternatively, we write that $\CNOT(c,t)$ changes $\ket{t}$ to $\ket{c \oplus t}$, where $\oplus$ denotes addition modulo 2.

\begin{figure}[!ht]
\renewcommand{\thefigure}{(a)}
\begin{subfigure}{.5\textwidth}
  \centering
  \[
\Qcircuit @C=1em @R=1.02em @!R {
          \lstick{\ket{1}} & \qw & \ctrl{1} & \qw & \targ & \qw & \qw & \rstick{\ket{2}}\\
          \lstick{\ket{2}} & \qw & \targ & \ctrl{1} & \ctrl{-1} & \qw & \qw & \rstick{\ket{1 \oplus 2}}\\
          \lstick{\ket{3}} & \ctrl{1} & \qw & \targ & \qw & \ctrl{1} & \qw & \rstick{\ket{1 \oplus 2 \oplus 3}}\\
          \lstick{\ket{4}} & \targ & \qw & \qw & \qw & \targ & \qw &  \rstick{\ket{1 \oplus 2 \oplus 4}}
  }
\]
\vspace{.1 cm}
  \caption{A CNOT circuit acting on $4$ qubits.}
  \label{fig:4qcircuit}
\end{subfigure}%
\renewcommand{\thefigure}{(b)}
\begin{subfigure}{.5\textwidth}
  \centering
\[
\vect{A} = \begin{blockarray}{ccccc}
 & 1' & 2' & 3' & 4' \\
\begin{block}{c(cccc)}
  1 & 0 & 1 & 1 & 1 \\
  2 & 1 & 1 & 1 & 1 \\
  3 & 0 & 0 & 1 & 0 \\
  4 & 0 & 0 & 0 & 1 \\
\end{block}
\end{blockarray}
 \]
  \caption{Parity matrix for figure (a).}
  \label{fig:paritymatrix}
\end{subfigure}
\renewcommand{\thefigure}{3}
\caption{The matrix representation of a $4$-qubit $\CNOT$ circuit.}
\label{fig:parityexample}
\end{figure}
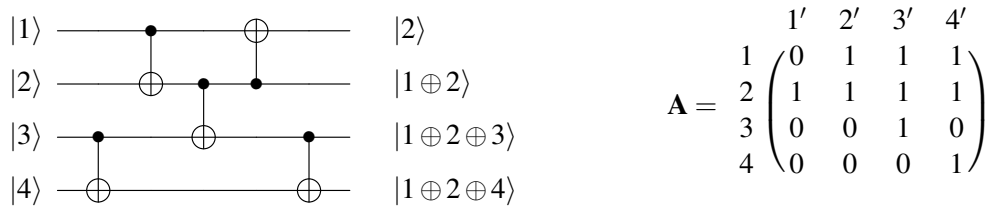

For a CNOT circuit, we can keep track of its state evolution by checking which qubits appear in the summation modulo $2$ at the circuit output. In \cref{fig:parityexample}.a, there are $5$ $\CNOT$s acting on $4$ qubits, whose overall behaviour is described by the sum of some logical qubits on each output wire. We call such a sum a \textit{parity term} because it keeps track of whether a logical qubit participates in the sum or not. As such, we can write a parity term as a binary string whose length is equal to the number of qubits in the circuit. In this representation, a $0$ means that the corresponding logical qubit is not present in the sum, and a $1$ means otherwise.

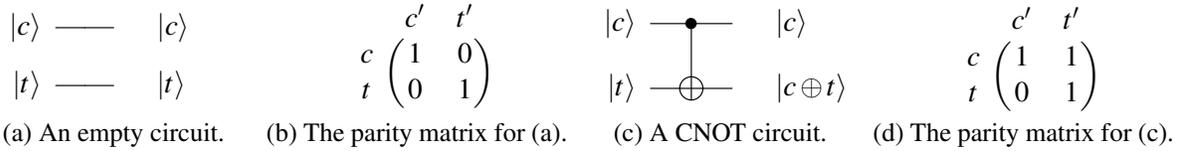
\begin{figure}[t!]
\renewcommand{\thefigure}{(a)}
\begin{subfigure}{.23\textwidth}
\[
\Qcircuit @C=1em @R=1em @!R {
          \lstick{\ket{c}} & \qw & \qw & \rstick{\ket{c}}\\
          & & & & \\
          \lstick{\ket{t}} & \qw & \qw & \rstick{\ket{t}}
  } 
\]
\caption{An empty circuit.}
\end{subfigure}
  \renewcommand{\thefigure}{(b)}
  \begin{subfigure}{.25\textwidth}
\[
\begin{blockarray}{ccc}
& c' & t' \\
\begin{block}{c(cc)}
  c & 1 & 0 \\
  t & 0 & 1 \\
\end{block}
\end{blockarray}
\]
\vspace{-1.9 em}
\caption{The parity matrix for (a).}
\end{subfigure}
\renewcommand{\thefigure}{(c)}
\begin{subfigure}{.23\textwidth}
\[
\Qcircuit @C=1em @R=.3em @!R {
          \lstick{\ket{c}} & \ctrl{2} & \qw & \rstick{\ket{c}}\\
          & & & & \\
          \lstick{\ket{t}} & \targ & \qw & \rstick{\ket{c \oplus t}}
  }
\]
\vspace{-.6 em}
\caption{A CNOT circuit.}
\end{subfigure}
  \renewcommand{\thefigure}{(d)}
  \begin{subfigure}{.25\textwidth}
\[
\begin{blockarray}{ccc}
& c' & t' \\
\begin{block}{c(cc)}
  c & 1 & 1 \\
  t & 0 & 1 \\
\end{block}
\end{blockarray}
\]
\vspace{-2 em}
\caption{The parity matrix for (c).}
\end{subfigure}
  \renewcommand{\thefigure}{4}\caption{Example constructions of parity matrices.}
\label{fig:parityproperty} 
\end{figure}

We can use the output parities of a CNOT circuit to create a square matrix where each column represents a parity term and each row represents an input qubit. This matrix is called a \textit{parity matrix} and its properties are well-studied in \cite{alber2001quantum,shende2003synthesis,2008PMH}. \cref{fig:parityproperty} shows two examples of constructing the parity matrix for the given CNOT circuits. Additionally, adding a CNOT to the circuit corresponds to a row operation on the parity matrix. That is, adding the row indexed by the target to the row indexed by the control, while keeping the target-indexed row unchanged (\cref{fig:convention}.b).

This means that we can extract a CNOT circuit from a parity matrix by adding rows of the parity matrix to other rows until we obtain the identity matrix. In the literature, Gaussian elimination is a commonly used algorithm for CNOT circuit synthesis~\cite{2008PMH,kissinger2020cnot,wu2023optimization,gheorghiu2022reducing}. Here, we refer readers to textbooks on linear algebra for more details about Gaussian elimination.

\subsection{Steiner tree}
\label{sec:steinertree}

We use Steiner trees to enforce the connectivity constraints when synthesizing a semantically equivalent CNOT circuit from the parity matrix. Note that Steiner trees are not the only approach to carry out the architecture-aware CNOT circuit synthesis procedure (e.g., see~\cite{2020dBBVMA} for an alternative), but it is the method we use here.

We start with the basics; a \textit{graph} is an order pair $G=(V_G,E_G)$, where $V_G$ is a set of \emph{vertices} and $E_G$ is a set of edges. Each \emph{edge} is defined as $e=(u,v)$ where $u,v\in V_G$. The \textit{degree} of a vertex is the number of edges that are incident to that vertex. Graphs also have a \textit{weight} assigned to each edge by a weight function $\omega_E: E_G \rightarrow \real$. 

The connectivity graph of a quantum computer is generally considered a \textit{simple} graph, 
meaning that it is an \emph{undirected} graph (i.e., $(u,v) \equiv (v,u)$) with all edge weights equal to $1$. It has at most one edge between two distinct vertices (i.e. $\forall (e,e')\in E_G: e\neq e'$), and no self-loops (i.e., $(u,u)\notin E_G$). 

Some graphs are \textit{connected}, meaning that for every vertex there exists a sequence of edges (a \textit{path}) from which we can go from that vertex to any other vertex in the graph. A graph that is not connected is \textit{disconnected}. The topology of a quantum computer needs to be connected if we want any pair of arbitrary qubits to interact with each other. A \textit{cut vertex} is a vertex that when removed, the graph will become disconnected. We use \textit{non-cutting vertex} to mean a vertex that is not a cut vertex.

A \textit{subgraph} $G'=(V_G^{'},E_G^{'})$ of $G$ is a graph that is wholly contained in $G$ such that $V_G^{'}\subseteq V_G$, $E_G^{'}\subseteq E_G$, and for all $(u,v)\in E_G^{'}$, $u, v\in V_G^{'}$. 

A \textit{tree} is an undirected connected graph that has no path which starts and ends at the same vertex. A tree is \textit{acyclic}. A \textit{minimum spanning tree} $T$ of a connected graph $G$ is a subgraph of $G$ with the same set of vertices $V_G$ and a subset of the edges $E_G$ such that the sum of the edge weights is minimal and $T$ is still connected. 

A Steiner tree is similar to a minimum spanning tree. It is defined as follows.
\begin{definition}[\textbf{Steiner tree}]
Given a graph $G=(V_G,E_G)$ with a weight function $\omega_E$ and a set of vertices $S \subseteq V_G$, a Steiner tree $T=(V_T,E_T)$ is a tree that is a subgraph of $G$ such that $S \subseteq V_T$ and the sum of edge weights in $E_T$ is minimized. The vertices in $S$ are called \emph{terminals} while those in $V_T\setminus S$ are called \emph{Steiner nodes}.
 \label{defn:steinerTree}
\end{definition}

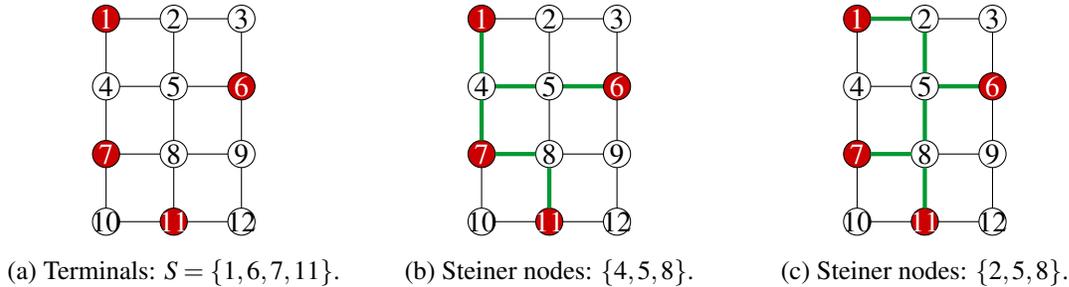
\begin{figure}[!ht]
\renewcommand{\thefigure}{(a)}
\begin{subfigure}{.3\textwidth}
 \centering
\begin{tikzpicture}[scale=0.9]
 \draw [fill=IMSRed] (1,1) circle [radius=0.2];
  \node [white] at (1,1) {$1$};
  \draw (1.2,1)--(1.8,1); 
  \draw (1.2,0)--(1.8,0); 
  \draw (2,1) circle [radius=0.2];
  \node at (2,1) {$2$};
  \draw (2.2,1)--(2.8,1); 
  \draw (2.2,0)--(2.8,0); 
  \draw (3,1) circle [radius=0.2];
  \node at (3,1) {$3$};
  \draw (1,0) circle [radius=0.2];
  \node at (1,0) {$4$};
  \draw (2,0) circle [radius=0.2];
  \node at (2,0) {$5$};
  \draw [fill=IMSRed] (3,0) circle [radius=0.2];
  \node [white] at (3,0) {$6$};
  \draw (1,0.8)--(1,0.2);    
  \draw (3,0.8)--(3,0.2);    
  \draw (2,0.8)--(2,0.2); 
  \draw [fill=IMSRed] (1,-1) circle [radius=0.2];
  \node [white] at (1,-1) {$7$};
  \draw (1,-.2)--(1,-.8);    
  \node at (2,-1) {$8$};
  \draw (2,-1) circle [radius=0.2];
  \draw (1.2,-1)--(1.8,-1);
  \draw (2,-.2)--(2,-.8);    
  \node at (3,-1) {$9$};
  \draw (3,-.2)--(3,-.8);    
  \draw (3,-1) circle [radius=0.2];
  \node at (1,-2) {$10$};
  \draw (1,-2) circle [radius=0.2];
  \draw (1,-1.2)--(1,-1.8);    
  \draw [fill=IMSRed] (2,-2) circle [radius=0.2];
  \node [white]at (2,-2) {$11$};
  \draw (2,-1.2)--(2,-1.8);    
  \draw (2.2,-1)--(2.8,-1);    
  \node at (3,-2) {$12$};
  \draw (3,-2) circle [radius=0.2];
  \draw (3,-1.2)--(3,-1.8);    
  \draw (1.2,-2)--(1.8,-2);    
  \draw (2.2,-2)--(2.8,-2);    
\end{tikzpicture}
\caption{Terminals: $S= \{1,6,7,11\}$.}
\label{fig5:a}
\end{subfigure}
  \renewcommand{\thefigure}{(b)}
  \begin{subfigure}{.3\textwidth}
  \centering
\begin{tikzpicture}[scale=0.9]
 \draw [fill=IMSRed] (1,1) circle [radius=0.2];
  \node [white] at (1,1) {$1$};
  \draw (1.2,1)--(1.8,1); 
  \draw [ultra thick,IMSGreen](1.2,0)--(1.8,0); 
  \draw (2,1) circle [radius=0.2];
  \node at (2,1) {$2$};
  \draw (2.2,1)--(2.8,1); 
  \draw [ultra thick,IMSGreen](2.2,0)--(2.8,0); 
  \draw (3,1) circle [radius=0.2];
  \node at (3,1) {$3$};
  \draw (1,0) circle [radius=0.2];
  \node at (1,0) {$4$};
  \draw (2,0) circle [radius=0.2];
  \node at (2,0) {$5$};
  \draw [fill=IMSRed] (3,0) circle [radius=0.2];
  \node [white] at (3,0) {$6$};
  \draw [ultra thick,IMSGreen](1,0.8)--(1,0.2);    
  \draw (3,0.8)--(3,0.2);    
  \draw (2,0.8)--(2,0.2); 
  \draw [fill=IMSRed] (1,-1) circle [radius=0.2];
  \node [white] at (1,-1) {$7$};
  \draw [ultra thick,IMSGreen](1,-.2)--(1,-.8);    
  \node at (2,-1) {$8$};
  \draw (2,-1) circle [radius=0.2];
  \draw [ultra thick,IMSGreen](1.2,-1)--(1.8,-1);
  \draw (2,-.2)--(2,-.8);    
  \node at (3,-1) {$9$};
  \draw (3,-.2)--(3,-.8);    
  \draw (3,-1) circle [radius=0.2];
  \node at (1,-2) {$10$};
  \draw (1,-2) circle [radius=0.2];
  \draw (1,-1.2)--(1,-1.8);    
  \draw [fill=IMSRed] (2,-2) circle [radius=0.2];
  \node [white]at (2,-2) {$11$};
  \draw [ultra thick,IMSGreen](2,-1.2)--(2,-1.8);    
  \draw (2.2,-1)--(2.8,-1);    
  \node at (3,-2) {$12$};
  \draw (3,-2) circle [radius=0.2];
  \draw (3,-1.2)--(3,-1.8);    
  \draw (1.2,-2)--(1.8,-2);    
  \draw (2.2,-2)--(2.8,-2);    
\end{tikzpicture}
\caption{Steiner nodes: $\{4,5,8\}$.}
\label{fig5:b}
\end{subfigure}
 \renewcommand{\thefigure}{(c)}
  \begin{subfigure}{.3\textwidth}
  \centering
\begin{tikzpicture}[scale=0.9]
 \draw [fill=IMSRed] (1,1) circle [radius=0.2];
  \node [white] at (1,1) {$1$};
  \draw [ultra thick,IMSGreen](1.2,1)--(1.8,1); 
  \draw (1.2,0)--(1.8,0); 
  \draw (2,1) circle [radius=0.2];
  \node at (2,1) {$2$};
  \draw (2.2,1)--(2.8,1); 
  \draw [ultra thick,IMSGreen](2.2,0)--(2.8,0); 
  \draw (3,1) circle [radius=0.2];
  \node at (3,1) {$3$};
  \draw (1,0) circle [radius=0.2];
  \node at (1,0) {$4$};
  \draw (2,0) circle [radius=0.2];
  \node at (2,0) {$5$};
  \draw [fill=IMSRed] (3,0) circle [radius=0.2];
  \node [white] at (3,0) {$6$};
  \draw (1,0.8)--(1,0.2);    
  \draw (3,0.8)--(3,0.2);    
  \draw [ultra thick,IMSGreen](2,0.8)--(2,0.2); 
  \draw [fill=IMSRed] (1,-1) circle [radius=0.2];
  \node [white] at (1,-1) {$7$};
  \draw (1,-.2)--(1,-.8);    
  \node at (2,-1) {$8$};
  \draw (2,-1) circle [radius=0.2];
  \draw [ultra thick,IMSGreen](1.2,-1)--(1.8,-1);
  \draw [ultra thick,IMSGreen](2,-.2)--(2,-.8);    
  \node at (3,-1) {$9$};
  \draw (3,-.2)--(3,-.8);    
  \draw (3,-1) circle [radius=0.2];
  \node at (1,-2) {$10$};
  \draw (1,-2) circle [radius=0.2];
  \draw (1,-1.2)--(1,-1.8);    
  \draw [fill=IMSRed] (2,-2) circle [radius=0.2];
  \node [white]at (2,-2) {$11$};
  \draw [ultra thick,IMSGreen](2,-1.2)--(2,-1.8);    
  \draw (2.2,-1)--(2.8,-1);    
  \node at (3,-2) {$12$};
  \draw (3,-2) circle [radius=0.2];
  \draw (3,-1.2)--(3,-1.8);    
  \draw (1.2,-2)--(1.8,-2);    
  \draw (2.2,-2)--(2.8,-2);    
\end{tikzpicture}
\caption{Steiner nodes: $\{2,5,8\}$.}
\label{fig5:c}
\end{subfigure}
  \renewcommand{\thefigure}{5}\caption{Solutions to the Steiner tree problem on the $12$-qubit grid $G$.}
\label{fig:steiner} 
\end{figure}

\cref{fig:steiner} demonstrates a solution to the Steiner tree problem on a $12$-qubit grid, with $S = \{1,6,7,11\}$. In \cref{fig:steiner}.a, nodes in $S$ are coloured in red. The edges of the Steiner tree $T$ are highlighted in green and the Steiner nodes can be read off from the graph. For example, in \cref{fig:steiner}.b, they are the unfilled node on the paths of $T$: $V_{T}\setminus S =\{4,5,8\}$. Note that the solution to a Steiner tree problem may not be unique. For the graph $G$ and a set of terminals $S$ in \cref{fig:steiner}.a, \cref{fig:steiner}.b and \cref{fig:steiner}.c provide two solutions. In this example and the remainder of this paper, we assume each edge is of unit weight without loss of generality. 

Computing Steiner trees is NP-hard and the related decision problem is NP-complete~\cite{1972K}. There are a number of heuristic algorithms that compute approximate Steiner trees~\cite{2005RZ, 2013BGRS, 1992HR}. There is a trade-off between the size of the approximate Steiner tree and the algorithm's runtime, so the choice of algorithm is determined by its application. 
Here, we create an approximate Steiner tree by building a minimum spanning tree over the terminals (we use Prim's algorithm~\cite{cormen2022introduction}), with the weights being the distance between the terminals. Whenever a terminal is added to the spanning tree, its entire path to the tree is added. The weights calculated in the next step involve the paths between the tree up until then and terminals not yet added to the spanning tree. For this, we use Floyd-Warshall's algorithm~\cite{cormen2022introduction} to calculate the shortest path between all qubits.

\subsection{Synthesizing CNOT circuits for specific topologies}
\label{subsec:synthesis}
Given the parity matrix of a CNOT circuit, we want to synthesize an equivalent CNOT circuit such that all CNOTs are allowed according to a connectivity graph. From \cref{subsec:paritymatrix}, we know that every $\CNOT$ corresponds to a row operation in the parity matrix. Moreover, we can use Gaussian elimination to turn the parity matrix into the identity matrix. The process of adding rows are called \textit{elimination}. If we keep track of which row operations are performed during the Gaussian elimination process, we obtain a CNOT circuit that is \emph{semantically equivalent}, which means that the parity matrix of the original circuit is equal to that of the synthesized circuit. Hence, these two circuits have the same input-output behaviour.

For the extracted CNOTs to adhere to the given connectivity constraints, we need to adjust our method to allow only row operations that correspond to the connected vertices in the topology. There are several methods to do this~\cite{kissinger2020cnot,nash2020quantum,2008PMH}. Our algorithm is based on the algorithm \textit{RowCol} \cite{wu2023optimization}, which reduces the input parity matrix into the identity matrix by eliminating a column and a row at each round of execution. 

It starts by selecting a qubit (i.e., a vertex in the connectivity graph). This determines the \emph{pivot column}, which is the column to be eliminated. To ensure the updated topology stays connected, the selected qubit must correspond to a non-cutting vertex. This means we can select qubits in an arbitrary order as long as the vertex removal does not disconnect the graph. Then we build Steiner trees to find the shortest paths over which the row additions are performed. As a result, the eliminations send the pivot column and a corresponding row to the basis vectors, after which they are removed from the parity matrix. Accordingly, the selected qubit is no longer needed and the vertex is removed from the graph. Next, the algorithm starts on a smaller instance of the problem. In what follows, we explain the column and row eliminations in more details.

To eliminate a column, we identify all 1s in the column. Then a Steiner tree $T$ is built with the diagonal as the root and the rows with a 1 as terminals. These rows should be added together so that we end up with an column equal to a basis vector. Due to the connectivity constraints, we use the Steiner nodes to "move" the $1$s to the terminals. This is done by traversing $T$ from the bottom up. When a Steiner node is reached, we add its child to itself, so the corresponding row in the pivot column is equal to $1$. Then, $T$ is traversed again, during which each row is added to its child in the Steiner tree. As a result, only the root row will be $1$, while other rows will be $0$ in the pivot column. Thus, the pivot column is eliminated. This procedure is also described by Algorithm \ref{alg:eliminatecolumn} in \cref{app:subroutines}.

A row is eliminated in a similar manner. For brevity, we call it a \emph{pivot row}. Compared to the column elimination, it is less straightforward to find which row operations to perform in order to eliminate the pivot row. According to \cite{wu2023optimization}, they are determined by solving a system of linear equations defined by the parity matrix. We can once more build a Steiner tree $T'$ with the diagonal as root and the rows added the pivot row as terminals. Then we traverse $T'$ top down from the root, adding every Steiner node to its parent. Next, we traverse $T'$ bottom up and add every node to its parent. As a result, the rows corresponding to the Steiner nodes are added twice to their parent. They do not participate in the sum since the row addition is performed modulo 2. Moreover, every terminal is added together and propagated to the root after the bottom-up traversal. This procedure is described as by Algorithm \ref{alg:eliminaterow} in \cref{app:subroutines}.


\subsection{Dynamic qubit mapping with permutation matrices}
\label{subsec:permutation}
The state-of-the-art CNOT circuit synthesis eliminates the parity matrix to an identity matrix, which is essentially a qubit map where logical qubit $i$ is stored in the physical qubit register $i$. This means that the exact synthesis preserves the circuit's semantics and moves the logical qubits back to their original registers. However, to route CNOTs in a topologically-constrained quantum circuit, it might not be necessary to stick to the parity matrix faithfully. Moreover, restoring the logical values in each physical qubit register adds extra CNOTs to the synthesis results. For example, in \cref{fig:swapTemplate}, the synthesized circuit (\cref{fig:swapTemplate}.d) have less CNOTs if we allow dynamic qubit maps (\cref{fig:swapTemplate}.c). Additionally, if the CNOT synthesis is done as part of a slice-and-build approach where each subcircuit is synthesized locally~\cite{gheorghiu2022reducing}, the cost of keeping the logical qubits in the same physical qubit registers grows linearly with the number of slices.

Recall that in the parity matrix, the $i$th column stores the parity term being output from the $i$th qubit of the CNOT circuit, $1 \leq i \leq n$, where $n$ is the number of qubits in the circuit. In other words, the order of the columns corresponds to the different physical qubit registers where these parity terms are stored. If we change which parity term ends up on each physical qubit register, we can equivalently synthesize a CNOT circuit where the parity matrix has its columns reordered with respect to the original parity matrix. In the case of routing CNOTs, this is exactly possible. Reordering the columns of the identity matrix corresponds to reading the logical qubits from different quantum registers that are defined by the new column order. Since reading the circuit output from different registers is a classical operation, it can be considered free when running a quantum circuit.


In this work, we remove the restriction on qubit maps in CNOT circuit synthesis and propose an algorithm that eliminates the parity matrix to the \emph{permutation matrix}, which is an identity matrix with reordered columns. Accordingly, a parity matrix $M$ that is a permutation matrix can be seen as a \textit{qubit map} where the logical qubit $i$ is stored in the physical qubit register $j$ iff $M_{i,j}=1$. 

However, it is not trivial to determine an optimal qubit remap. In \cite{kissinger2020cnot}, a generic algorithm \textit{Steiner-Gauss} was proposed to find a better qubit map, but it is unfortunately unscalable. In the next section, we explain how to do this in a scalable way.

\subsection{Reverse traversal strategy}
\label{subsection:reversetraversal}
\textit{Reverse Traversal Strategy} (RT)~\cite{li2019tackling} leverages the reversibility of quantum circuits. We define a reverse circuit to be the circuit that is the original circuit in reverse. Since CNOTs are self-inverse, this is equivalent to the inverse circuit. 
Reverse Traversal iteratively improves the initial qubit mapping using any routing procedure that optimizes the circuit and remaps the qubits. It does this by using the output qubit mapping as input qubit mapping for the reverse circuit and reapplying the routing procedure. Because the routing procedure optimizes the circuit in the process, we might find a smaller circuit than what we started out with. Then, we can repeat this processes of routing the reverse-reverse circuit to obtain a new output mapping and possibly find a better circuit. We refer the reader to the orignal paper~\cite{li2019tackling} for a more detailed explanation with pictures.

This technique can only be used in a routing method that remaps the qubits. Therefore, RT could not be used for Steiner-tree based methods until now.

\section{The PermRowCol Algorithm}
\label{sec:methods}

Here, we introduce the algorithm \textit{PermRowCol}. It is a Steiner-tree based synthesis algorithm that dynamically remaps the logical qubits to physical registers while routing CNOTs. To this end, we build on the algorithm \textit{RowCol}~\cite{wu2023optimization} described in \cref{subsec:synthesis}. This algorithm is executed iteratively on the input parity matrix until it is eliminated to an identity. At each round, \textit{RowCol} picks a new logical qubit and eliminates the corresponding row and column such that they can be removed from the problem. 

Our adjustment is described by algorithm \ref{alg:permrowcol}, where we lift the restrictions for the row and column to be removed such that they don't necessarily intersect at the diagonal. Specifically, we pick the logical qubit corresponding to the row (i.e., the pivot row), and a column (i.e., the pivot column) to be the new register for that logical qubit. Then, we can eliminate both the pivot column and row through a sequence of row operations such that they become basis vectors. In the meantime, \textit{PermRowCol} uses subroutines described in \cref{app:subroutines}, whose behaviour is articulated below.

Given a parity matrix $M$ to synthesize over a connectivity graph $G$, the vertices of $G$ correspond to the numbering of rows in $M$. Before the synthesis, the original qubit map indicates that the logical qubit $i$ is stored in the physical qubit register $i$, which corresponds to the vertex $i$. 

At the start of each round, pick the logical qubit $i$ that we want to remove from the problem.  The only constraint is that it needs to be non-cutting for $G$. Among the non-cutting qubits, we use Algorithm \ref{alg:chooserow} to determine which row to eliminate. For our purpose, a simple heuristic is used: selecting the row with the lowest hamming weight in $M$. This gives us the pivot row $i$.


Next, we use Algorithm \ref{alg:choosecolumn} to choose the pivot column to eliminate. Suppose column $j$ is picked for row $i$. This means logical qubit $i$ will be stored in the physical qubit register $j$ after this round of elimination. Any column should work as long as it has a $1$ on row $i$ and does not have a qubit assigned to it yet. Among all the candidate columns, here we choose the column with the lowest hamming weight.


Note that in principle, we can choose any arbitrary row and column, as long as the vertex corresponding to the pivot row does not disconnect $G$ after it is removed. Moreover, the pivot row and column should not have been picked before. This means Algorithms \ref{alg:chooserow} and \ref{alg:choosecolumn} could be replaced by other heuristics so long as the above constraints are satisfied. We discuss the implications of these choices in \cref{sec:conclusion} and a possible improvement in \cref{sec:future}.

With the pivot row and column, we can add rows together such that the pivot row and column only have a 1 at their intersection in $M$, and $0$'s everywhere else. This means that they are eliminated. More precisely, in Algorithm \ref{alg:eliminatecolumn}, we start with the pivot column and gather all its rows that are non-zero. Let it be $S$. Then, build a Steiner tree with the pivot row as the root and the nodes in $S$ as terminals. Starting at the leafs, for each Steiner node in the tree, add its child to it. By construction, the Steiner nodes correspond to the rows that are zero in the pivot column. Thus, this method will make all Steiner nodes equal to 1. Next, traverse the tree again from the leafs to the root and add every parent to its child. This results in a matrix with all zeros in the pivot column except for the pivot row. 

Afterwards, Algorithm \ref{alg:eliminaterow} carries out a similar elimination process for the pivot row. First, we need to find which rows to add together such that the entire pivot row is filled with $0$'s except for the pivot column. This could be done by solving a system of linear equations defined by $M$. Let the set of rows to be added be $T$. Then, we can construct a new Steiner tree with the pivot row as the root and the rows in $T$ as terminals. Again we traverse the tree twice: once top-down while adding Steiner nodes to their parents, and once bottom-up while adding all nodes to their parents. As a result, all rows in $T$ are added to the pivot row, and thus it is turned into a basis vector. 

After column $j$ and row $i$ are eliminated to basis vectors, they are removed from the parity matrix. Accordingly, vertex $i$ is removed from the connectivity graph. We can update the output qubit map to indicate that logical qubit $i$ is now stored in the physical qubit register $j$ after this round of elimination. As discussed in \cref{subsec:synthesis}, Algorithms \ref{alg:eliminatecolumn} and \ref{alg:eliminaterow} are adapted from the algorithm \emph{RowCol} in \cite{wu2023optimization}.

In summary, Algorithm \ref{alg:permrowcol} constructs Steiner trees based on the rows of $M$ and generates CNOT based up row operations. Once a row in $M$ becomes an identity row (i.e. only contains a single 1), this row is eliminated and we can remove the corresponding vertex from the connectivity graph. Then we restart the algorithm on a smaller problem, until the updated graph has no vertex. This ensures the termination of the algorithm.


The time complexity of Algorithm \ref{alg:permrowcol}-\emph{PermRowCol} depends on the choice of heuristics in Algorithms \ref{alg:chooserow} and \ref{alg:choosecolumn}, as well as the choice of the (approximate) Steiner tree algorithm when eliminating a row and column in Algorithms \ref{alg:eliminatecolumn} and \ref{alg:eliminaterow}. The remaining time complexity is dominated by calculating the non-cutting vertices. Thus, given $N$ qubits and a connectivity graph with $E$ edges, the time complexity for \emph{PermRowCol} is as follows.

\scriptsize
\begin{alignat}{1}
    O(PermRowCol) = O(N)\big(&O(NonCuttingVertices) + O(ChooseRow) + O(ChooseColumn)\nonumber + O(EliminateColumn) + O(EliminateRow)\big) \nonumber\\
    \hspace{40pt} = O(N)\big(&O(NonCuttingVertices) + O(ChooseRow) + O(ChooseColumn)\nonumber + 2*O(Steiner tree)\big)\nonumber
\end{alignat}

\normalsize
The suggested \emph{ChooseRow} and \emph{ChooseColumn} heuristics both have time complexity $O(N^2)$, but these can be replaced by other heuristics with improved complexities. In the meantime, the complexity of the Steiner tree algorithm depends heavily on the choice of the algorithm. Building Steiner trees is NP-hard but the topologies being benchmarked are sparse enough that approximate methods perform well. The algorithm we use is based on Prim's and Floyd-Warshall's algorithms and it has time complexity $O(N^3)$. Thus, the time complexity for our specific implementation of \emph{PermRowCol} is

$$ O(PermRowCol) = O(N)\big(O(N^2+EN) + 2*O(N^2) + 2*O(N^3)\big) = O(N^4).$$

\SetKwComment{Comment}{/* }{ */}
\SetKwComment{comment}{// }{}
\SetKwInOut{Input}{Input}
\SetKwInOut{Output}{Output}
\RestyleAlgo{ruled}
\begin{algorithm}[!ht]
\caption{\textit{PermRowCol}}
\label{alg:permrowcol}
\Input{Parity matrix $M$ and topology $G(V_G, E_G)$ with labels corresponding to the rows of $M$}
\Output{CNOT circuit $C$ and output qubit map $P$}
$P \gets [-1\dots -1]$ \Comment*[r]{$|V_G|$ times}
$C \gets $ New empty circuit\;
\While{$|V_G| > 1$}{
    \comment{Find non-cutting qubits $Vs$ that can (still) be removed $G$.}
    $Vs \gets NonCuttingVertices(G)$\; 
    $r \gets \textbf{ChooseRow}(Vs, M)$\;
    \comment{Choose a physical qubit register to map $r$ to.}
    $c \gets \textbf{ChooseColumn}(M, r, [i:i\in [1\dots |P|]$ where $P[i]=-1])$\;
    $Nodes \gets [i:i\in V_G$ where $M_{i,c}=1 ]$\;
    $C.add(\textbf{EliminateColumn}(M, G, r, Nodes))$\;
    \comment{Reduce the row if it is not yet eliminated.}
    \If{$\sum_{j\in 1\dots |P|} M_{r,j} > 1$}{ 
    $A \gets M$ without row $r$ and without column $c$\;
    $B \gets M[r]$ without column $c$\;
    $X \gets A^{-1}B$ \Comment*[r]{Find rows to add to eliminate row $r$}
    $Nodes \gets [i : i \in V_G$ where $i=r$ or $X[Index(i)]=1]$\;
    $C.add(\textbf{EliminateRow}(M, G, r, Nodes))$\;
    }
    \comment{Update the output qubit map.}
    $P[c] \gets r$\;
    $G \gets $ subgraph of $G$ with vertex $r$ and connecting edges removed\;
  }
  \comment{The loop ends with 1 qubit in $G$.}
  \comment{Update the map for the last output qubit.}
  $i \gets [i:i\in [1\dots |P|]$ where $P[i]=-1]$\Comment*[r]{Find index with -1}
  $P[i] \gets V_G[0]$\;
  \Return{$C, P$}
\end{algorithm}

\section{Benchmarking Results}
\label{sec:results}

We benchmark our algorithm against \textit{Steiner-Gauss}~\cite{kissinger2020cnot} and \textit{RowCol}~\cite{wu2023optimization} to demonstrate the advantage of dynamically remapping qubits during synthesis, both with and without \textit{Reverse Traversal} (RT). To do this, we generate CNOT circuits with $q$ qubits and $d$ CNOTs that are sampled uniformly at random. The final dataset consists of 100 such CNOT circuits per $(q,d)$-pair. 
The implementation of our proposed algorithm, the benchmark algorithms, the CNOT circuit generation script, unit tests, as well as the dataset of random CNOT circuits can be found on GitHub\footnote{\url{https://github.com/Aerylia/pyzx/tree/rowcol}}. The specific script that ran our experiments can also be found there\footnote{\url{https://github.com/Aerylia/pyzx/blob/rowcol/demos/PermRowCol\%20results.ipynb}}.

\begin{figure}[!ht]
     \centering
     \begin{subfigure}[b]{0.49\textwidth}
         \centering
         \includegraphics[width=\textwidth]{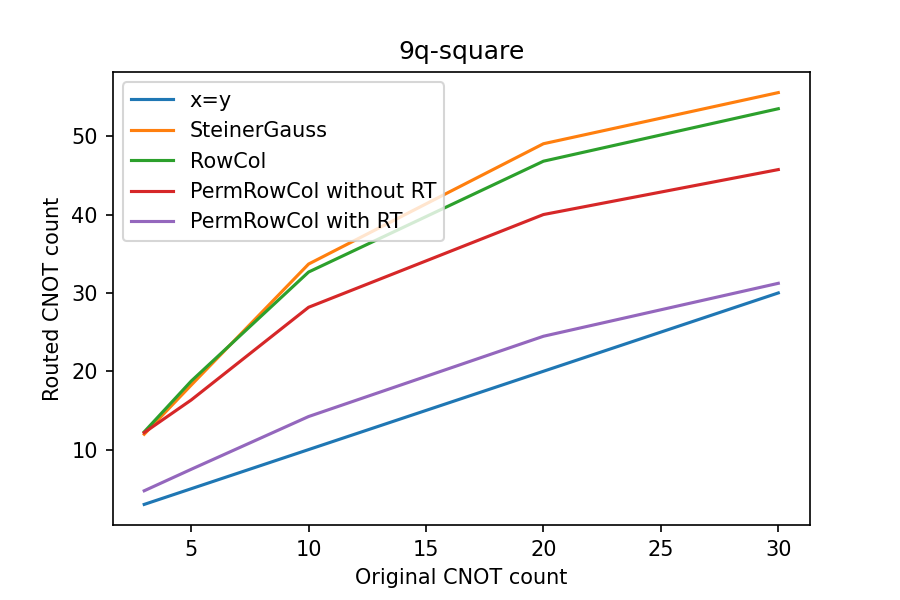}
         \label{fig:9qsquare}
     \end{subfigure}
     \hfill
     \begin{subfigure}[b]{0.49\textwidth}
         \centering
         \includegraphics[width=\textwidth]{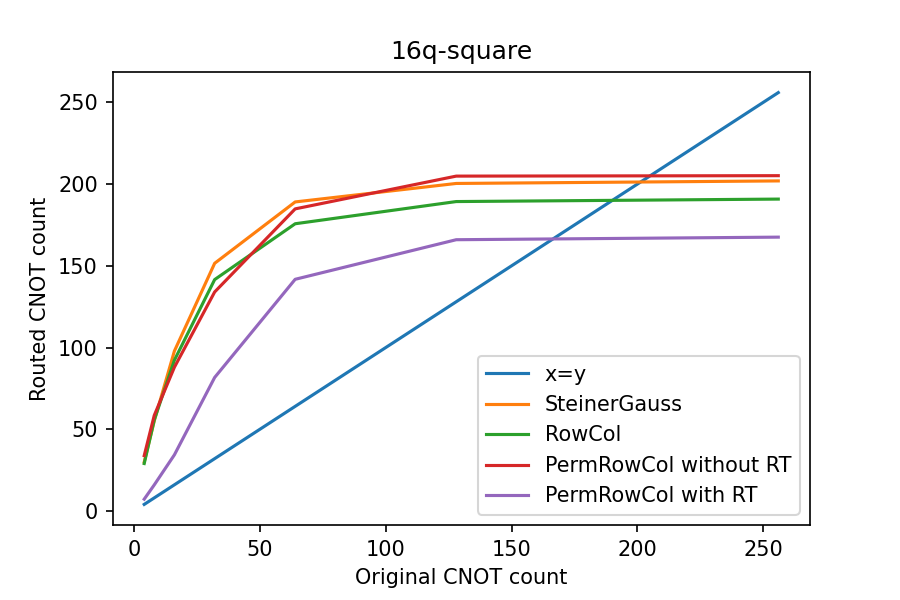}
         \label{fig:16qsquare}
     \end{subfigure}
     \vspace{-1 em}
        \caption{Compare the number of CNOTs generated by \textit{Steiner-Gauss}~\cite{kissinger2020cnot} (orange), \textit{RowCol}~\cite{wu2023optimization} (green), and \textit{PermRowCol} without \textit{Reverse Traversal} (RT) (red) and with RT (purple) for different fictitious square grid topologies: $9$-qubit square grid (left) and $16$-qubit square grid (right). The blue line $x=y$ serves as the baseline to compare the routing overhead of different algorithms. If a point is above the blue line, the routed circuit requires more CNOTs than the original circuit. If a point is below the blue line, then otherwise.}
    \label{fig:faketopogies}
\end{figure}

\begin{figure}[!ht]
     \centering
     \begin{subfigure}[b]{0.49\textwidth}
         \centering
         \includegraphics[width=\textwidth]{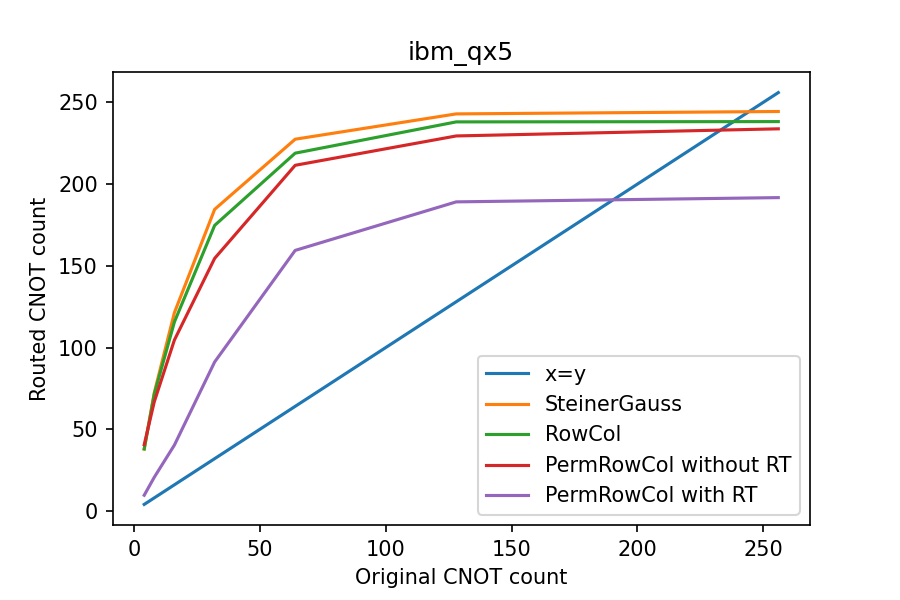}
         \label{fig:qx5}
     \end{subfigure}
     \hfill
     \begin{subfigure}[b]{0.49\textwidth}
         \centering
         \includegraphics[width=\textwidth]{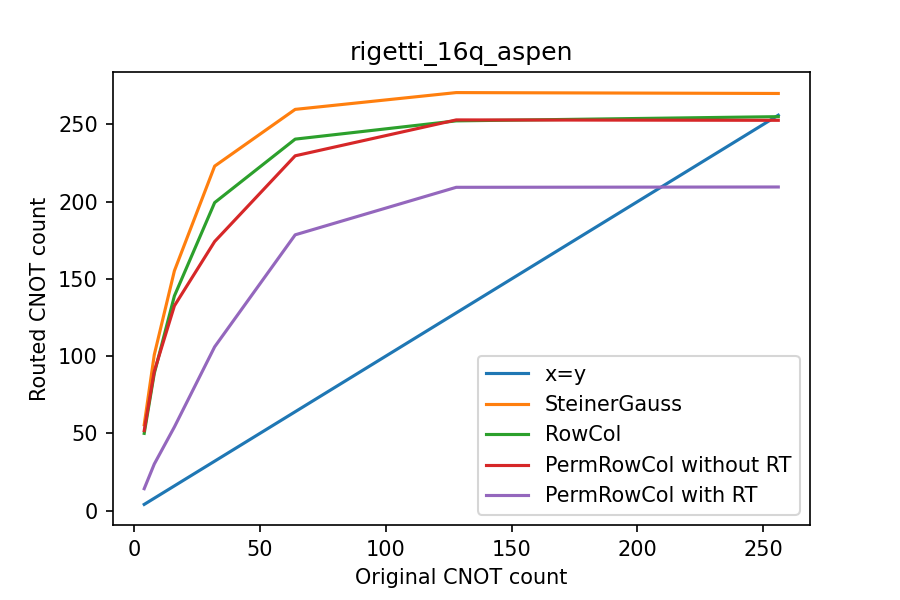}
         \label{fig:aspen}
     \end{subfigure}
     \hfill
     \begin{subfigure}[b]{.6\textwidth}
         \centering
         \includegraphics[width=\textwidth]{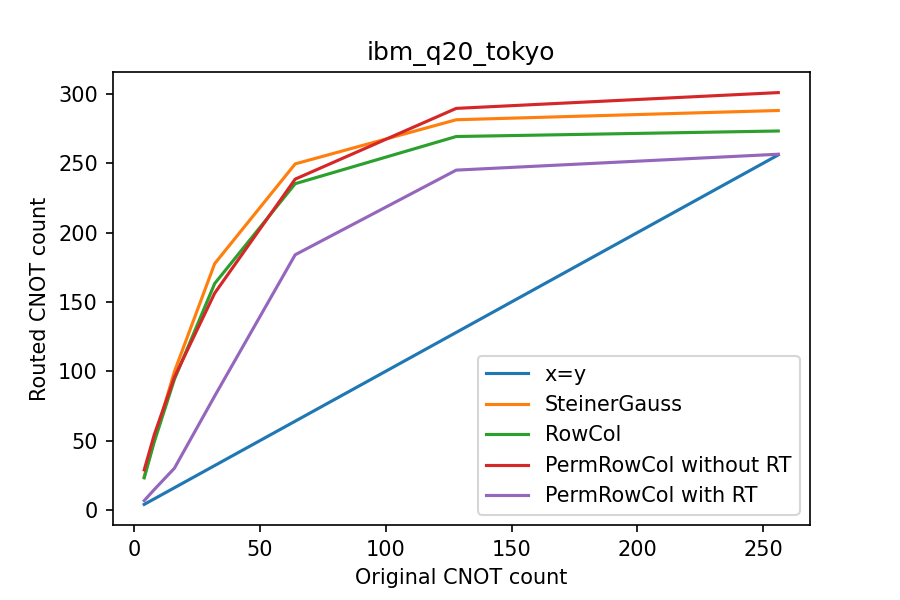}
         \label{fig:tokyo}
     \end{subfigure}
      \vspace{-1 em}
        \caption{Compare the number of CNOTs generated by \textit{Steiner-Gauss}~\cite{kissinger2020cnot} (orange), \textit{RowCol}~\cite{wu2023optimization} (green), and \textit{PermRowCol} without \textit{Reverse Traversal} (RT) (red) and with RT (purple) for different real topologies: $16$-qubit IBM QX5 (left), $16$-qubit Rigetti Aspen (right), and $20$-qubit IBM Tokyo (middle). The blue line $x=y$ serves as the baseline to compare the routing overhead of different algorithms. If a point is above the blue line, the routed circuit requires more CNOTs than the original circuit. If a point is below the blue line, then otherwise.}
        \label{fig:realtopogies}
\end{figure}

The number of CNOTs in the routed circuit versus the number of CNOTs in the original circuit are plotted in \cref{fig:faketopogies,fig:realtopogies,fig:unrestricted,fig:A*results}. The blue line $x=y$ serves as the baseline to compare the routing overhead of different algorithms. If a point is above the blue line, the routed circuit requires more CNOTs than the original circuit. This is expected in practice, but we want as few CNOTs as possible. On the other hand, when the original CNOT circuit has many CNOTs, it is possible for some algorithm to synthesize a circuit with fewer CNOTs than that of the original circuit. This implies that the original circuit contains redundant CNOTs which are removed by the algorithm. This happens because after a certain amount of CNOTs, the parity matrix representing the circuit becomes a random matrix and synthesizing a random parity matrix requires a constant amount of CNOTs, as discussed in \cite{2008PMH}.

\cref{fig:faketopogies} compares the performance of the three algorithms on two fictitious square grid topologies, whereas \cref{fig:realtopogies} compares their performance on three real device topologies. The corresponding connectivity graphs are shown in \cref{fig:graphs}. Overall, the proposed \textit{PermRowCol} algorithm outperforms the other algorithms when the CNOT count is small. For a large number of CNOTs, it depends on the topology whether the proposed algorithm still outperforms the others. Possible reasons for this are discussed in \cref{sec:conclusion}.

When the algorithm \textit{PermRowCol} is combined with the \textit{Reverse Traversal} (RT) strategy, it outperforms the other algorithms on all architectures, as shown by the purple line in \cref{fig:faketopogies,fig:realtopogies}. Note that because algorithms \emph{SteinerGauss} and \emph{RowCol} does not remap the qubits, RT would have no effect on them.



\section{Discussion and Conclusion}
\label{sec:conclusion}

In this paper, we propose the algorithm \emph{PermRowCol} that synthesizes CNOT circuits from a parity matrix while respecting the connectivity constraints of a quantum computer. It dynamically remaps logical qubits to different physical qubit registers. This technique is designed for the global optimization of quantum circuits. Therefore, our technique can add improvements that cannot be found with local optimization methods. Our work is based on the observation that allowing dynamic qubit maps during the circuit synthesis may reduce the output CNOT count. We provide a recipe to construct the improved algorithm whose performance has been confirmed by the benchmarking results. We show that in some but not all cases, the qubit remap results in a smaller $\CNOT$ overhead compared to the other state-of-the-art algorithms. Moreover, adding the Reverse Traversal (RT) technique improves the results noticeably.

Looking ahead, many problems remain open. For example, in some cases, the \textit{PermRowCol} performs worse than the original \textit{RowCol}, which differs from the \textit{PermRowCol} because we pick a different row and column to eliminate rather than the ones intersecting at the diagonal. This flexibility results in the remap of qubits. If the \textit{PermRowCol} produces more $\CNOT$ overhead than that of \textit{RowCol}, we may have picked an inefficient remap. Thus, the simple heuristics used in our algorithm may not be optimal when synthesizing a random parity matrix. This makes sense because our heuristic relies on the number of $1$s in each row and column; but in a random parity matrix, the number of $1$s in each row and column is approximately the same. Therefore, the \textit{PermRowCol} with our choice of heuristic may decide to eliminate a random row and column that requires more $\CNOT$s later in the synthesis process.


Besides, there seems to be some connections between the underlying topologies and different algorithms' performance. In \cref{fig:graphs}, the 16-qubit square grid machine, the IBM QX5 machine, and the Rigetti Aspen machine all consist of $16$ qubits, but have distinct connectivity constraints. In \cref{fig:faketopogies}, to synthesize large CNOT circuits ($\geq 200$ \CNOT s) given the $16$-qubit square grid (\cref{fig:graphs}.c), \textit{PermRowCol} without RT generates the most CNOTs. For a circuit of arbitrary size and for all topologies in \cref{fig:graphs}, \textit{PermRowCol} with RT generates the least CNOTs (\cref{fig:realtopogies}). In the meantime, \cref{table:analysis} shows that the topologies of the real devices are sparser than their fictitious counterparts. This means that when removing random non-cutting vertices from the connectivity graph, our options are much more limited when working with real devices. Conversely, removing random non-cutting vertices from a square grid will restrict our options less because the vertices are connected through more paths. This makes it less likely to pick an inefficient option. However, when synthesizing circuits without any connectivity constraints, \cref{fig:unrestricted} shows that \emph{PermRowCol} outperforms other algorithms even when it is not combined with RT. It seems that picking an inefficient qubit allocation is less detrimental when CNOTs are allowed between arbitrary pairs of qubits. 


\begin{table}[h!]
\begin{tabular}{l|ll|l|l}
\hline
Qubit Count & \multicolumn{2}{c|}{Topology}                                                  & Avg. Graph Distance & Average Degree \\ \hline
\multirow{4}{*}{16} & \multicolumn{1}{l|}{\multirow{2}{*}{Fictitious devices}} & Square          & $2.5$                  & $4$            \\
                    & \multicolumn{1}{l|}{}                                    & Fully-Connected & $1$                    & $15$           \\ \cline{2-5} 
                    & \multicolumn{1}{l|}{\multirow{2}{*}{Real devices}}       & Rigetti Aspen   & $3.25$                 & $2.25$         \\
                    & \multicolumn{1}{l|}{}                                    & IBM QX5             & $3.125$                & $2.75$         \\ \hline
\end{tabular}
\caption{Different topologies in \cref{fig:graphs} have distinct average graph distance and average vertex degree. The topologies of real devices (i.e., Rigetti Aspen and IBM QX5) have greater average graph distance and smaller average vertex degree than those of fictitious devices (i.e., Square and Fully-Connected). In our example, the topologies of real devices are sparser than their fictitious counterparts.}
\label{table:analysis}
\end{table}

Based upon the above observations, we expect that the heuristics for picking the pivot row and column can be further improved. Since we need to pick the pivot row and column in an optimal order, this is a combinatorial search space. For completeness, we have implemented an A* algorithm~\cite{russell2002artificial} for the choice of pivots that tries all possible pivots following Dijkstra's shortest path algorithm. These results and their technical details can be found in \cref{app:a*results}. Since this method scales exponentially with respect to the number of qubits, we do not think this is a sensible approach. This is why it is not in the main body of this paper. The A* algorithm is only added to indicate the effect of a ``perfect`` heuristic. 

In \cref{app:a*results}, we show that the A* algorithm performs marginally worse than RT on its own on $5$-qubit devices. This seems to indicate that it is equally important for \textit{PermRowCol} to have a good initial qubit mapping as having a good heuristic. 
This effect is diminished when \textit{PermRowCol} is applied on multiple CNOT slices in a circuit. Because the initial qubit mapping cascades throughout the repeated applications of PermRowCol that uses the heuristic for every slice.

Additionally, it is likely that the performance of alternative heuristics depends heavily on the given topology and parity matrix. Therefore, we encourage the reader to think about what heuristics would work well for their specific use case.

In conclusion, we need a better method to determine the dynamic qubit maps than the heuristic we have used in this paper. This is also why we have not yet compared the performance of \textit{PermRowCol} with those in the established quantum compilers such as Qiskit, TKET, and SABRE. We will discuss possible directions for improvements in the next section.

\section{Future Work}
\label{sec:future}

Our \textit{PermRowCol} algorithm is shown to be promising when combined with Reverse Traversal (RT). In \cref{sec:improvements}, we discuss ways to improve the \textit{PermRowCol}. In \cref{sec:extension}, we illustrate how to leverage the \textit{PermRowCol} to synthesize an arbitrary quantum circuit.

\subsection{Improve the PermRowCol algorithm}
\label{sec:improvements}

In \cref{sec:conclusion}, we conclude that a better heuristic is needed for choosing the row and column to eliminate. Moreover, the quality of the heuristics may depend on the given topology and the original CNOT circuit. Beyond this, to achieve the full-stack quantum circuit compilation, it is important to extend our method by accounting for different gate error rates and the difficulty to couple two qubits. When constructing a Steiner tree, we can define a non-uniform weight function based on the quality of the gates acting on each qubit and the prevalent error model in the chosen architecture. This insight may inspire some interesting error-mitigation techniques. Moreover, due to the dynamic qubit maps in the \textit{PermRowCol}, we might redesign the algorithm to take into account the quality of the single qubit gates that are executed after the CNOT circuit. Under such restrictions, some qubits cannot be mapped to certain registers. These kinds of selection rules can then be added to the \textit{PermRowCol} by changing the heuristic for choosing which column to eliminate.

Additionally, the \textit{PermRowCol} can be improved by adding a blockwise elimination method~\cite{2008PMH}. In \cite{wu2023optimization}, in addition to the algorithm \textit{RowCol}, the \textit{size-block elimination} (SBE) was introduced. It uses a similar strategy as did in~\cite{2008PMH} but with Steiner trees and Gray codes. Eliminating blocks of rows and columns might not be very efficient on sparse graphs, but it might be interesting as an alternative to the unconstrained Gaussian elimination tasks where a permutation matrix is accepted as a valid solution (e.g. ZX-diagram extraction~\cite{backens2021there}).

\subsection{Extension PermRowCol to synthesize arbitrary quantum circuits}\label{sec:extension}

To achieve the universality for quantum computing, we need to work with unitaries beyond just CNOT gate. Therefore, a natural next step for us is to extend the \textit{PermRowCol} algorithm such that we can route and map qubits in any quantum circuit. There are two potential approaches: (1) synthesize the full circuit from a flexible representation; (2) cut the circuit into pieces so that we can synthesize each subcircuit separately and glue the synthesized pieces back together. In what follows, we discuss these strategies by building upon our \textit{PermRowCol} algorithm to work with more general quantum circuits until we end up with the set of universal quantum circuits.

Since the \emph{PermRowCol} can only synthesize circuits over CNOT, we start by extending it to work with the class of circuits over CNOT and $R_z$ rotations. These circuit can be characterized by \textit{phase polynomials}, which are described by the set of parity terms where each $R_z$ occurs, along with a parity matrix describing the output parity terms of the quantum circuit \cite{amy2014polynomial,amy2018controlled}. This notion is also known the \textit{sum-over-paths}. Various Steiner-tree-based methods have been proposed to synthesize the parity term of each $R_z$ gate~\cite{nash2020quantum,meijer-vandegriend2020architecture,gheorghiu2022reducing,vandaele2022phase}. The remaining parity matrix can then be synthesized by \textit{PermRowCol}. To extend the phase polynomials to arbitrary quantum circuits, we need to add the $H$ gate which these methods cannot synthesize. To this end, we can cut the circuit into subcircuits at the locality of $H$ gates, and each subcircuit is composed of CNOT and $R_z$ gates. Then our problem is reduced to synthesizing the phase polynomial of each subcircuit and then glue them back together. This method is known as the \textit{slice-and-build}, and it is proposed in \cite{gheorghiu2022reducing}. \textit{PermRowCol} can make a difference because the dynamic qubit maps allows this algorithm to move the $H$ gates so that they could act on different but more convenient quantum registers.


Moreover, the presence of a NOT gate is compatible with the current characterization because a NOT gate on some physical qubit introduces an $1$ modulo $2$ to the corresponding parity term. Thus, we can further extend the above method to work with circuits over NOT, CNOT, and $R_z$ rotations by adding an extra row to the parity matrix to represent the participation of a NOT gate. Then we can proceed as before using the phase-polynomial network synthesis and the slice-and-build to synthesize a circuit over NOT, CNOT, and $R_z$ rotations \cite{gheorghiu2022reducing}. This is much simpler than first synthesizing the phase polynomial without the NOT gates, replacing each NOT by $H R_z(\pi) H$, and then synthesizing the latter phase polynomial. Accordingly, we have generalized \emph{PermRowCol} to synthesize the Clifford+T circuits, which is a family of circuits that are well-suited for universal quantum computation.


Furthermore, it is worth investigating how to combine \textit{PermRowCol} with a generalized notion of the sum-over-paths: \textit{Pauli exponentials}~\cite{cowtan2020generic}. Like phase polynomials, the Pauli exponential keeps track of the parity terms where each $R_z$ rotation occurs. This is done by using \textit{Pauli strings} over $I, X, Y, Z$ rather than using the binary strings.
In \cite{cowtan2020generic}, an algorithm is proposed to extract a circuit from the Pauli exponential form. This is done by adding Clifford gates to the circuit until the Pauli exponential is reduced to a phase polynomial. 
Although the algorithm is not architecture-aware, it is possible to adjust the algorithm such that it only generates CNOTs that are allowed by the target topology, and the phase polynomial can be synthesized using the methods described above. Additionally, the algorithm was created for synthesizing particular quantum chemistry circuits, but it can be argued that the Pauli exponentials should serve as an important primitive for quantum computation in general~\cite{li2022paulihedral}.

Lastly, we can use \textit{PermRowCol} in the extraction of ZX diagrams~\cite{backens2021there}. ZX calculus is universal for quantum computation, so any quantum circuit can be expressed as a ZX diagram. Then, we can make the diagram into a normal form from which to extract an optimized circuit. The Gaussian elimination is used in the extraction procedure, and it can be replaced by the \textit{PermRowCol}. Even though the extraction procedure doesn't take any topologies into account, it may be interesting to use the \textit{PermRowCol} with a fully-connected graph. In ZX-calculus, only connectivity matters, so reordering the outputs of the extracted circuit is equivalent to bending wires and therefore free. Thus, it is better to end up with crossing wires than to extract CNOTs. In \cref{fig:unrestricted}, we show that the \textit{PermRowCol} results in less CNOTs compared to the Gaussian elimination and the \textit{RowCol}. Therefore, the \textit{PermRowCol} could be beneficial to improve the overhead of quantum circuit synthesis.

\subsection*{Acknowledgements}
The authors would like to thank Jukka K. Nurminen, Douwe van Gijn, and Dustin Meijer for proof-reading. They also wish to thank Matthew Amy, Neil J. Ross, and John van de Wetering for helpful feedback on choosing the paper's title.

\bibliographystyle{eptcs}
\bibliography{prc}

\begin{thebibliography}{10}
\providecommand{\bibitemdeclare}[2]{}
\providecommand{\surnamestart}{}
\providecommand{\surnameend}{}
\providecommand{\urlprefix}{Available at }
\providecommand{\url}[1]{\texttt{#1}}
\providecommand{\href}[2]{\texttt{#2}}
\providecommand{\urlalt}[2]{\href{#1}{#2}}
\providecommand{\doi}[1]{doi:\urlalt{https://doi.org/#1}{#1}}
\providecommand{\eprint}[1]{arXiv:\urlalt{https://arxiv.org/abs/#1}{#1}}
\providecommand{\bibinfo}[2]{#2}

\bibitemdeclare{article}{alber2001quantum}
\bibitem{alber2001quantum}
\bibinfo{author}{Gernot \surnamestart Alber\surnameend},
  \bibinfo{author}{Thomas \surnamestart Beth\surnameend},
  \bibinfo{author}{Micha{\l} \surnamestart Horodecki\surnameend},
  \bibinfo{author}{Pawe{\l} \surnamestart Horodecki\surnameend},
  \bibinfo{author}{Ryszard \surnamestart Horodecki\surnameend},
  \bibinfo{author}{Martin \surnamestart R{\"o}tteler\surnameend},
  \bibinfo{author}{Harald \surnamestart Weinfurter\surnameend},
  \bibinfo{author}{Reinhard \surnamestart Werner\surnameend},
  \bibinfo{author}{Anton \surnamestart Zeilinger\surnameend},
  \bibinfo{author}{Thomas \surnamestart Beth\surnameend} et~al.
  (\bibinfo{year}{2001}): \emph{\bibinfo{title}{Quantum algorithms: Applicable
  algebra and quantum physics}}.
\newblock {\slshape \bibinfo{journal}{Quantum information: an introduction to
  basic theoretical concepts and experiments}}, pp. \bibinfo{pages}{96--150},
  \doi{10.1007/3-540-44678-8_4}.

\bibitemdeclare{article}{amy2018controlled}
\bibitem{amy2018controlled}
\bibinfo{author}{Matthew \surnamestart Amy\surnameend},
  \bibinfo{author}{Parsiad \surnamestart Azimzadeh\surnameend} \&
  \bibinfo{author}{Michele \surnamestart Mosca\surnameend}
  (\bibinfo{year}{2018}): \emph{\bibinfo{title}{On the controlled-NOT
  complexity of controlled-NOT-phase circuits}}.
\newblock {\slshape \bibinfo{journal}{Quantum Science and Technology}}
  \bibinfo{volume}{4}(\bibinfo{number}{1}),
  \doi{10.1088/2058-9565/aad8ca}.

\bibitemdeclare{article}{amy2014polynomial}
\bibitem{amy2014polynomial}
\bibinfo{author}{Matthew \surnamestart Amy\surnameend}, \bibinfo{author}{Dmitri
  \surnamestart Maslov\surnameend} \& \bibinfo{author}{Michele \surnamestart
  Mosca\surnameend} (\bibinfo{year}{2014}):
  \emph{\bibinfo{title}{Polynomial-time T-depth optimization of Clifford+T
  circuits via matroid partitioning}}.
\newblock {\slshape \bibinfo{journal}{IEEE Transactions on Computer-Aided
  Design of Integrated Circuits and Systems}}
  \bibinfo{volume}{33}(\bibinfo{number}{10}), pp. \bibinfo{pages}{1476--1489},
  \doi{10.1109/TCAD.2014.2341953}.

\bibitemdeclare{article}{arute2019quantum}
\bibitem{arute2019quantum}
\bibinfo{author}{Frank \surnamestart Arute\surnameend}, \bibinfo{author}{Kunal
  \surnamestart Arya\surnameend}, \bibinfo{author}{Ryan \surnamestart
  Babbush\surnameend}, \bibinfo{author}{Dave \surnamestart Bacon\surnameend},
  \bibinfo{author}{Joseph~C \surnamestart Bardin\surnameend},
  \bibinfo{author}{Rami \surnamestart Barends\surnameend},
  \bibinfo{author}{Rupak \surnamestart Biswas\surnameend},
  \bibinfo{author}{Sergio \surnamestart Boixo\surnameend},
  \bibinfo{author}{Fernando~GSL \surnamestart Brandao\surnameend},
  \bibinfo{author}{David~A \surnamestart Buell\surnameend} et~al.
  (\bibinfo{year}{2019}): \emph{\bibinfo{title}{Quantum supremacy using a
  programmable superconducting processor}}.
\newblock {\slshape \bibinfo{journal}{Nature}}
  \bibinfo{volume}{574}(\bibinfo{number}{7779}), pp. \bibinfo{pages}{505--510},
  \doi{10.1038/s41586-019-1666-5}.

\bibitemdeclare{article}{backens2021there}
\bibitem{backens2021there}
\bibinfo{author}{Miriam \surnamestart Backens\surnameend},
  \bibinfo{author}{Hector \surnamestart Miller-Bakewell\surnameend},
  \bibinfo{author}{Giovanni \surnamestart de~Felice\surnameend},
  \bibinfo{author}{Leo \surnamestart Lobski\surnameend} \&
  \bibinfo{author}{John \surnamestart van~de Wetering\surnameend}
  (\bibinfo{year}{2021}): \emph{\bibinfo{title}{There and back again: A circuit
  extraction tale}}.
\newblock {\slshape \bibinfo{journal}{Quantum}} \bibinfo{volume}{5},
\doi{10.22331/q-2021-03-25-421}.

\bibitemdeclare{inproceedings}{2020dBBVMA}
\bibitem{2020dBBVMA}
\bibinfo{author}{Timoth{\'e}e Goubault~de \surnamestart
  Brugi{\`e}re\surnameend}, \bibinfo{author}{Marc \surnamestart
  Baboulin\surnameend}, \bibinfo{author}{Beno{\^\i}t \surnamestart
  Valiron\surnameend}, \bibinfo{author}{Simon \surnamestart Martiel\surnameend}
  \& \bibinfo{author}{Cyril \surnamestart Allouche\surnameend}
  (\bibinfo{year}{2020}): \emph{\bibinfo{title}{Quantum CNOT circuits synthesis
  for NISQ architectures using the syndrome decoding problem}}.
\newblock In: {\slshape \bibinfo{booktitle}{Quantum CNOT circuits synthesis for
  NISQ architectures using the syndrome decoding problem}},
  \bibinfo{organization}{Springer}, pp. \bibinfo{pages}{189--205},
  \doi{10.1007/978-3-030-52482-1_11}.

\bibitemdeclare{article}{2013BGRS}
\bibitem{2013BGRS}
\bibinfo{author}{Jaros{\l}aw \surnamestart Byrka\surnameend},
  \bibinfo{author}{Fabrizio \surnamestart Grandoni\surnameend},
  \bibinfo{author}{Thomas \surnamestart Rothvo{\ss}\surnameend} \&
  \bibinfo{author}{Laura \surnamestart Sanit{\`a}\surnameend}
  (\bibinfo{year}{2013}): \emph{\bibinfo{title}{Steiner tree approximation via
  iterative randomized rounding}}.
\newblock {\slshape \bibinfo{journal}{Journal of the ACM (JACM)}}
  \bibinfo{volume}{60}(\bibinfo{number}{1}),
  \doi{10.1145/2432622.2432628}.

\bibitemdeclare{book}{cormen2022introduction}
\bibitem{cormen2022introduction}
\bibinfo{author}{Thomas~H \surnamestart Cormen\surnameend},
  \bibinfo{author}{Charles~E \surnamestart Leiserson\surnameend},
  \bibinfo{author}{Ronald~L \surnamestart Rivest\surnameend} \&
  \bibinfo{author}{Clifford \surnamestart Stein\surnameend}
  (\bibinfo{year}{2022}): \emph{\bibinfo{title}{Introduction to algorithms}}.
\newblock \bibinfo{publisher}{MIT press}.

\bibitemdeclare{article}{cowtan2020generic}
\bibitem{cowtan2020generic}
\bibinfo{author}{Alexander \surnamestart Cowtan\surnameend},
  \bibinfo{author}{Will \surnamestart Simmons\surnameend} \&
  \bibinfo{author}{Ross \surnamestart Duncan\surnameend}
  (\bibinfo{year}{2020}): \emph{\bibinfo{title}{A generic compilation strategy
  for the unitary coupled cluster ansatz}}.
\newblock {\slshape \bibinfo{journal}{arXiv preprint}}.
\newblock \eprint{2007.10515},
\doi{10.48550/arXiv.2007.10515}.

\bibitemdeclare{article}{gheorghiu2022reducing}
\bibitem{gheorghiu2022reducing}
\bibinfo{author}{Vlad \surnamestart Gheorghiu\surnameend},
  \bibinfo{author}{Jiaxin \surnamestart Huang\surnameend},
  \bibinfo{author}{Sarah~Meng \surnamestart Li\surnameend},
  \bibinfo{author}{Michele \surnamestart Mosca\surnameend} \&
  \bibinfo{author}{Priyanka \surnamestart Mukhopadhyay\surnameend}
  (\bibinfo{year}{2022}): \emph{\bibinfo{title}{Reducing the CNOT count for
  Clifford+ T circuits on NISQ architectures}}.
\newblock {\slshape \bibinfo{journal}{IEEE Transactions on Computer-Aided
  Design of Integrated Circuits and Systems}},
  \doi{10.1109/TCAD.2022.3213210}.

\bibitemdeclare{article}{meijer-vandegriend2020architecture}
\bibitem{meijer-vandegriend2020architecture}
\bibinfo{author}{Arianne Meijer-van \surnamestart de~Griend\surnameend} \&
  \bibinfo{author}{Ross \surnamestart Duncan\surnameend}
  (\bibinfo{year}{2020}): \emph{\bibinfo{title}{Architecture-aware synthesis of
  phase polynomials for NISQ devices}}.
\newblock {\slshape \bibinfo{journal}{arXiv preprint}}.
\newblock \eprint{2004.06052},
\doi{10.48550/arXiv.2004.06052}.

\bibitemdeclare{article}{1992HR}
\bibitem{1992HR}
\bibinfo{author}{Frank~K \surnamestart Hwang\surnameend} \&
  \bibinfo{author}{Dana~S \surnamestart Richards\surnameend}
  (\bibinfo{year}{1992}): \emph{\bibinfo{title}{Steiner tree problems}}.
\newblock {\slshape \bibinfo{journal}{Networks}}
  \bibinfo{volume}{22}(\bibinfo{number}{1}),
  \doi{10.1007/0-306-48332-7_489}.

\bibitemdeclare{incollection}{1972K}
\bibitem{1972K}
\bibinfo{author}{Richard~M \surnamestart Karp\surnameend}
  (\bibinfo{year}{1972}): \emph{\bibinfo{title}{Reducibility among
  combinatorial problems}}.
\newblock In: {\slshape \bibinfo{booktitle}{Complexity of computer
  computations}}, \bibinfo{publisher}{Springer}, pp. \bibinfo{pages}{85--103},
  \doi{10.1007/978-1-4684-2001-2_9}.

\bibitemdeclare{article}{kissinger2020cnot}
\bibitem{kissinger2020cnot}
\bibinfo{author}{Aleks \surnamestart Kissinger\surnameend} \&
  \bibinfo{author}{Arianne~Meijer \surnamestart van~de Griend\surnameend}
  (\bibinfo{year}{2020}): \emph{\bibinfo{title}{CNOT circuit extraction for
  topologically-constrained quantum memories}}.
\newblock {\slshape \bibinfo{journal}{Quantum Information and Computation}}
  \bibinfo{volume}{20}(\bibinfo{number}{7-8}).
\newblock \eprint{1904.00633},
\doi{10.48550/arXiv.1904.00633}.

\bibitemdeclare{inproceedings}{li2019tackling}
\bibitem{li2019tackling}
\bibinfo{author}{Gushu \surnamestart Li\surnameend}, \bibinfo{author}{Yufei
  \surnamestart Ding\surnameend} \& \bibinfo{author}{Yuan \surnamestart
  Xie\surnameend} (\bibinfo{year}{2019}): \emph{\bibinfo{title}{Tackling the
  qubit mapping problem for NISQ-era quantum devices}}.
\newblock In: {\slshape \bibinfo{booktitle}{Proceedings of the Twenty-Fourth
  International Conference on Architectural Support for Programming Languages
  and Operating Systems}}, pp. \bibinfo{pages}{1001--1014}.
\newblock \eprint{1809.02573},
\doi{10.1145/3297858.3304023}.

\bibitemdeclare{inproceedings}{li2022paulihedral}
\bibitem{li2022paulihedral}
\bibinfo{author}{Gushu \surnamestart Li\surnameend}, \bibinfo{author}{Anbang
  \surnamestart Wu\surnameend}, \bibinfo{author}{Yunong \surnamestart
  Shi\surnameend}, \bibinfo{author}{Ali \surnamestart
  Javadi-Abhari\surnameend}, \bibinfo{author}{Yufei \surnamestart
  Ding\surnameend} \& \bibinfo{author}{Yuan \surnamestart Xie\surnameend}
  (\bibinfo{year}{2022}): \emph{\bibinfo{title}{Paulihedral: a generalized
  block-wise compiler optimization framework for Quantum simulation kernels}}.
\newblock In: {\slshape \bibinfo{booktitle}{Proceedings of the 27th ACM
  International Conference on Architectural Support for Programming Languages
  and Operating Systems}}, pp. \bibinfo{pages}{554--569},
  \doi{10.1145/3503222.3507715}.

\bibitemdeclare{article}{nash2020quantum}
\bibitem{nash2020quantum}
\bibinfo{author}{Beatrice \surnamestart Nash\surnameend}, \bibinfo{author}{Vlad
  \surnamestart Gheorghiu\surnameend} \& \bibinfo{author}{Michele \surnamestart
  Mosca\surnameend} (\bibinfo{year}{2020}): \emph{\bibinfo{title}{Quantum
  circuit optimizations for NISQ architectures}}.
\newblock {\slshape \bibinfo{journal}{Quantum Science and Technology}}
  \bibinfo{volume}{5}(\bibinfo{number}{2}).
\newblock \eprint{1904.01972},
\doi{10.1088/2058-9565/ab79b1}.

\bibitemdeclare{article}{nielsen2001quantum}
\bibitem{nielsen2001quantum}
\bibinfo{author}{Michael~A \surnamestart Nielsen\surnameend} \&
  \bibinfo{author}{Isaac~L \surnamestart Chuang\surnameend}
  (\bibinfo{year}{2001}): \emph{\bibinfo{title}{Quantum computation and quantum
  information}}.
\newblock {\slshape \bibinfo{journal}{Phys. Today}}
  \bibinfo{volume}{54}(\bibinfo{number}{2}), p.~\bibinfo{pages}{60},
  \doi{10.1017/CBO9780511976667}.

\bibitemdeclare{article}{2008PMH}
\bibitem{2008PMH}
\bibinfo{author}{Ketan~N \surnamestart Patel\surnameend},
  \bibinfo{author}{Igor~L \surnamestart Markov\surnameend} \&
  \bibinfo{author}{John~P \surnamestart Hayes\surnameend}
  (\bibinfo{year}{2004}): \emph{\bibinfo{title}{Optimal synthesis of linear
  reversible circuits}}.
\newblock {\slshape \bibinfo{journal}{Quantum Information \& Computation}}
  \bibinfo{volume}{8}(\bibinfo{number}{3}),
  \doi{10.48550/arXiv.quant-ph/0302002}.

\bibitemdeclare{article}{2005RZ}
\bibitem{2005RZ}
\bibinfo{author}{Gabriel \surnamestart Robins\surnameend} \&
  \bibinfo{author}{Alexander \surnamestart Zelikovsky\surnameend}
  (\bibinfo{year}{2005}): \emph{\bibinfo{title}{Tighter bounds for graph
  Steiner tree approximation}}.
\newblock {\slshape \bibinfo{journal}{SIAM Journal on Discrete Mathematics}}
  \bibinfo{volume}{19}(\bibinfo{number}{1}),
  \doi{10.1137/S0895480101393155}.

\bibitemdeclare{book}{russell2002artificial}
\bibitem{russell2002artificial}
\bibinfo{author}{Stuart \surnamestart Russell\surnameend} \&
  \bibinfo{author}{Peter \surnamestart Norvig\surnameend}
  (\bibinfo{year}{2010}): \emph{\bibinfo{title}{Artificial Intelligence: A
  Modern Approach}}, \bibinfo{edition}{3} edition.
\newblock \bibinfo{publisher}{Prentice Hall}.

\bibitemdeclare{article}{shende2003synthesis}
\bibitem{shende2003synthesis}
\bibinfo{author}{Vivek~V \surnamestart Shende\surnameend},
  \bibinfo{author}{Aditya~K \surnamestart Prasad\surnameend},
  \bibinfo{author}{Igor~L \surnamestart Markov\surnameend} \&
  \bibinfo{author}{John~P \surnamestart Hayes\surnameend}
  (\bibinfo{year}{2003}): \emph{\bibinfo{title}{Synthesis of reversible logic
  circuits}}.
\newblock {\slshape \bibinfo{journal}{IEEE Transactions on Computer-Aided
  Design of Integrated Circuits and Systems}}
  \bibinfo{volume}{22}(\bibinfo{number}{6}), pp. \bibinfo{pages}{710--722},
  \doi{10.1109/TCAD.2003.811448}.

\bibitemdeclare{article}{sivarajah2020tket}
\bibitem{sivarajah2020tket}
\bibinfo{author}{Seyon \surnamestart Sivarajah\surnameend},
  \bibinfo{author}{Silas \surnamestart Dilkes\surnameend},
  \bibinfo{author}{Alexander \surnamestart Cowtan\surnameend},
  \bibinfo{author}{Will \surnamestart Simmons\surnameend},
  \bibinfo{author}{Alec \surnamestart Edgington\surnameend} \&
  \bibinfo{author}{Ross \surnamestart Duncan\surnameend}
  (\bibinfo{year}{2020}): \emph{\bibinfo{title}{{t$|$ket$>$}: a retargetable
  compiler for NISQ devices}}.
\newblock {\slshape \bibinfo{journal}{Quantum Science and Technology}}
  \bibinfo{volume}{6}(\bibinfo{number}{1}).
\newblock \eprint{2003.10611},
\doi{10.1088/2058-9565/ab8e92}.

\bibitemdeclare{article}{stassi2020scalable}
\bibitem{stassi2020scalable}
\bibinfo{author}{Roberto \surnamestart Stassi\surnameend},
  \bibinfo{author}{Mauro \surnamestart Cirio\surnameend} \&
  \bibinfo{author}{Franco \surnamestart Nori\surnameend}
  (\bibinfo{year}{2020}): \emph{\bibinfo{title}{Scalable quantum computer with
  superconducting circuits in the ultrastrong coupling regime}}.
\newblock {\slshape \bibinfo{journal}{npj Quantum Information}}
  \bibinfo{volume}{6}(\bibinfo{number}{1}).
\newblock \eprint{1910.14478},
\doi{10.1038/s41534-020-00294-x}.

\bibitemdeclare{article}{vandaele2022phase}
\bibitem{vandaele2022phase}
\bibinfo{author}{Vivien \surnamestart Vandaele\surnameend},
  \bibinfo{author}{Simon \surnamestart Martiel\surnameend} \&
  \bibinfo{author}{Timoth{\'e}e~Goubault \surnamestart
  de~Brugi{\`e}re\surnameend} (\bibinfo{year}{2022}):
  \emph{\bibinfo{title}{Phase polynomials synthesis algorithms for NISQ
  architectures and beyond}}.
\newblock {\slshape \bibinfo{journal}{Quantum Science and Technology}}.
\newblock \eprint{2104.00934},
\doi{10.1088/2058-9565/ac5a0e}.

\bibitemdeclare{inproceedings}{qiskit}
\bibitem{qiskit}
\bibinfo{author}{Robert \surnamestart Wille\surnameend}, \bibinfo{author}{Rod
  \surnamestart Van~Meter\surnameend} \& \bibinfo{author}{Yehuda \surnamestart
  Naveh\surnameend} (\bibinfo{year}{2019}): \emph{\bibinfo{title}{IBM's Qiskit
  tool chain: Working with and developing for real quantum computers}}.
\newblock In: {\slshape \bibinfo{booktitle}{2019 Design, Automation \& Test in
  Europe Conference \& Exhibition (DATE)}}, \bibinfo{organization}{IEEE}, pp.
  \bibinfo{pages}{1234--1240}, 
  \doi{10.23919/DATE.2019.8715261}.

\bibitemdeclare{article}{wu2023optimization}
\bibitem{wu2023optimization}
\bibinfo{author}{Bujiao \surnamestart Wu\surnameend}, \bibinfo{author}{Xiaoyu
  \surnamestart He\surnameend}, \bibinfo{author}{Shuai \surnamestart
  Yang\surnameend}, \bibinfo{author}{Lifu \surnamestart Shou\surnameend},
  \bibinfo{author}{Guojing \surnamestart Tian\surnameend},
  \bibinfo{author}{Jialin \surnamestart Zhang\surnameend} \&
  \bibinfo{author}{Xiaoming \surnamestart Sun\surnameend}
  (\bibinfo{year}{2023}): \emph{\bibinfo{title}{Optimization of CNOT circuits
  on limited-connectivity architecture}}.
\newblock {\slshape \bibinfo{journal}{Physical Review Research}}
  \bibinfo{volume}{5}(\bibinfo{number}{1}), p. \bibinfo{pages}{013065},
  \doi{10.1103/PhysRevResearch.5.013065}.

\bibitemdeclare{article}{mcts}
\bibitem{mcts}
\bibinfo{author}{Xiangzhen \surnamestart Zhou\surnameend},
  \bibinfo{author}{Yuan \surnamestart Feng\surnameend} \&
  \bibinfo{author}{Sanjiang \surnamestart Li\surnameend}
  (\bibinfo{year}{2022}): \emph{\bibinfo{title}{A Monte Carlo Tree Search Framework for Quantum Circuit Transformation}}.
\newblock {\slshape \bibinfo{journal}{ACM Transactions on Design Automation of
  Electronic Systems (TODAES)}}.
\newblock \eprint{2008.09331},
\doi{10.1145/3514239}.

\bibitemdeclare{article}{zhu2022quantum}
\bibitem{zhu2022quantum}
\bibinfo{author}{Qingling \surnamestart Zhu\surnameend}, \bibinfo{author}{Sirui
  \surnamestart Cao\surnameend}, \bibinfo{author}{Fusheng \surnamestart
  Chen\surnameend}, \bibinfo{author}{Ming-Cheng \surnamestart Chen\surnameend},
  \bibinfo{author}{Xiawei \surnamestart Chen\surnameend},
  \bibinfo{author}{Tung-Hsun \surnamestart Chung\surnameend},
  \bibinfo{author}{Hui \surnamestart Deng\surnameend}, \bibinfo{author}{Yajie
  \surnamestart Du\surnameend}, \bibinfo{author}{Daojin \surnamestart
  Fan\surnameend}, \bibinfo{author}{Ming \surnamestart Gong\surnameend} et~al.
  (\bibinfo{year}{2022}): \emph{\bibinfo{title}{Quantum computational advantage
  via 60-qubit 24-cycle random circuit sampling}}.
\newblock {\slshape \bibinfo{journal}{Science bulletin}}
  \bibinfo{volume}{67}(\bibinfo{number}{3}), pp. \bibinfo{pages}{240--245},
  \doi{10.1016/j.scib.2021.10.017}.

\end{thebibliography}

\appendix
\FloatBarrier
\section{Subroutines}
\label{app:subroutines}
For reference, this section contains the pseudo-code of all subroutines used in Algorithm \ref{alg:permrowcol}.

\begin{algorithm}
\caption{\textbf{ChooseRow}: Subroutine for choosing which row to eliminate.}
\label{alg:chooserow}
\Input{Parity matrix $M$, candidate vertices $Vs$ that correspond to rows in $M$}
\Output{The row to eliminate: the pivot row}
\comment{Pick the row with the least 1s in $M$.}
$row \gets argmin_{v \in Vs}(\sum M[v])$\;
\Return{pivot row}
\end{algorithm}

\begin{algorithm}
\caption{\textbf{ChooseColumn}: Subroutine for choosing which column to eliminate.}
\label{alg:choosecolumn}
\Input{Parity matrix $M$, candidate columns in $M$, the pivot row}
\Output{The column to eliminate: the pivot column}
\comment{For columns with an $1$ in the pivot row, pick the one with the least 1s.}
$row \gets argmin_{c \in ColsToEliminate}(\sum M[:][c]$ if $M[row][c]=1$ else $|ColsToEliminate|)$\;
\Return{pivot column}
\end{algorithm}

\begin{algorithm}[!ht]
\caption{\textbf{EliminateColumn}: Subroutine for eliminating a column. Here, $SteinerTree$ is a routine which creates a Steiner tree over graph $G$ with root $root$ and nodes $Terminals$. $BottomUpTraversal$ performs a post-order traversal on the given tree and returns the edges rather than the vertices.}
\label{alg:eliminatecolumn}
\Input{Parity matrix $M$, graph $G$ representing the connectivity constraints, $root$ vertex, $Terminals$ to build the Steiner tree over}
\Output{List of CNOTs $C$ to add to the circuit.}
$C \gets []$\;
$Tree \gets SteinerTree(G, root, Terminals)$\;
\For{$edge \in BottomUpTraversal(Tree)$}{ 
    \If{$M[edge[0]][col] = 0$}{
        $C.add(CNOT(edge[0], edge[1]))$ \Comment*[r]{Make Steiner nodes into 1s}
        $M[edge[0]] \gets M[edge[0]] + M[edge[1]]$ mod $2$\;
    }
}
\For{$edge \in BottomUpTraversal(Tree)$}{
    $C.add(CNOT(edge[1], edge[0]))$ \Comment*[r]{Make the column into identity}
    $M[edge[1]] \gets M[edge[0]] + M[edge[1]]$ mod $2$\;
}
\Return{$C$}
\end{algorithm}

\begin{algorithm}
\caption{\textbf{EliminateRow}: Subroutine for eliminating a row. Here, $SteinerTree$ is a routine which creates a Steiner tree over graph $G$ with root $root$ and nodes $Terminals$. $TopDownTraversal$ performs a pre-order traversal on the given tree and returns the edges rather than the vertices. Similarly, $BottomUpTraversal$ performs a post-order traversal on the tree and returns the edges when traversed for the second time.} 
\label{alg:eliminaterow}
\Input{Parity matrix $M$, graph $G$ representing the connectivity constraints, $root$ vertex, $Terminals$ to build the Steiner tree over}
\Output{List of CNOTs $C$ to add to the circuit.}
$C \gets []$\;
$Tree \gets SteinerTree(G, root, Terminals)$\;
\For{$edge \in TopDownTraversal(Tree)$}{ 
    \If{$edge[1] \notin Nodes$}{
        $C.add(CNOT(edge[0], edge[1]))$ 
        $M[edge[0]] \gets M[edge[0]] + M[edge[1]]$ mod $2$\;
    }
}
\For{$edge \in BottomUpTraversal(Tree)$}{
        $C.add(CNOT(edge[0], edge[1]))$ 
        $M[edge[0]] \gets M[edge[0]] + M[edge[1]]$ mod $2$\;
}
\Return{$C$}
\end{algorithm}
\FloatBarrier
\section{Unconstrained performance}
\label{app:unconstrained}

\begin{figure}[!ht]
     \centering
     \begin{subfigure}[b]{0.49\textwidth}
         \centering
         \includegraphics[width=\textwidth]{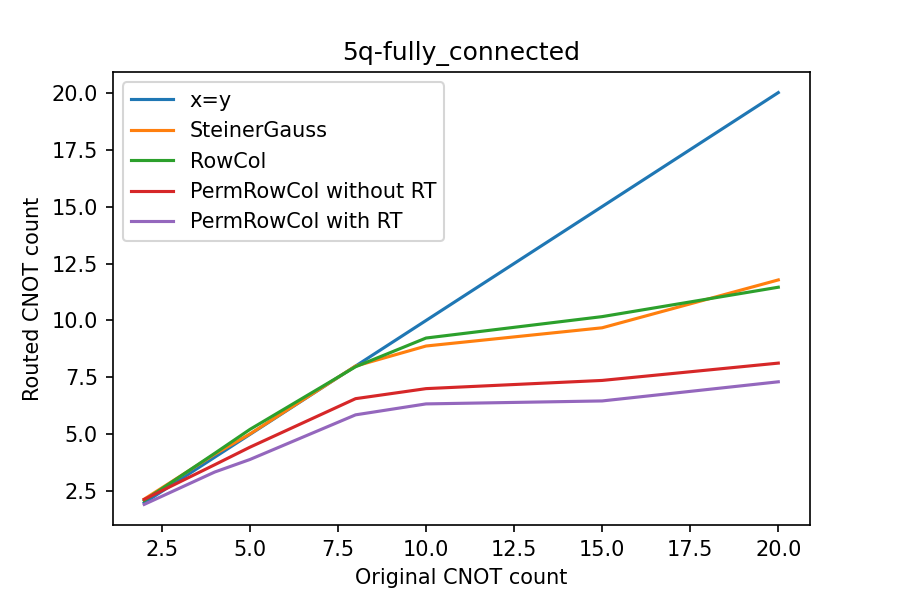}
         \caption{Algorithms' performance for the 5-qubit fully connected graph.}
         \label{fig:5qfully}
     \end{subfigure}
     \hfill
     \begin{subfigure}[b]{0.49\textwidth}
         \centering
         \includegraphics[width=\textwidth]{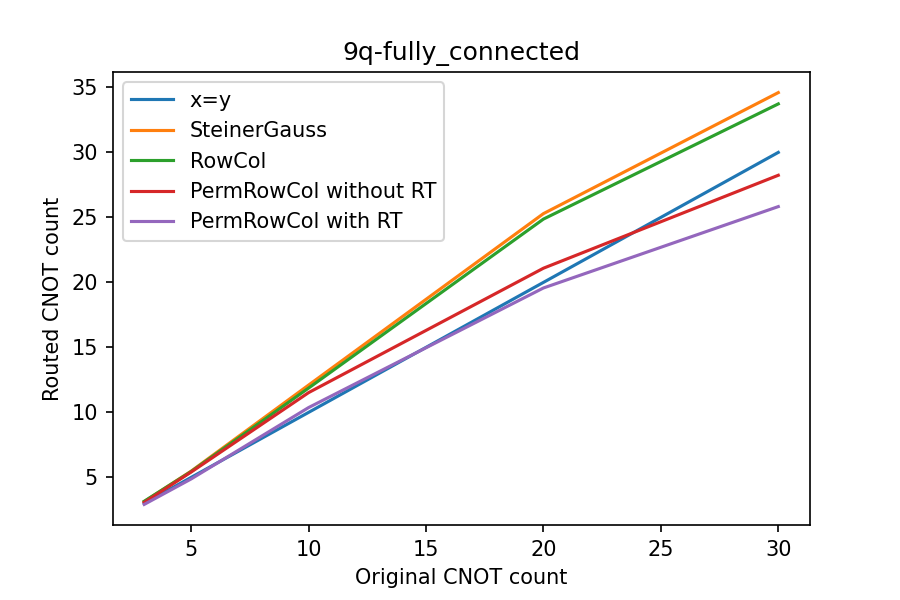}
         \caption{Algorithms' performance for the 9-qubit fully connected graph.}
         \label{fig:9qfully}
     \end{subfigure}
     \hfill
     \begin{subfigure}[b]{0.49\textwidth}
         \centering
         \includegraphics[width=\textwidth]{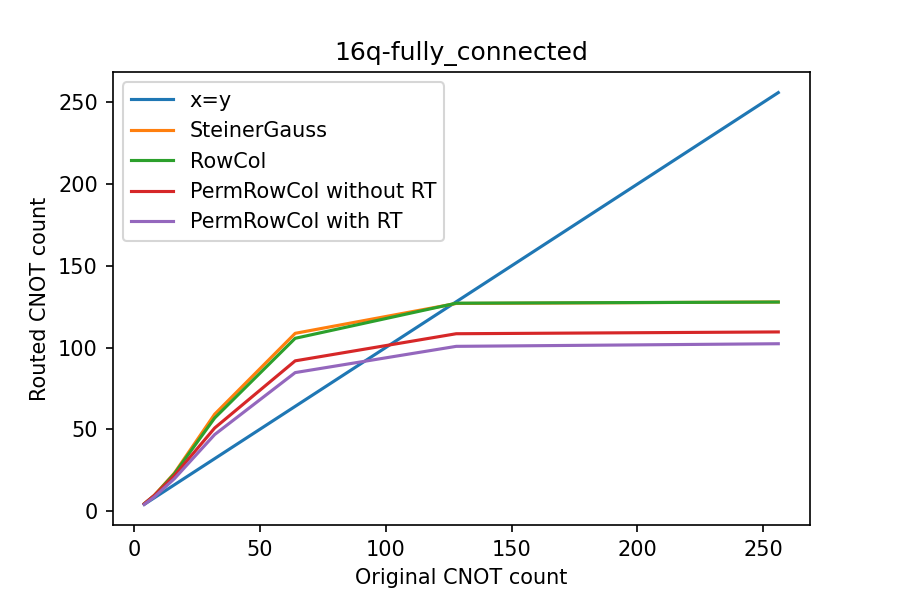}
         \caption{Algorithms' performance for the 16-qubit fully connected graph.}
         \label{fig:16qfully}
     \end{subfigure}
     \hfill
     \begin{subfigure}[b]{0.49\textwidth}
         \centering
         \includegraphics[width=\textwidth]{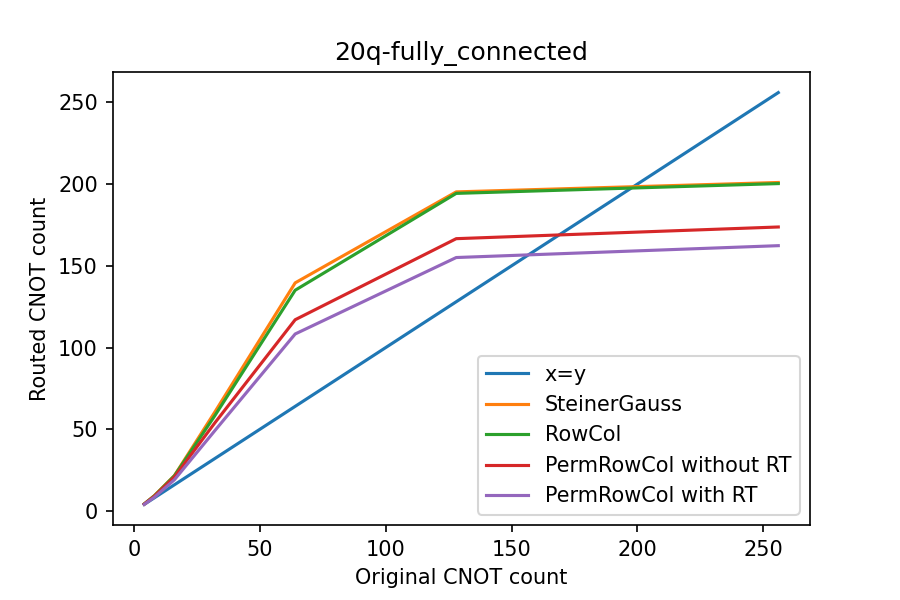}
         \caption{Algorithms' performance for the 20-qubit fully connected graph.}
         \label{fig:20qfully}
     \end{subfigure}
        \caption{These figures show the number of CNOTs generated by \textit{Steiner-Gauss}~\cite{kissinger2020cnot} (orange), \textit{RowCol}~\cite{wu2023optimization} (green), and \textit{PermRowCol} (proposed, red) for the unconstrained case, i.e. a complete graph of $5$, $9$, $16$, and $20$ qubits where every pair of distinct vertices is connected by a unique edge. The blue $x=y$-line is used to infer the CNOT overhead.}
    \label{fig:unrestricted}
\end{figure}

As additional information, \cref{fig:unrestricted} compares the algorithm performance in the unconstrained case, i.e. synthesizing a CNOT circuit given a fully connected topology. Although this use case does not include the routing of CNOTs, there are two use cases for which this is interesting. Both of these cases need to allow reallocation of qubits to make use of \textit{PermRowCol}. Restricting \textit{PermRowCol} to a fixed allocation is equivalent to the \textit{RowCol} algorithm.

First of all, we can use \textit{PermRowCol} without topological restrictions in case we have a circuit with many CNOTs (with respect to the number of qubits) to optimize the number CNOTs.

Secondly, \textit{PermRowCol} can be used in cases when the CNOT circuit is not known beforehand and needs to be synthesized from a given parity matrix. 
For example, this can be done as part of the \textit{GraySynth}~\cite{amy2018controlled} algorithm that is used in the \textit{PauliSynth}~\cite{cowtan2020generic} algorithm for generating UCCSD circuits in quantum chemistry. Alternatively, it can be used in place of Gaussian elimination as part of ZX-diagram extraction~\cite{backens2021there}. As we have already discussed in \cref{sec:extension}.

\FloatBarrier
\section{Topologies of real devices}
\label{app:topologies}
\begin{figure}[!ht]
\centering
\includegraphics[width=18cm]{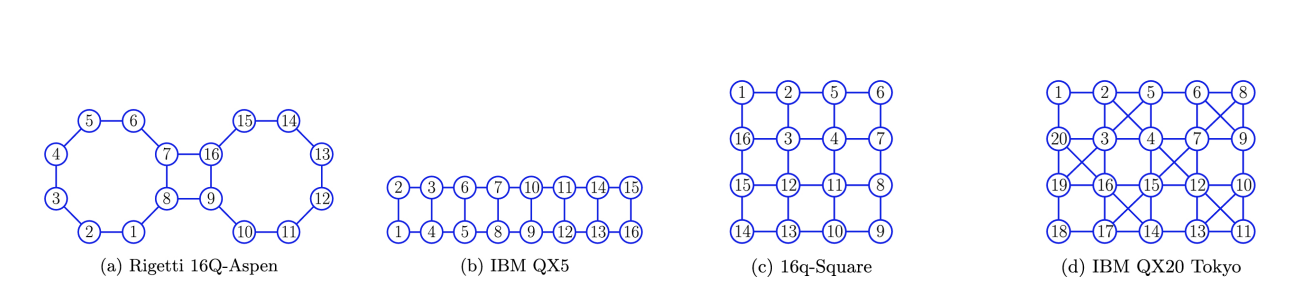}
\caption{Topologies of existing quantum computers that we use for testing our algorithm. Images are taken from \cite{2020dBBVMA}.}
\label{fig:graphs}
\end{figure}

We show in \cref{fig:graphs} the different topologies for the real quantum computers that we used for the connectivity constraints in our experiments.

\FloatBarrier
\section{Results in table form}
\label{app:resulttable}
For completeness, we show the results from \cref{fig:faketopogies} and \ref{fig:realtopogies} in the form of \cref{table:results}.

\begin{table}
\centering
\sisetup{
round-mode = places,
round-precision = 2
}
 \begin{tabular}{|l|l||c|c|c|c|}%
 \hline
    \bfseries \small CNOT & \bfseries \small Topology & \bfseries \small{\textit{Steiner-Gauss}} & \bfseries \small{\textit{RowCol}} & \bfseries \small{\textit{PermRowCol}} & \bfseries \small{\textit{PermRowCol+RT}}
    \begin{filecontents*}{"PermRowColResults.txt"}
,,,SteinerGauss,SteinerGauss,RowCol,RowCol,PermRowCol without RT,PermRowCol without RT,PermRowCol with RT,PermRowCol with RT
,,,count,overhead (%),count,overhead (%),count,overhead (%),count,overhead (%)
Original,Architecture,Qubits,,,,,,,,
3,9q-square,9,11.96,298.6666666666667,12.26,308.6666666666667,12.17,305.6666666666667,4.74,58.0
5,9q-square,9,18.23,264.6,18.74,274.8,16.33,226.6,7.48,49.6
10,9q-square,9,33.69,236.9,32.67,226.7,28.17,181.7,14.22,42.2
20,9q-square,9,49.05,145.25,46.82,134.1,40.01,100.05,24.47,22.35
30,9q-square,9,55.58,85.26666666666667,53.52,78.4,45.75,52.5,31.23,4.100000000000002
4,16q-square,16,29.9,647.5,29.02,625.5,33.86,746.5,7.21,80.25
8,16q-square,16,55.14,589.25,55.7,596.25,58.49,631.125,15.96,99.5
16,16q-square,16,97.85,511.5625,92.24,476.5,87.95,449.6875,34.34,114.625
32,16q-square,16,151.54,373.5625,141.57,342.40625,133.99,318.71875,81.68,155.25
64,16q-square,16,189.09,195.453125,175.74,174.59375,184.84,188.8125,141.75,121.484375
128,16q-square,16,200.4,56.5625,189.32,47.90625,204.89,60.0703125,165.97,29.6640625
256,16q-square,16,201.95,-21.11328125,190.83,-25.45703125,205.16,-19.859375,167.55,-34.55078125
4,rigetti 16q aspen,16,55.59,1289.75,50.0,1150.0,51.77,1194.25,14.17,254.25
8,rigetti 16q aspen,16,100.85,1160.625,88.78,1009.75,90.12,1026.5,30.13,276.625
16,rigetti 16q aspen,16,155.25,870.3125,138.99,768.6875,132.58,728.625,54.15,238.4375
32,rigetti 16q aspen,16,223.03,596.96875,199.41,523.15625,174.21,444.40625,106.04,231.375
64,rigetti 16q aspen,16,259.77,305.890625,240.51,275.796875,229.71,258.921875,178.55,178.984375
128,rigetti 16q aspen,16,270.64,111.4375,252.33,97.1328125,252.95,97.6171875,209.31,63.5234375
256,rigetti 16q aspen,16,270.07,5.49609375,255.06,-0.3671875,252.69,-1.29296875,209.52,-18.15625
4,ibm qx5,16,37.87,846.75,37.96,849.0,40.54,913.5,9.62,140.5
8,ibm qx5,16,72.34,804.25,71.43,792.875,66.81,735.125,20.62,157.75
16,ibm qx5,16,121.33,658.3125,115.63,622.6875,104.6,553.75,40.31,151.9375
32,ibm qx5,16,184.57,476.78125,174.74,446.0625,154.56,383.0,91.17,184.90625
64,ibm qx5,16,227.48,255.4375,218.93,242.078125,211.51,230.484375,159.43,149.109375
128,ibm qx5,16,242.96,89.8125,238.07,85.9921875,229.47,79.2734375,189.13,47.7578125
256,ibm qx5,16,244.43,-4.51953125,238.25,-6.93359375,233.83,-8.66015625,191.73,-25.10546875
4,ibm q20 tokyo,20,24.04,501.0,23.14,478.5,28.88,622.0,6.71,67.75
8,ibm q20 tokyo,20,50.58,532.25,49.09,513.625,54.29,578.625,14.72,84.0
16,ibm q20 tokyo,20,99.7,523.125,94.17,488.5625,95.66,497.875,30.08,88.0
32,ibm q20 tokyo,20,177.58,454.9375,163.22,410.0625,156.28,388.375,82.09,156.53125
64,ibm q20 tokyo,20,249.51,289.859375,235.23,267.546875,238.52,272.6875,183.99,187.484375
128,ibm q20 tokyo,20,281.34,119.796875,269.25,110.3515625,289.53,126.1953125,245.02,91.421875
256,ibm q20 tokyo,20,288.01,12.50390625,273.23,6.73046875,300.92,17.546875,256.48,0.1875
\end{filecontents*}
\csvreader[filter expr={test{\ifnumgreater{\thecsvinputline}{3}}}]
        {"PermRowColResults.txt"} 
        {1=\original,2=\architecture,3=\qubits, 4=\SGc,5=\SGo,6=\RCc,7=\RCo, 8=\PRCc, 9=\PRCo, 10=\RTc, 11=\RTo}
        {\\\small \original & \small \architecture & \small{\num{\SGc}\hfill \textcolor{gray}{(\num{\SGo} $\%$)}} & \small{$\num{\RCc}$\hfill \textcolor{gray}{($\num{\RCo}$ $\%$)}} & \small{$\num{\PRCc}$\hfill \textcolor{gray}{($\num{\PRCo}$ $\%$)}} & \small{$\num{\RTc}$\hfill \textcolor{gray}{($\num{\RTo}$ $\%$)}}}
        \\\hline
    \end{tabular}
    \caption{This table shows the performance of \textit{Steiner-Gauss}~\cite{kissinger2020cnot,nash2020quantum}, \textit{RowCol}~\cite{wu2023optimization}, \textit{PermRowCol} (proposed), and \textit{PermRowCol} with \textit{Reverse Traversal strategy} (proposed) for different topologies. The shown numbers represent the average CNOT count over 100 circuits and the average CNOT overhead with respect to the original circuit in brackets behind it.}
    \label{table:results}
    \end{table}

\FloatBarrier
\section{A* results}
\label{app:a*results}
In this appendix, we show the results for using the A* algorithm~\cite{russell2002artificial} for choosing the row and column in the iterations of \textit{PermRowCol}. We do this for different 5-qubit topologies because the overhead of A* becomes too long for larger topologies. 

We added the A* algorithm into \textit{PermRowCol} using a priority queue. Initially, we push the original problem into the queue with priority equal to 0. Then, while the queue is not empty, we remove an instance from the queue, reduce the matrix for each combination of chosen row and column and push the resulting smaller problem to the queue with the size of the circuit until now as priority. We continue this process until a solution has been found. By construction, the resulting solution will be the smallest circuit that can be found with \textit{PermRowCol} regardless of the choice in ChooseRow and ChooseColumn heuristics, but the complexity is exponential in the number of qubits.

To reduce the runtime of the algorithm, we restrict the number of new problem instances created at each iteration by a parameter called \textit{choiceWidth}. Instead of expanding each possible combination of row and column, we only expand the top \textit{choiceWidth} options where we rank the options using the original ChooseRow and ChooseColumn heuristics. Where the ChooseRow has priority. Additionally, we limit the size of the queue such that problem instances low in the queue are removed from memory since they probably don't need to be expanded before a solution is found. For these results, we used \textit{choiceWidth}$=4$ and \textit{max_size}$=10$. Resulting in an algorithm with time complexity $O(choiceWidth^{O(PermRowCol)} = O(4^{n^4})$.

\begin{figure}[!ht]
     \centering
     \begin{subfigure}[b]{0.49\textwidth}
         \centering
         \includegraphics[width=\textwidth]{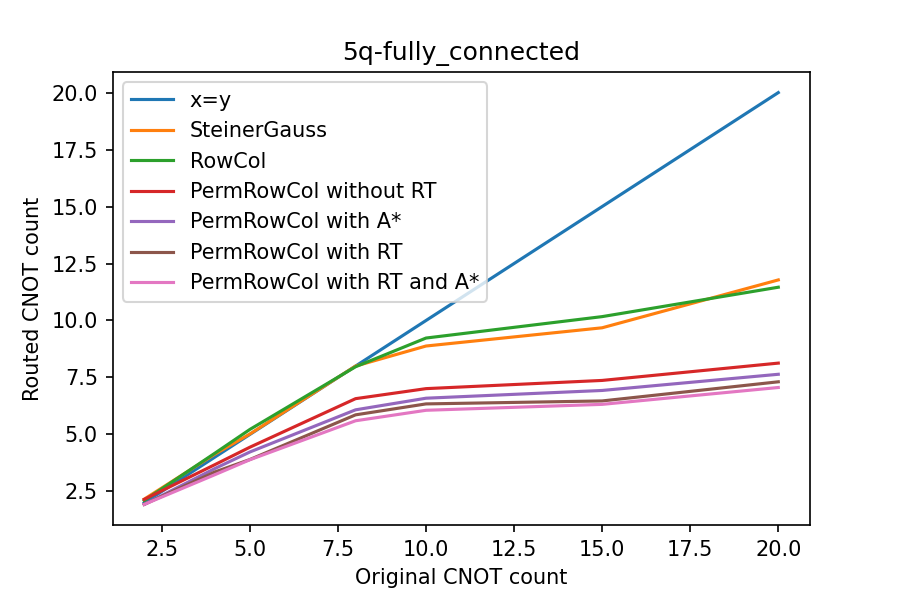}
         \caption{Algorithms' performance for the 5-qubit fully connected graph.}
         \label{fig:A5qfully}
     \end{subfigure}
     \hfill
     \begin{subfigure}[b]{0.49\textwidth}
         \centering
         \includegraphics[width=\textwidth]{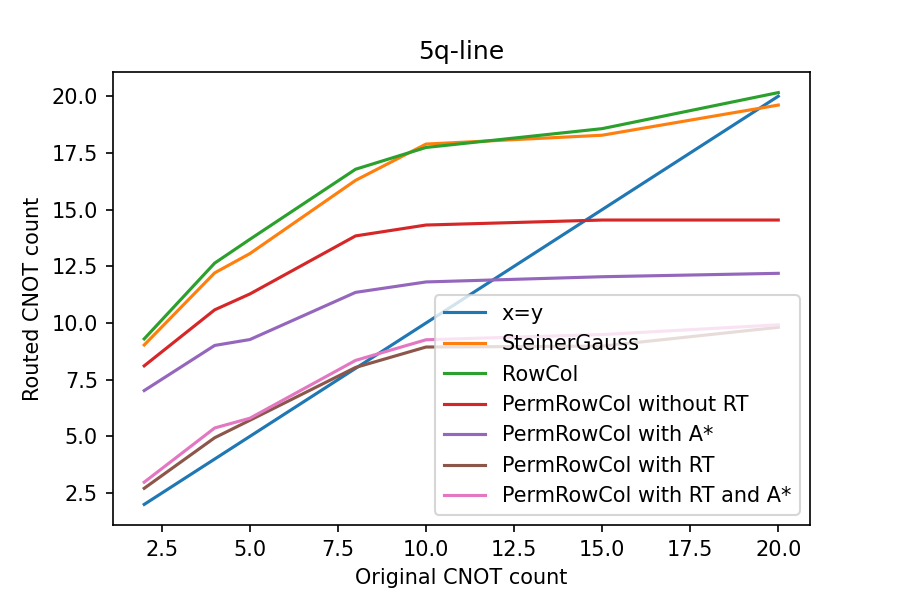}
         \caption{Algorithms' performance for the 5-qubit line graph.}
         \label{fig:A5qline}
     \end{subfigure}
     \hfill
     \begin{subfigure}[b]{0.49\textwidth}
         \centering
         \includegraphics[width=\textwidth]{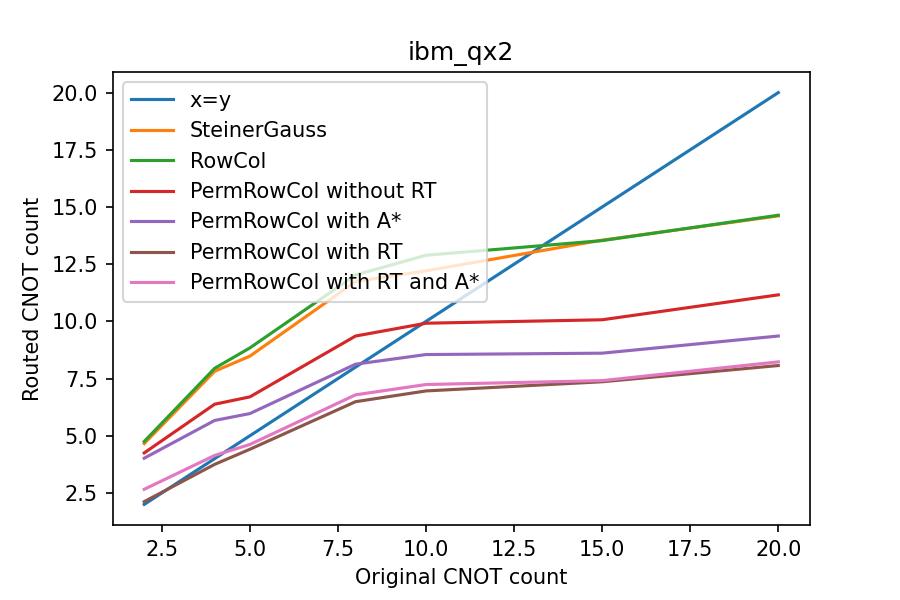}
         \caption{Algorithms' performance for the 5-qubit IBM QX2 device.}
         \label{fig:A5qQX2}
     \end{subfigure}
     \hfill
     \begin{subfigure}[b]{0.49\textwidth}
         \centering
         \includegraphics[width=\textwidth]{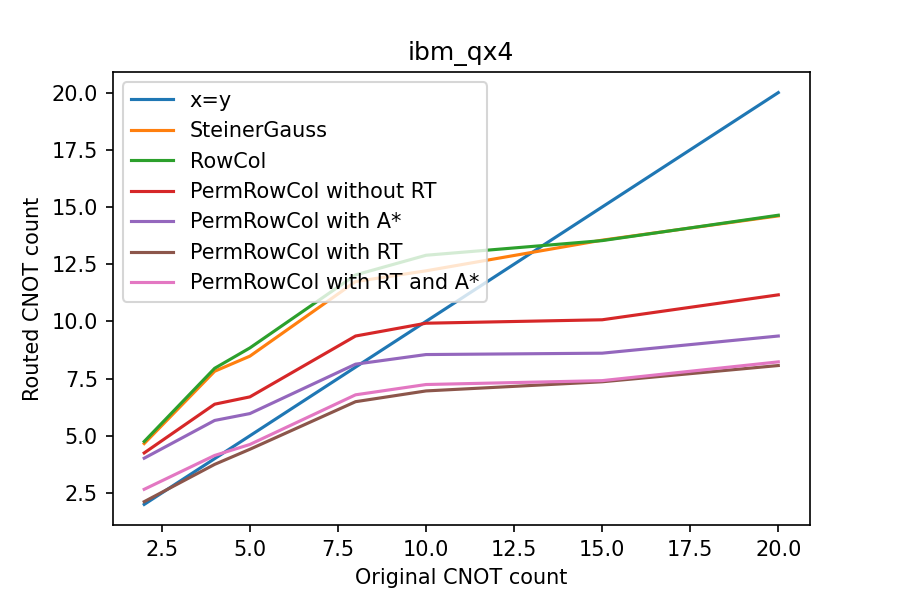}
         \caption{Algorithms' performance for the 5-qubit IBM QX4 device.}
         \label{fig:A5qQX4}
     \end{subfigure}
        \caption{These figures show the number of CNOTs generated by \textit{Steiner-Gauss}~\cite{kissinger2020cnot} (orange), \textit{RowCol}~\cite{wu2023optimization} (green), and different variants of \textit{PermRowCol} (proposed) on different 5-qubit topologies. The different variants of \textit{PermRowCol} are the original (red), with A* algorithm (purple), with Reverse Traversal (RT) (brown), and with both A* and RT (pink). The blue $x=y$-line is used to infer the CNOT overhead.}
    \label{fig:A*results}
\end{figure}

The results are shown in \cref{fig:A*results} where we see that A* does not perform better than than the Reverse Traversal (RT) strategy, even when combining the two algorithms.

\FloatBarrier

\section{Example execution of \textit{PermRowCol}}
\label{app:example}
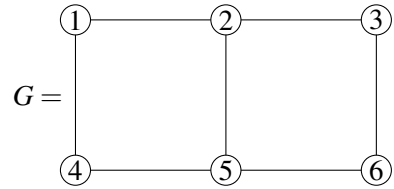
\begin{figure}[!ht]
  \renewcommand{\thefigure}{(a)}
  \begin{subfigure}[b]{1\textwidth}
  \[
\Qcircuit @C=.9em @R=.9em @!R {
          \lstick{\ket{1}} & \ctrl{1} & \qw & \qw & \qw & \qw & \qw & \qw & \qw & \targ & \ctrl{2} & \qw &  \rstick{\ket{2 \oplus 3 \oplus 4 \oplus 5 \oplus 6}}\\
          \lstick{\ket{2}} & \targ & \ctrl{4} & \targ & \ctrl{3} & \ctrl{2} & \qw & \qw & \qw & \qw & \qw & \qw & \rstick{\ket{1 \oplus 2 \oplus 4}}\\
          \lstick{\ket{3}} & \qw & \qw & \qw & \qw & \qw & \qw & \ctrl{3} & \qw & \qw & \targ & \qw &  \rstick{\ket{2 \oplus 4 \oplus 5 \oplus 6}}\\
          \lstick{\vect{C} \: =\:\ket{4}} & \qw & \qw & \ctrl{-2} & \qw & \targ & \ctrl{2} & \qw & \qw & \qw & \qw & \qw & \rstick{\ket{1 \oplus 2}}\\
          \lstick{\ket{5}} & \qw & \qw & \qw & \targ & \qw & \qw & \qw & \targ & \ctrl{-4} & \qw & \qw &  \rstick{\ket{1 \oplus 2 \oplus 3 \oplus 4 \oplus 5 \oplus 6}}\\
          \lstick{\ket{6}} & \qw & \targ & \qw & \qw & \qw & \targ & \targ & \ctrl{-1} & \qw & \qw & \qw &  \rstick{\ket{3 \oplus 6}}
  }
\]
\label{appG-a}
\caption{A CNOT circuit $\vect{C}$ composed of $6$ qubits.}
\end{subfigure}
  \renewcommand{\thefigure}{(b)}
  \begin{subfigure}[b]{0.5\textwidth}
      \centering
 \[
\vect{A} = \begin{blockarray}{ccccccc}
 & \color{IMSGreen}1' & \color{IMSGreen}2' & \color{IMSGreen}3' & \color{IMSGreen}4' & \color{IMSGreen}5' & \color{IMSGreen}6'\\
\begin{block}{c(cccccc)}
  \color{IMSGreen}1 & 0 & 1 & 0 & 1 & 1 & 0 \\
  \color{IMSGreen}2 & 1 & 1 & 1 & 1 & 1 & 0 \\
  \color{IMSGreen}3 & 1 & 0 & 0 & 0 & 1 & 1 \\
  \color{IMSGreen}4 & 1 & 1 & 1 & 0 & 1 & 0 \\
  \color{IMSGreen}5 & 1 & 0 & 1 & 0 & 1 & 0 \\
  \color{IMSGreen}6 & 1 & 0 & 1 & 0 & 1 & 1 \\
\end{block}
\end{blockarray}
 \]
\label{appG-b}
\caption{The parity matrix of $\vect{C}$.}
\end{subfigure}
  \renewcommand{\thefigure}{(c)}
  \begin{subfigure}[b]{0.5\textwidth}
\centering
\begin{tikzpicture}
\node at (0.5,2) {$G=$};
 \draw (1,3) circle [radius=0.2];
  \node at (1,3) {$1$};
  \draw (1.2,3)--(2.8,3); 
  \draw (1.2,1)--(2.8,1); 
  \draw (3,3) circle [radius=0.2];
  \node at (3,3) {$2$};
  \draw (3.2,3)--(4.8,3); 
  \draw (3.2,1)--(4.8,1); 
  \draw (5,3) circle [radius=0.2];
  \node at (5,3) {$3$};
  \draw (1,1) circle [radius=0.2];
  \node at (1,1) {$4$};
  \draw (3,1) circle [radius=0.2];
  \node at (3,1) {$5$};
  \draw (5,1) circle [radius=0.2];
  \node at (5,1) {$6$};
  \node at (1,0) {$ $};
  \draw (1,2.8)--(1,1.2);    
  \draw (5,2.8)--(5,1.2);    
  \draw (3,2.8)--(3,1.2); 
\end{tikzpicture}
\label{appG-c}
\caption{The 6-qubit square grid $G$.}
\end{subfigure}
  \renewcommand{\thefigure}{11}\caption{The CNOT circuit $\vect{C}$ in figure (a) can be exactly represented by the $6 \times 6$ parity matrix $\vect{A}$ in figure (b). Under the constrained topology $G$ in figure (c), the Steiner-tree based algorithm \textit{PermRowCol} accounts for $G$ and re-synthesizes $\vect{C}$.}
\label{fig:initial} 
\end{figure}

In the section, we execute algorithm \textit{PermRowCol} with inputs in \cref{fig:initial}. We start with a parity matrix $\vect{A}$ to synthesize over the topology graph $G$, and reduce the problem to eliminating $\vect{A}$ to a permutation matrix. The labelling of vertices in $G$ corresponds to the numbering of rows in $\vect{A}$.

\FloatBarrier
\subsection{Notations}
We declare notations that will be used in \cref{subsection:paritymatrix,subsection:algowalkthrough}.

According to algorithm \textit{PermRowCol}, the chosen logical qubit corresponds to a row in $\vect{A}$. Its index is denoted by $r$. The chosen new physical register for $r$ corresponds to a column in $\vect{A}$. Its index is denoted by $c$. $V_s$ is a set of non-cutting vertices of $G$ under which $\vect{A}$ is synthesized. $S$ is the set of terminal nodes corresponding the indices of all non-zero entries in a row or column. $R(i,j)$ denotes a row operation on $\vect{A}$ such that row $i$ is added to row $j$, while row $i$ remains unchanged.

The output qubit allocation \cref{table:rcallocation} keeps track of the row and column being selected at each elimination step.

\begin{table}[!ht]
\centering
\begin{tabular}{|l|l|l|l|l|l|l|}
\hline
Logical qubit/r     & 1 & 2 & 3 & 4 & 5 & 6 \\ \hline
Physical register/c &   &   &   &   &   &   \\ \hline
\end{tabular}
\caption{The output qubit allocation table}
\label{table:rcallocation}
\end{table}

\subsection{Parity matrix for a CNOT circuit}
\label{subsection:paritymatrix}
Consider a CNOT circuit composed of $n$ qubits, where $\CNOT(c,t)$ has control $c$ and target $t$. Based on the specification in \cref{subsec:paritymatrix}, $\CNOT(c,t)$ corresponds to a row operation $R(t,c)$ such that row $t$ is added to row $c$, while row $t$ remains unchanged. 

In addition to the parity matrix defined in this paper, we note an alternative characterization for CNOT circuits \cite{2008PMH,nash2020quantum,gheorghiu2022reducing}. Let's call it as the \textit{alt-parity matrix} for a CNOT circuit. Similar to the parity matrix characterization, a square matrix is constructed to record the output parities of the circuit. Each row represents a parity term, and each column represents the input qubit. By construction, given a CNOT circuit, its alt-parity matrix is the transpose of its parity matrix. Accordingly, $\CNOT(c,t)$ corresponds to a row operation $R(c,t)$ such that row $c$ is added to row $t$, while row $c$ remains unchanged. However, the synthesis procedure constructs the circuit in reverse. 

For example, in \cref{fig:convention}, we compare the differences between the two notations and the corresponding CNOT synthesis algorithms. 
In \cref{fig:convention}.c, we see that the column-wise representation of the parities (\cref{fig:convention}.b) results in taking row operations that correspond to the CNOTs in order, while adding the target to the control: $R(2,1)R(3,2) \sim CNOT(1,2)CNOT(2,3)$.
Alternatively, in \cref{fig:convention}.e, we see that the row-wise representation of the parities (\cref{fig:convention}.d) results in taking row operations that correspond to the CNOTs in reverse order, while adding the control to the target: $R(2,3)R(1,2) \sim CNOT(1,2)CNOT(2,3)$.

\begin{figure}[!ht]
  \renewcommand{\thefigure}{(a)}
  \begin{subfigure}[b]{1\textwidth}
    \[
\Qcircuit @C=1em @R=1em @!R {
          \lstick{\ket{1}} & \qw & \ctrl{1} & \qw & \qw & \qw & \qw & \rstick{\ket{1}}\\
          \lstick{\vect{C}=\ket{2}} & \qw & \targ & \qw & \ctrl{1} &  \qw & \qw & \rstick{\ket{1\oplus 2}}\\
          \lstick{\ket{3}} & \qw & \qw & \qw & \targ \qw & \qw & \qw  & \rstick{\ket{1 \oplus 2 \oplus 3}}\\
  }
\]
\label{appF-a}
\caption{A CNOT circuit $\vect{C}$ composed of $3$ qubits and $2$ CNOTs. The labels of input qubits are denoted on the left of the circuit. The output parities are denoted on the right of the circuit.}
\end{subfigure}
  \renewcommand{\thefigure}{(b)}
  \begin{subfigure}[b]{1\textwidth}
\centering
  \[
\vect{A} = \begin{blockarray}{cccc}
 & 1' & 2' & 3' \\
\begin{block}{c(ccc)}
  1 & 1 & 1 & 1 \\
  2 & 0 & 1 & 1 \\
  3 & 0 & 0 & 1 \\
\end{block}
\end{blockarray}
 \]
\label{appF-b}
\caption{$\vect{A}$ is the parity matrix of $\vect{C}$. Each column represents an output parity term and each row represents an input qubit.}
\end{subfigure}
  \renewcommand{\thefigure}{(c)}
  \begin{subfigure}[b]{1\textwidth}
   \[
   \vect{A} = \begin{blockarray}{cccc}
 & 1' & 2' & 3' \\
\begin{block}{c(ccc)}
  1 & 1 & 1 & 1 \\
  2 & 0 & 1 & 1 \\
  3 & 0 & 0 & 1 \\
\end{block}
\end{blockarray}\xrightarrow{R(2,1)}
\vect{A}' = \begin{blockarray}{cccc}
 & 1' & 2' & 3' \\
\begin{block}{c(ccc)}
  1 & \color{IMSRed}1 & \color{IMSRed}0 & \color{IMSRed}0 \\
  2 & 0 & 1 & 1 \\
  3 & 0 & 0 & 1 \\
\end{block}
\end{blockarray}\xrightarrow{R(3,2)}
I = \begin{blockarray}{cccc}
 & 1' & 2' & 3' \\
\begin{block}{c(ccc)}
  1 & 1 & 0 & 0 \\
  2 & \color{IMSRed}0 & \color{IMSRed}1 & \color{IMSRed}0 \\
  3 & 0 & 0 & 1 \\
\end{block}
\end{blockarray}
\]
\label{appF-c}
\caption{Given $\vect{A}$, $\CNOT(1,2)$ corresponds to the row operation $R(2,1)$ and $\CNOT(2,3)$ corresponds to the row operation $R(3,2)$.}
\end{subfigure}
      \renewcommand{\thefigure}{(d)}
  \begin{subfigure}[b]{1\textwidth}
      \centering
  \[
\vect{B} = \begin{blockarray}{cccc}
 & 1 & 2 & 3 \\
\begin{block}{c(ccc)}
  1' & 1 & 0 & 0 \\
  2' & 1 & 1 & 0 \\
  3' & 1 & 1 & 1 \\
\end{block}
\end{blockarray}
 \]
\label{appF-d}
\caption{$\vect{B}$ is the alt-parity matrix of $\vect{C}$. Each row represents an output parity term and each column represents an input qubit.}
\end{subfigure}
  \renewcommand{\thefigure}{(e)}
  \begin{subfigure}[b]{1\textwidth}
  \[
\vect{B} = \begin{blockarray}{cccc}
 & 1 & 2 & 3 \\
\begin{block}{c(ccc)}
  1' & 1 & 0 & 0 \\
  2' & 1 & 1 & 0 \\
  3' & 1 & 1 & 1 \\
\end{block}
\end{blockarray}\xrightarrow{R(2,3)} \vect{B}' = \begin{blockarray}{cccc}
 & 1 & 2 & 3 \\
\begin{block}{c(ccc)}
  1' & 1 & 0 & 0 \\
  2' & 1 & 1 & 0 \\
  3' & \color{IMSRed}0 & \color{IMSRed}0 & \color{IMSRed}1 \\
\end{block}
\end{blockarray}\xrightarrow{R(1,2)}I = \begin{blockarray}{cccc}
 & 1 & 2 & 3 \\
\begin{block}{c(ccc)}
  1' & 1 & 0 & 0 \\
  2' & \color{IMSRed}0 & \color{IMSRed}1 & \color{IMSRed}0 \\
  3' & 0 & 0 & 1 \\
\end{block}
\end{blockarray}  
\]
\label{appF-e}
\caption{Given $\vect{B}$, $\CNOT(2,3)$ corresponds to the row operation $R(2,3)$ and $\CNOT(1,2)$ corresponds to the row operation $R(1,2)$.}
\end{subfigure}
  \renewcommand{\thefigure}{12}\caption{Circuit $\vect{C}$ in (a) is composed of $\CNOT(1,2)$ and $\CNOT(2,3)$. Its output parities are described by the parity matrix $\vect{A}$ in (b), or equivalently, by the alt-parity matrix $\vect{B}$ in (d). $\vect{B}^\top = \vect{A}$.}
\label{fig:convention} 
\end{figure}

\FloatBarrier
\subsection{\textit{PermRowCol} walkthrough}
\label{subsection:algowalkthrough}


\begin{description}
    \item[Elimination step 1] Before the first elimination step, the parity matrix $\vect{A}$ and the constrained topology $G$ are shown in \cref{fig:initial}.\\

    \begin{description}
        \item[Choose the row and column to eliminate:] The set of non-cutting vertices of $G$ is $V_s = \{1,2,3,4,5,6\}.$ Then $r = 1$ and $c = 4$ since
        
        \[
        \begin{blockarray}{ccccccccc}
            & \color{IMSGreen}1' & \color{IMSGreen}2' & \color{IMSGreen}3' & \color{IMSGreen}4' & \color{IMSGreen}5' & \color{IMSGreen}6' & \color{IMSOrange}\text{Sum} & \color{IMSRed}\text{Row}\\
            \begin{block}{c(cccccc)cc}
                \color{IMSGreen}1 & \color{IMSRed}0 & \color{IMSRed}1 & \color{IMSRed}0 & \color{IMSRed}1 & \color{IMSRed}1 & \color{IMSRed}0 & \color{IMSOrange}3 & \color{IMSRed}\checkmark\\
                \color{IMSGreen}2 & 1 & 1 & 1 & 1 & 1 & 0 & \color{IMSOrange}5 & \\
                \color{IMSGreen}3 & 1 & 0 & 0 & 0 & 1 & 1 & \color{IMSOrange}3 & \\
                \color{IMSGreen}4 & 1 & 1 & 1 & 0 & 1 & 0 & \color{IMSOrange}4 & \\
                \color{IMSGreen}5 & 1 & 0 & 1 & 0 & 1 & 0 & \color{IMSOrange}3 & \\
                \color{IMSGreen}6 & 1 & 0 & 1 & 0 & 1 & 1 & \color{IMSOrange}4 & \\
            \end{block}
        \end{blockarray} \Rightarrow \begin{blockarray}{ccccccc}
            & \color{IMSGreen}1' & \color{IMSGreen}2' & \color{IMSGreen}3' & \color{IMSGreen}4' & \color{IMSGreen}5' & \color{IMSGreen}6'\\
            \begin{block}{c(cccccc)}
                \color{IMSGreen}1 & \color{IMSRed}0 & \color{IMSRed}\mathbf{1} & \color{IMSRed}0 & \color{IMSRed}\mathbf{1} & \color{IMSRed}\mathbf{1} & \color{IMSRed}0\\
                \color{IMSGreen}2 & 1 & 1 & 1 & \color{IMSRed}1 & 1 & 0 \\
                \color{IMSGreen}3 & 1 & 0 & 0 & \color{IMSRed}0 & 1 & 1 \\
                \color{IMSGreen}4 & 1 & 1 & 1 & \color{IMSRed}0 & 1 & 0 \\
                \color{IMSGreen}5 & 1 & 0 & 1 & \color{IMSRed}0 & 1 & 0 \\
                \color{IMSGreen}6 & 1 & 0 & 1 & \color{IMSRed}0 & 1 & 1 \\
            \end{block}
                \color{IMSOrange}\text{Sum} & \color{IMSOrange}\setminus & \color{IMSOrange}3 & \color{IMSOrange}\setminus & \color{IMSOrange}2 & \color{IMSOrange}6 & \color{IMSOrange}\setminus \\
                \color{IMSRed}\text{Column} & & & & \color{IMSRed}\checkmark &
        \end{blockarray}
        \]
        \item[Eliminate the chosen row and column:] We start by eliminating column $4'$ to $e_1^{\intercal}$, then $S = \{1,2\}$. The Steiner tree $\mathcal{T}_{1,S}$ has root $1$ and a set of terminals $S$:
        \begin{center}        
            \begin{tikzpicture}
                \node at (0,2) {$\mathcal{T}_{1,S}=$};
                \draw [fill=black] (1,2) circle [radius=0.2];
                \node [white] at (1,2) {$1$};
                \draw (1.2,2)--(2.8,2); 
                \draw [ultra thick] (3,2) circle [radius=0.2];
                \node at (3,2) {$2$};
            \end{tikzpicture}
        \end{center}
        Thus, algorithm \textit{PermRowCol} assigns $\CNOT(2,1)$. It follows that
        \begin{align}
            \vect{A} = \begin{blockarray}{ccccccc}
                & \color{IMSGreen}1' & \color{IMSGreen}2' & \color{IMSGreen}3' & \color{IMSGreen}4' & \color{IMSGreen}5' & \color{IMSGreen}6'\\
            \begin{block}{c(cccccc)}
                \color{IMSGreen}1 & \color{IMSRed}0 & \color{IMSRed}1 & \color{IMSRed}0 & \color{IMSRed}1 & \color{IMSRed}1 & \color{IMSRed}0 \\
                \color{IMSGreen}2 & 1 & 1 & 1 & \color{IMSRed}1 & 1 & 0 \\
                \color{IMSGreen}3 & 1 & 0 & 0 & \color{IMSRed}0 & 1 & 1 \\
                \color{IMSGreen}4 & 1 & 1 & 1 & \color{IMSRed}0 & 1 & 0 \\
                \color{IMSGreen}5 & 1 & 0 & 1 & \color{IMSRed}0 & 1 & 0 \\
                \color{IMSGreen}6 & 1 & 0 & 1 & \color{IMSRed}0 & 1 & 1 \\
            \end{block}
        \end{blockarray} \xrightarrow{R(1,2)} \begin{blockarray}{ccccccc}
            & \color{IMSGreen}1' & \color{IMSGreen}2' & \color{IMSGreen}3' & \color{IMSGreen}4' & \color{IMSGreen}5' & \color{IMSGreen}6'\\
            \begin{block}{c(cccccc)}
                \color{IMSGreen}1 & \color{IMSRed}0 & \color{IMSRed}1 & \color{IMSRed}0 & \color{IMSBlue}1 & \color{IMSRed}1 & \color{IMSRed}0 \\
                \color{IMSGreen}2 & 1 & 0 & 1 & \color{IMSBlue}0 & 0 & 0 \\
                \color{IMSGreen}3 & 1 & 0 & 0 & \color{IMSBlue}0 & 1 & 1 \\
                \color{IMSGreen}4 & 1 & 1 & 1 & \color{IMSBlue}0 & 1 & 0 \\
                \color{IMSGreen}5 & 1 & 0 & 1 & \color{IMSBlue}0 & 1 & 0 \\
                \color{IMSGreen}6 & 1 & 0 & 1 & \color{IMSBlue}0 & 1 & 1 \\
            \end{block}
            \end{blockarray}. 
        \end{align}
        Next, we eliminate row $1$ to $e_4$. From solving the system of linear equations while emitting column $4'$, row $1$ is formed by rows $2$ and $4$. Thus $S = \{1,2,4\}$. The Steiner tree $\mathcal{T}_{1,S}$ has root $1$ and a set of terminals $S$:
        \begin{center}        
            \begin{tikzpicture}
                \node at (0,2) {$\mathcal{T}_{1,S}=$};
                \draw [fill=black] (1,2) circle [radius=0.2];
                \node [white] at (1,2) {$1$};
                \draw (1.2,2)--(2.8,2); 
                \draw [ultra thick] (3,2) circle [radius=0.2];
                \node at (3,2) {$2$};
                \draw (1,1.8)--(1,0.4); 
                \draw [ultra thick] (1,0.2) circle [radius=0.2];
                \node at (1,0.2) {$4$};
            \end{tikzpicture}
        \end{center}
        Thus, algorithm \textit{PermRowCol} assigns $\CNOT(1,2)\CNOT(1,4)$. It follows that
        \begin{align}
            (1) = \begin{blockarray}{ccccccc}
                & \color{IMSGreen}1' & \color{IMSGreen}2' & \color{IMSGreen}3' & \color{IMSGreen}4' & \color{IMSGreen}5' & \color{IMSGreen}6'\\
                \begin{block}{c(cccccc)}
                    \color{IMSGreen}1 & \color{IMSRed}0 & \color{IMSRed}1 & \color{IMSRed}0 & \color{IMSBlue}1 & \color{IMSRed}1 & \color{IMSRed}0 \\
                    \color{IMSGreen}2 & 1 & 0 & 1 & \color{IMSBlue}0 & 0 & 0 \\
                    \color{IMSGreen}3 & 1 & 0 & 0 & \color{IMSBlue}0 & 1 & 1 \\
                    \color{IMSGreen}4 & 1 & 1 & 1 & \color{IMSBlue}0 & 1 & 0 \\
                    \color{IMSGreen}5 & 1 & 0 & 1 & \color{IMSBlue}0 & 1 & 0 \\
                    \color{IMSGreen}6 & 1 & 0 & 1 & \color{IMSBlue}0 & 1 & 1 \\
                \end{block}
                \end{blockarray}&\xrightarrow{R(2,1)}\begin{blockarray}{ccccccc}
                    & \color{IMSGreen}1' & \color{IMSGreen}2' & \color{IMSGreen}3' & \color{IMSGreen}4' & \color{IMSGreen}5' & \color{IMSGreen}6'\\
                \begin{block}{c(cccccc)}
                    \color{IMSGreen}1 & \color{IMSRed}1 & \color{IMSRed}1 & \color{IMSRed}1 & \color{IMSBlue}1 & \color{IMSRed}1 & \color{IMSRed}0 \\
                    \color{IMSGreen}2 & 1 & 0 & 1 & \color{IMSBlue}0 & 0 & 0 \\
                    \color{IMSGreen}3 & 1 & 0 & 0 & \color{IMSBlue}0 & 1 & 1 \\
                    \color{IMSGreen}4 & 1 & 1 & 1 & \color{IMSBlue}0 & 1 & 0 \\
                    \color{IMSGreen}5 & 1 & 0 & 1 & \color{IMSBlue}0 & 1 & 0 \\
                    \color{IMSGreen}6 & 1 & 0 & 1 & \color{IMSBlue}0 & 1 & 1 \\
                \end{block}
                \end{blockarray}\\
                &\xrightarrow{R(4,1)}\begin{blockarray}{ccccccc}
                    & \color{IMSGreen}1' & \color{IMSGreen}2' & \color{IMSGreen}3' & \color{IMSGreen}4' & \color{IMSGreen}5' & \color{IMSGreen}6'\\
                \begin{block}{c(cccccc)}
                    \color{IMSGreen}1 & \color{IMSBlue}0 & \color{IMSBlue}0 & \color{IMSBlue}0 & \color{IMSBlue}1 & \color{IMSBlue}0 & \color{IMSBlue}0 \\
                    \color{IMSGreen}2 & 1 & 0 & 1 & \color{IMSBlue}0 & 0 & 0 \\
                    \color{IMSGreen}3 & 1 & 0 & 0 & \color{IMSBlue}0 & 1 & 1 \\
                    \color{IMSGreen}4 & 1 & 1 & 1 & \color{IMSBlue}0 & 1 & 0 \\
                    \color{IMSGreen}5 & 1 & 0 & 1 & \color{IMSBlue}0 & 1 & 0 \\
                    \color{IMSGreen}6 & 1 & 0 & 1 & \color{IMSBlue}0 & 1 & 1 \\
                \end{block}
                \end{blockarray}
            \end{align}
 
        \item[Update the output qubit allocation:]The output qubit allocation after eliminating row $1$ and column $4'$ is updated in \cref{table:rcallocationstep1}.
        
        \begin{table}[!ht]
        \centering
            \begin{tabular}{|l|l|l|l|l|l|l|}
            \hline
            Logical qubit/r     & \color{IMSRed}1 & 2 & 3 & 4 & 5 & 6 \\ \hline
            Physical register/c & \color{IMSRed}4 &   &   &   &   &   \\ \hline
            \end{tabular}
            \caption{Logical qubit $1$ is stored in the physical register $4$.}
            \label{table:rcallocationstep1}
        \end{table}
    \end{description}
    
    \item[Elimination step 2] After the first elimination step, the parity matrix $\vect{A}'$ and the constrained topology $G'$ are shown in \cref{fig:afterstep1}.\\
    \begin{figure}[!ht]
        \renewcommand{\thefigure}{(a)}
        \begin{subfigure}{.5\textwidth}
        \centering
        \[
            \vect{A}' = \begin{blockarray}{ccccccc}
            & \color{IMSGreen}1' & \color{IMSGreen}2' & \color{IMSGreen}3' & \color{IMSGreen}4' & \color{IMSGreen}5' & \color{IMSGreen}6'\\
            \begin{block}{c(cccccc)}
                \color{IMSGreen}1 & \color{IMSBlue}0 & \color{IMSBlue}0 & \color{IMSBlue}0 & \color{IMSBlue}1 & \color{IMSBlue}0 & \color{IMSBlue}0 \\
                \color{IMSGreen}2 & 1 & 0 & 1 & \color{IMSBlue}0 & 0 & 0 \\
                \color{IMSGreen}3 & 1 & 0 & 0 & \color{IMSBlue}0 & 1 & 1 \\
                \color{IMSGreen}4 & 1 & 1 & 1 & \color{IMSBlue}0 & 1 & 0 \\
                \color{IMSGreen}5 & 1 & 0 & 1 & \color{IMSBlue}0 & 1 & 0 \\
                \color{IMSGreen}6 & 1 & 0 & 1 & \color{IMSBlue}0 & 1 & 1 \\
            \end{block}
            \end{blockarray}
        \]
        \caption{The updated parity matrix $\vect{A}'$.}
        \end{subfigure}
        \renewcommand{\thefigure}{(b)}
        \begin{subfigure}{.5\textwidth}
        \centering
        \begin{tikzpicture}
            \node at (0.5,2) {$G'=$};
            \draw (1.2,1)--(2.8,1); 
            \draw (3,3) circle [radius=0.2];
            \node at (3,3) {$2$};
            \draw (3.2,3)--(4.8,3); 
            \draw (3.2,1)--(4.8,1); 
            \draw (5,3) circle [radius=0.2];
            \node at (5,3) {$3$};
            \draw (1,1) circle [radius=0.2];
            \node at (1,1) {$4$};
            \node at (1,4.2) {$ $};
            \draw (3,1) circle [radius=0.2];
            \node at (3,1) {$5$};
            \draw (5,1) circle [radius=0.2];
            \node at (5,1) {$6$};
            \node at (1,0) {$ $};
            \draw (5,2.8)--(5,1.2);    
            \draw (3,2.8)--(3,1.2); 
        \end{tikzpicture}
        \caption{The $5$-qubit grid $G'$.}
        \end{subfigure}
        \renewcommand{\thefigure}{13}\caption{ After elimination step 1, \textit{PermRowCol} eliminates the parity matrix $\vect{A}$ to $\vect{A}'$ in (a) under the constrained topology $G$. Accordingly, $G$ is reduced to $G'$ in (b).}
        \label{fig:afterstep1} 
    \end{figure}
    \begin{description}
        \item[Choose the row and column to eliminate:] The set of non-cutting vertices of $G'$ is $V_s = \{2,3,4,6\}.$ Then $r = 2$ and $c = 3$ since
        
        \[
        \vect{A}' = \begin{blockarray}{ccccccccc}
            & \color{IMSGreen}1' & \color{IMSGreen}2' & \color{IMSGreen}3' & \color{IMSGreen}4' & \color{IMSGreen}5' & \color{IMSGreen}6' & \color{IMSOrange}\text{Sum} & \color{IMSRed}\text{Row}\\
            \begin{block}{c(cccccc)cc}
                \color{IMSGreen}1 & \color{IMSBlue}0 & \color{IMSBlue}0 & \color{IMSBlue}0 & \color{IMSBlue}1 & \color{IMSBlue}0 & \color{IMSBlue}0 & \color{IMSOrange}\setminus & \\
                \color{IMSGreen}2 & \color{IMSRed}1 & \color{IMSRed}0 & \color{IMSRed}1 & \color{IMSBlue}0 & \color{IMSRed}0 & \color{IMSRed}0 & \color{IMSOrange}2 & \color{IMSRed}\checkmark\\
                \color{IMSGreen}3 & 1 & 0 & 0 & \color{IMSBlue}0 & 1 & 1 &\color{IMSOrange}3 & \\
                \color{IMSGreen}4 & 1 & 1 & 1 & \color{IMSBlue}0 & 1 & 0 &\color{IMSOrange}4 & \\
                \color{IMSGreen}5 & 1 & 0 & 1 & \color{IMSBlue}0 & 1 & 0 & \color{IMSOrange}\setminus & \\
                \color{IMSGreen}6 & 1 & 0 & 1 & \color{IMSBlue}0 & 1 & 1 & \color{IMSOrange}4 & \\
            \end{block}
            \end{blockarray} \Rightarrow \begin{blockarray}{ccccccc}
            & \color{IMSGreen}1' & \color{IMSGreen}2' & \color{IMSGreen}3' & \color{IMSGreen}4' & \color{IMSGreen}5' & \color{IMSGreen}6' \\
            \begin{block}{c(cccccc)}
                \color{IMSGreen}1 & \color{IMSBlue}0 & \color{IMSBlue}0 & \color{IMSBlue}0 & \color{IMSBlue}1 & \color{IMSBlue}0 & \color{IMSBlue}0 \\
                \color{IMSGreen}2 & \color{IMSRed}\mathbf{1} & \color{IMSRed}0 & \color{IMSRed}\mathbf{1} & \color{IMSBlue}0 & \color{IMSRed}0 & \color{IMSRed}0 \\
                \color{IMSGreen}3 & 1 & 0 & \color{IMSRed}0 & \color{IMSBlue}0 & 1 & 1 \\
                \color{IMSGreen}4 & 1 & 1 & \color{IMSRed}1 & \color{IMSBlue}0 & 1 & 0 \\
                \color{IMSGreen}5 & 1 & 0 & \color{IMSRed}1 & \color{IMSBlue}0 & 1 & 0 \\
                \color{IMSGreen}6 & 1 & 0 & \color{IMSRed}1 & \color{IMSBlue}0 & 1 & 1 \\
            \end{block}
            \color{IMSOrange}\text{Sum} & \color{IMSOrange}5 & \color{IMSOrange}\setminus & \color{IMSOrange}4 & \color{IMSOrange}\setminus & \color{IMSOrange}\setminus & \color{IMSOrange}\setminus \\
                \color{IMSRed}\text{Column} & & & \color{IMSRed}\checkmark & &
            \end{blockarray}
        \]
        \item[Eliminate the chosen row and column:] We start by eliminating column $3'$ to $e_2^{\intercal}$, then $S = \{2,4,5,6\}$. The Steiner tree $\mathcal{T}_{2,S}$ has root $2$ and a set of terminals $S$:
        \begin{center}        
            \begin{tikzpicture}
                \node at (0,2) {$\mathcal{T}_{2,S}=$};
                \draw (1.2,1)--(2.8,1); 
                \draw [fill=black](3,3) circle [radius=0.2];
                \node [white] at (3,3) {$2$};
                \draw (3.2,1)--(4.8,1); 
                \draw [ultra thick] (1,1) circle [radius=0.2];
                \node at (1,1) {$4$};
                \draw [ultra thick] (3,1) circle [radius=0.2];
                \node at (3,1) {$5$};
                \draw [ultra thick] (5,1) circle [radius=0.2];
                \node at (5,1) {$6$};
                \node at (1,0) {$ $};
                \draw (3,2.8)--(3,1.2); 
            \end{tikzpicture}
        \end{center}
        Thus, algorithm \textit{PermRowCol} assigns $\CNOT(4,5)\CNOT(6,5)\CNOT(5,2)$. It follows that
        \begin{align}
            \vect{A}' = \begin{blockarray}{ccccccc}
            & \color{IMSGreen}1' & \color{IMSGreen}2' & \color{IMSGreen}3' & \color{IMSGreen}4' & \color{IMSGreen}5' & \color{IMSGreen}6' \\
            \begin{block}{c(cccccc)}
                \color{IMSGreen}1 & \color{IMSBlue}0 & \color{IMSBlue}0 & \color{IMSBlue}0 & \color{IMSBlue}1 & \color{IMSBlue}0 & \color{IMSBlue}0 \\
                \color{IMSGreen}2 & \color{IMSRed}1 & \color{IMSRed}0 & \color{IMSRed}1 & \color{IMSBlue}0 & \color{IMSRed}0 & \color{IMSRed}0 \\
                \color{IMSGreen}3 & 1 & 0 & \color{IMSRed}0 & \color{IMSBlue}0 & 1 & 1 \\
                \color{IMSGreen}4 & 1 & 1 & \color{IMSRed}1 & \color{IMSBlue}0 & 1 & 0 \\
                \color{IMSGreen}5 & 1 & 0 & \color{IMSRed}1 & \color{IMSBlue}0 & 1 & 0 \\
                \color{IMSGreen}6 & 1 & 0 & \color{IMSRed}1 & \color{IMSBlue}0 & 1 & 1 \\
            \end{block}
            \end{blockarray} &\xrightarrow{R(5,4)} \begin{blockarray}{ccccccc}
            & \color{IMSGreen}1' & \color{IMSGreen}2' & \color{IMSGreen}3' & \color{IMSGreen}4' & \color{IMSGreen}5' & \color{IMSGreen}6' \\
            \begin{block}{c(cccccc)}
                \color{IMSGreen}1 & \color{IMSBlue}0 & \color{IMSBlue}0 & \color{IMSBlue}0 & \color{IMSBlue}1 & \color{IMSBlue}0 & \color{IMSBlue}0 \\
                \color{IMSGreen}2 & \color{IMSRed}1 & \color{IMSRed}0 & \color{IMSRed}1 & \color{IMSBlue}0 & \color{IMSRed}0 & \color{IMSRed}0 \\
                \color{IMSGreen}3 & 1 & 0 & \color{IMSRed}0 & \color{IMSBlue}0 & 1 & 1 \\
                \color{IMSGreen}4 & 0 & 1 & \color{IMSRed}0 & \color{IMSBlue}0 & 0 & 0 \\
                \color{IMSGreen}5 & 1 & 0 & \color{IMSRed}1 & \color{IMSBlue}0 & 1 & 0 \\
                \color{IMSGreen}6 & 1 & 0 & \color{IMSRed}1 & \color{IMSBlue}0 & 1 & 1 \\
            \end{block}
            \end{blockarray}\\
            &\xrightarrow{R(5,6)} \begin{blockarray}{ccccccc}
            & \color{IMSGreen}1' & \color{IMSGreen}2' & \color{IMSGreen}3' & \color{IMSGreen}4' & \color{IMSGreen}5' & \color{IMSGreen}6' \\
            \begin{block}{c(cccccc)}
                \color{IMSGreen}1 & \color{IMSBlue}0 & \color{IMSBlue}0 & \color{IMSBlue}0 & \color{IMSBlue}1 & \color{IMSBlue}0 & \color{IMSBlue}0 \\
                \color{IMSGreen}2 & \color{IMSRed}1 & \color{IMSRed}0 & \color{IMSRed}1 & \color{IMSBlue}0 & \color{IMSRed}0 & \color{IMSRed}0 \\
                \color{IMSGreen}3 & 1 & 0 & \color{IMSRed}0 & \color{IMSBlue}0 & 1 & 1 \\
                \color{IMSGreen}4 & 0 & 1 & \color{IMSRed}0 & \color{IMSBlue}0 & 0 & 0 \\
                \color{IMSGreen}5 & 1 & 0 & \color{IMSRed}1 & \color{IMSBlue}0 & 1 & 0 \\
                \color{IMSGreen}6 & 0 & 0 & \color{IMSRed}0 & \color{IMSBlue}0 & 0 & 1 \\
            \end{block}
            \end{blockarray}\\
            &\xrightarrow{R(2,5)} \begin{blockarray}{ccccccc}
            & \color{IMSGreen}1' & \color{IMSGreen}2' & \color{IMSGreen}3' & \color{IMSGreen}4' & \color{IMSGreen}5' & \color{IMSGreen}6' \\
            \begin{block}{c(cccccc)}
                \color{IMSGreen}1 & \color{IMSBlue}0 & \color{IMSBlue}0 & \color{IMSBlue}0 & \color{IMSBlue}1 & \color{IMSBlue}0 & \color{IMSBlue}0 \\
                \color{IMSGreen}2 & \color{IMSRed}1 & \color{IMSRed}0 & \color{IMSBlue}1 & \color{IMSBlue}0 & \color{IMSRed}0 & \color{IMSRed}0 \\
                \color{IMSGreen}3 & 1 & 0 & \color{IMSBlue}0 & \color{IMSBlue}0 & 1 & 1 \\
                \color{IMSGreen}4 & 0 & 1 & \color{IMSBlue}0 & \color{IMSBlue}0 & 0 & 0 \\
                \color{IMSGreen}5 & 0 & 0 & \color{IMSBlue}0 & \color{IMSBlue}0 & 1 & 0 \\
                \color{IMSGreen}6 & 0 & 0 & \color{IMSBlue}0 & \color{IMSBlue}0 & 0 & 1 \\
            \end{block}
            \end{blockarray}.
        \end{align}
        Next, we eliminate row $2$ to $e_3$. From solving the system of linear equations while emitting columns $3'$ and $4'$, row $2$ is formed by rows $3$, $5$, and $6$. Thus $S = \{2,3,5,6\}$. The Steiner tree $\mathcal{T}_{2,S}$ has root $2$ and a set of terminals $S$:
        \begin{center}        
            \begin{tikzpicture}
            \node at (1.5,2) {$\mathcal{T}_{2,S}=$};
            \draw [fill=black] (3,3) circle [radius=0.2];
            \node [white] at (3,3) {$2$};
            \draw (3.2,3)--(4.8,3); 
            \draw [ultra thick] (5,3) circle [radius=0.2];
            \node at (5,3) {$3$};
            \draw [ultra thick] (3,1) circle [radius=0.2];
            \node at (3,1) {$5$};
            \draw [ultra thick] (5,1) circle [radius=0.2];
            \node at (5,1) {$6$};
            \node at (1,0) {$ $};
            \draw (5,2.8)--(5,1.2);    
            \draw (3,2.8)--(3,1.2); 
        \end{tikzpicture}
        \end{center}
        Thus, algorithm \textit{PermRowCol} assigns $\CNOT(3,6)\CNOT(2,3)\CNOT(2,5)$. It follows that
        \begin{align}
            (6) = \begin{blockarray}{ccccccc}
            & \color{IMSGreen}1' & \color{IMSGreen}2' & \color{IMSGreen}3' & \color{IMSGreen}4' & \color{IMSGreen}5' & \color{IMSGreen}6' \\
            \begin{block}{c(cccccc)}
                \color{IMSGreen}1 & \color{IMSBlue}0 & \color{IMSBlue}0 & \color{IMSBlue}0 & \color{IMSBlue}1 & \color{IMSBlue}0 & \color{IMSBlue}0 \\
                \color{IMSGreen}2 & \color{IMSRed}1 & \color{IMSRed}0 & \color{IMSBlue}1 & \color{IMSBlue}0 & \color{IMSRed}0 & \color{IMSRed}0 \\
                \color{IMSGreen}3 & 1 & 0 & \color{IMSBlue}0 & \color{IMSBlue}0 & 1 & 1 \\
                \color{IMSGreen}4 & 0 & 1 & \color{IMSBlue}0 & \color{IMSBlue}0 & 0 & 0 \\
                \color{IMSGreen}5 & 0 & 0 & \color{IMSBlue}0 & \color{IMSBlue}0 & 1 & 0 \\
                \color{IMSGreen}6 & 0 & 0 & \color{IMSBlue}0 & \color{IMSBlue}0 & 0 & 1 \\
            \end{block}
            \end{blockarray}&\xrightarrow{R(6,3)}\begin{blockarray}{ccccccc}
            & \color{IMSGreen}1' & \color{IMSGreen}2' & \color{IMSGreen}3' & \color{IMSGreen}4' & \color{IMSGreen}5' & \color{IMSGreen}6' \\
            \begin{block}{c(cccccc)}
                \color{IMSGreen}1 & \color{IMSBlue}0 & \color{IMSBlue}0 & \color{IMSBlue}0 & \color{IMSBlue}1 & \color{IMSBlue}0 & \color{IMSBlue}0 \\
                \color{IMSGreen}2 & \color{IMSRed}1 & \color{IMSRed}0 & \color{IMSBlue}1 & \color{IMSBlue}0 & \color{IMSRed}0 & \color{IMSRed}0 \\
                \color{IMSGreen}3 & 1 & 0 & \color{IMSBlue}0 & \color{IMSBlue}0 & 1 & 0 \\
                \color{IMSGreen}4 & 0 & 1 & \color{IMSBlue}0 & \color{IMSBlue}0 & 0 & 0 \\
                \color{IMSGreen}5 & 0 & 0 & \color{IMSBlue}0 & \color{IMSBlue}0 & 1 & 0 \\
                \color{IMSGreen}6 & 0 & 0 & \color{IMSBlue}0 & \color{IMSBlue}0 & 0 & 1 \\
            \end{block}
            \end{blockarray}\\
                &\xrightarrow{R(3,2)}\begin{blockarray}{ccccccc}
            & \color{IMSGreen}1' & \color{IMSGreen}2' & \color{IMSGreen}3' & \color{IMSGreen}4' & \color{IMSGreen}5' & \color{IMSGreen}6' \\
            \begin{block}{c(cccccc)}
                \color{IMSGreen}1 & \color{IMSBlue}0 & \color{IMSBlue}0 & \color{IMSBlue}0 & \color{IMSBlue}1 & \color{IMSBlue}0 & \color{IMSBlue}0 \\
                \color{IMSGreen}2 & \color{IMSRed}0 & \color{IMSRed}0 & \color{IMSBlue}1 & \color{IMSBlue}0 & \color{IMSRed}1 & \color{IMSRed}0 \\
                \color{IMSGreen}3 & 1 & 0 & \color{IMSBlue}0 & \color{IMSBlue}0 & 1 & 0 \\
                \color{IMSGreen}4 & 0 & 1 & \color{IMSBlue}0 & \color{IMSBlue}0 & 0 & 0 \\
                \color{IMSGreen}5 & 0 & 0 & \color{IMSBlue}0 & \color{IMSBlue}0 & 1 & 0 \\
                \color{IMSGreen}6 & 0 & 0 & \color{IMSBlue}0 & \color{IMSBlue}0 & 0 & 1 \\
            \end{block}
            \end{blockarray}\\
            &\xrightarrow{R(5,2)}\begin{blockarray}{ccccccc}
            & \color{IMSGreen}1' & \color{IMSGreen}2' & \color{IMSGreen}3' & \color{IMSGreen}4' & \color{IMSGreen}5' & \color{IMSGreen}6' \\
            \begin{block}{c(cccccc)}
                \color{IMSGreen}1 & \color{IMSBlue}0 & \color{IMSBlue}0 & \color{IMSBlue}0 & \color{IMSBlue}1 & \color{IMSBlue}0 & \color{IMSBlue}0 \\
                \color{IMSGreen}2 & \color{IMSBlue}0 & \color{IMSBlue}0 & \color{IMSBlue}1 & \color{IMSBlue}0 & \color{IMSBlue}0 & \color{IMSBlue}0 \\
                \color{IMSGreen}3 & 1 & 0 & \color{IMSBlue}0 & \color{IMSBlue}0 & 1 & 0 \\
                \color{IMSGreen}4 & 0 & 1 & \color{IMSBlue}0 & \color{IMSBlue}0 & 0 & 0 \\
                \color{IMSGreen}5 & 0 & 0 & \color{IMSBlue}0 & \color{IMSBlue}0 & 1 & 0 \\
                \color{IMSGreen}6 & 0 & 0 & \color{IMSBlue}0 & \color{IMSBlue}0 & 0 & 1 \\
            \end{block}
            \end{blockarray}
            \end{align}
 
        \item[Update the output qubit allocation:]The output qubit allocation after eliminating row $2$ and column $3'$ is updated in \cref{table:rcallocationstep2}.
        
        \begin{table}[!ht]
        \centering
            \begin{tabular}{|l|l|l|l|l|l|l|}
            \hline
            Logical qubit/r     & 1 & \color{IMSRed}2 & 3 & 4 & 5 & 6 \\ \hline
            Physical register/c & 4 & \color{IMSRed}3  &   &   &   &   \\ \hline
            \end{tabular}
            \caption{Logical qubit $2$ is stored in the physical register $3$.}
            \label{table:rcallocationstep2}
        \end{table}
    \end{description}
    \item[Elimination step 3] After the second elimination step, the parity matrix $\vect{A}''$ and the constrained topology $G''$ are shown in \cref{fig:afterstep2}.\\
    \begin{figure}[!ht]
        \renewcommand{\thefigure}{(a)}
        \begin{subfigure}{.5\textwidth}
        \centering
        \[
            \vect{A}'' = \begin{blockarray}{ccccccc}
            & \color{IMSGreen}1' & \color{IMSGreen}2' & \color{IMSGreen}3' & \color{IMSGreen}4' & \color{IMSGreen}5' & \color{IMSGreen}6' \\
            \begin{block}{c(cccccc)}
                \color{IMSGreen}1 & \color{IMSBlue}0 & \color{IMSBlue}0 & \color{IMSBlue}0 & \color{IMSBlue}1 & \color{IMSBlue}0 & \color{IMSBlue}0 \\
                \color{IMSGreen}2 & \color{IMSBlue}0 & \color{IMSBlue}0 & \color{IMSBlue}1 & \color{IMSBlue}0 & \color{IMSBlue}0 & \color{IMSBlue}0 \\
                \color{IMSGreen}3 & 1 & 0 & \color{IMSBlue}0 & \color{IMSBlue}0 & 1 & 0 \\
                \color{IMSGreen}4 & 0 & 1 & \color{IMSBlue}0 & \color{IMSBlue}0 & 0 & 0 \\
                \color{IMSGreen}5 & 0 & 0 & \color{IMSBlue}0 & \color{IMSBlue}0 & 1 & 0 \\
                \color{IMSGreen}6 & 0 & 0 & \color{IMSBlue}0 & \color{IMSBlue}0 & 0 & 1 \\
            \end{block}
            \end{blockarray}
        \]
        \caption{The updated parity matrix $\vect{A}''$.}
        \end{subfigure}
        \renewcommand{\thefigure}{(b)}
        \begin{subfigure}{.5\textwidth}
        \centering
        \begin{tikzpicture}
            \node at (0.5,2) {$G''=$};
            \draw (1.2,1)--(2.8,1); 
            \draw (3.2,1)--(4.8,1); 
            \draw (5,3) circle [radius=0.2];
            \node at (5,3) {$3$};
            \draw (1,1) circle [radius=0.2];
            \node at (1,1) {$4$};
            \node at (1,4.2) {$ $};
            \draw (3,1) circle [radius=0.2];
            \node at (3,1) {$5$};
            \draw (5,1) circle [radius=0.2];
            \node at (5,1) {$6$};
            \node at (1,0) {$ $};
            \draw (5,2.8)--(5,1.2);    
        \end{tikzpicture}
        \caption{The $4$-qubit line $G''$.}
        \end{subfigure}
        \renewcommand{\thefigure}{14}\caption{After elimination step 2, \textit{PermRowCol} eliminates the parity matrix $\vect{A}'$ to $\vect{A}''$ in (a) under the constrained topology $G'$. Accordingly, $G'$ is reduced to $G''$ in (b).}
        \label{fig:afterstep2} 
    \end{figure}
    
    \FloatBarrier
    
    \begin{description}
        \item[Choose the row and column to eliminate:] The set of non-cutting vertices of $G''$ is $V_s = \{3,4,5,6\}.$ Then $r = 4$ and $c = 2$ since
        
        \[
        \vect{A}'' = \begin{blockarray}{ccccccccc}
            & \color{IMSGreen}1' & \color{IMSGreen}2' & \color{IMSGreen}3' & \color{IMSGreen}4' & \color{IMSGreen}5' & \color{IMSGreen}6' & \color{IMSOrange}\text{Sum} & \color{IMSRed}\text{Row}\\
            \begin{block}{c(cccccc)cc}
                \color{IMSGreen}1 & \color{IMSBlue}0 & \color{IMSBlue}0 & \color{IMSBlue}0 & \color{IMSBlue}1 & \color{IMSBlue}0 & \color{IMSBlue}0 & \color{IMSOrange}\setminus & \\
                \color{IMSGreen}2 & \color{IMSBlue}0 & \color{IMSBlue}0 & \color{IMSBlue}1 & \color{IMSBlue}0 & \color{IMSBlue}0 & \color{IMSBlue}0 & \color{IMSOrange}\setminus & \\
                \color{IMSGreen}3 & 1 & 0 & \color{IMSBlue}0 & \color{IMSBlue}0 & 1 & 0 & \color{IMSOrange}2 & \\
                \color{IMSGreen}4 & \color{IMSRed}0 & \color{IMSRed}1 & \color{IMSBlue}0 & \color{IMSBlue}0 & \color{IMSRed}0 & \color{IMSRed}0 & \color{IMSOrange}1 & \color{IMSRed}\checkmark\\
                \color{IMSGreen}5 & 0 & 0 & \color{IMSBlue}0 & \color{IMSBlue}0 & 1 & 0 & \color{IMSOrange}1 & \\
                \color{IMSGreen}6 & 0 & 0 & \color{IMSBlue}0 & \color{IMSBlue}0 & 0 & 1 & \color{IMSOrange}1 & \\
            \end{block}
            \end{blockarray} \Rightarrow \begin{blockarray}{ccccccc}
            & \color{IMSGreen}1' & \color{IMSGreen}2' & \color{IMSGreen}3' & \color{IMSGreen}4' & \color{IMSGreen}5' & \color{IMSGreen}6' \\
            \begin{block}{c(cccccc)}
                \color{IMSGreen}1 & \color{IMSBlue}0 & \color{IMSBlue}0 & \color{IMSBlue}0 & \color{IMSBlue}1 & \color{IMSBlue}0 & \color{IMSBlue}0 \\
                \color{IMSGreen}2 & \color{IMSBlue}0 & \color{IMSBlue}0 & \color{IMSBlue}1 & \color{IMSBlue}0 & \color{IMSBlue}0 & \color{IMSBlue}0 \\
                \color{IMSGreen}3 & 1 & \color{IMSRed}0 & \color{IMSBlue}0 & \color{IMSBlue}0 & 1 & 0 \\
                \color{IMSGreen}4 & \color{IMSRed}0 & \color{IMSRed}\mathbf{1} & \color{IMSBlue}0 & \color{IMSBlue}0 & \color{IMSRed}0 & \color{IMSRed}0 \\
                \color{IMSGreen}5 & 0 & \color{IMSRed}0 & \color{IMSBlue}0 & \color{IMSBlue}0 & 1 & 0 \\
                \color{IMSGreen}6 & 0 & \color{IMSRed}0 & \color{IMSBlue}0 & \color{IMSBlue}0 & 0 & 1 \\
            \end{block}
            \color{IMSOrange}\text{Sum} & \color{IMSOrange}\setminus & \color{IMSOrange}1 & \color{IMSOrange}\setminus & \color{IMSOrange}\setminus & \color{IMSOrange}\setminus & \color{IMSOrange}\setminus \\
            \color{IMSRed}\text{Column} & & \color{IMSRed}\checkmark & & &
            \end{blockarray}
        \]
        
        \item[Eliminate the chosen row and column:] In fact, column $2'$ and row $4$ is $e_4^{\intercal}$ and $e_2$ respectively, this step is complete.
 
        \item[Update the output qubit allocation:]The output qubit allocation after eliminating row $4$ and column $2'$ is updated in \cref{table:rcallocationstep3}.
        
        \begin{table}[!ht]
        \centering
            \begin{tabular}{|l|l|l|l|l|l|l|}
            \hline
            Logical qubit/r     & 1 & 2 & 3 & \color{IMSRed}4 & 5 & 6 \\ \hline
            Physical register/c & 4 & 3  &   &  \color{IMSRed}2 &   &   \\ \hline
            \end{tabular}
            \caption{Logical qubit $4$ is stored in the physical register $2$.}
            \label{table:rcallocationstep3}
        \end{table}
    \end{description}
    \item[Elimination step 4] After the third elimination step, the parity matrix $\vect{A}'''$ and the constrained topology $G'''$ are shown in \cref{fig:afterstep3}.\\
    \begin{figure}[!ht]
        \renewcommand{\thefigure}{(a)}
        \begin{subfigure}{.5\textwidth}
        \centering
        \[
            \vect{A}''' = \begin{blockarray}{ccccccc}
            & \color{IMSGreen}1' & \color{IMSGreen}2' & \color{IMSGreen}3' & \color{IMSGreen}4' & \color{IMSGreen}5' & \color{IMSGreen}6' \\
            \begin{block}{c(cccccc)}
                \color{IMSGreen}1 & \color{IMSBlue}0 & \color{IMSBlue}0 & \color{IMSBlue}0 & \color{IMSBlue}1 & \color{IMSBlue}0 & \color{IMSBlue}0 \\
                \color{IMSGreen}2 & \color{IMSBlue}0 & \color{IMSBlue}0 & \color{IMSBlue}1 & \color{IMSBlue}0 & \color{IMSBlue}0 & \color{IMSBlue}0 \\
                \color{IMSGreen}3 & 1 & \color{IMSBlue}0 & \color{IMSBlue}0 & \color{IMSBlue}0 & 1 & 0 \\
                \color{IMSGreen}4 & \color{IMSBlue}0 & \color{IMSBlue}1 & \color{IMSBlue}0 & \color{IMSBlue}0 & \color{IMSBlue}0 & \color{IMSBlue}0 \\
                \color{IMSGreen}5 & 0 & \color{IMSBlue}0 & \color{IMSBlue}0 & \color{IMSBlue}0 & 1 & 0 \\
                \color{IMSGreen}6 & 0 & \color{IMSBlue}0 & \color{IMSBlue}0 & \color{IMSBlue}0 & 0 & 1 \\
            \end{block}
            \end{blockarray}
        \]
        \caption{The updated parity matrix $\vect{A}'''$.}
        \end{subfigure}
        \renewcommand{\thefigure}{(b)}
        \begin{subfigure}{.5\textwidth}
        \centering
        \begin{tikzpicture}
            \node at (1.5,2) {$G'''=$};
            \draw (3.2,1)--(4.8,1); 
            \draw (5,3) circle [radius=0.2];
            \node at (5,3) {$3$};
            \draw (3,1) circle [radius=0.2];
            \node at (3,1) {$5$};
            \node at (1,4.2) {$ $};
            \draw (5,1) circle [radius=0.2];
            \node at (5,1) {$6$};
            \node at (1,0) {$ $};
            \draw (5,2.8)--(5,1.2);    
        \end{tikzpicture}
        \caption{The $3$-qubit line $G'''$.}
        \end{subfigure}
        \renewcommand{\thefigure}{15}\caption{After elimination step 3, \textit{PermRowCol} eliminates the parity matrix $\vect{A}''$ to $\vect{A}'''$ in (a) under the constrained topology $G''$. Accordingly, $G''$ is reduced to $G'''$ in (b).}
        \label{fig:afterstep3} 
    \end{figure}
    
    \FloatBarrier
    
    \begin{description}
        \item[Choose the row and column to eliminate:] The set of non-cutting vertices of $G'''$ is $V_s = \{3,5\}.$ Then $r = c = 5$ since
        
        \[
            \vect{A}''' = \begin{blockarray}{ccccccccc}
                & \color{IMSGreen}1' & \color{IMSGreen}2' & \color{IMSGreen}3' & \color{IMSGreen}4' & \color{IMSGreen}5' & \color{IMSGreen}6' & \color{IMSOrange}\text{Sum} & \color{IMSRed}\text{Row}\\
                \begin{block}{c(cccccc)cc}
                    \color{IMSGreen}1 & \color{IMSBlue}0 & \color{IMSBlue}0 & \color{IMSBlue}0 & \color{IMSBlue}1 & \color{IMSBlue}0 & \color{IMSBlue}0 & \color{IMSOrange}\setminus &\\
                    \color{IMSGreen}2 & \color{IMSBlue}0 & \color{IMSBlue}0 & \color{IMSBlue}1 & \color{IMSBlue}0 & \color{IMSBlue}0 & \color{IMSBlue}0 & \color{IMSOrange}\setminus &\\
                    \color{IMSGreen}3 & 1 & \color{IMSBlue}0 & \color{IMSBlue}0 & \color{IMSBlue}0 & 1 & 0 & \color{IMSOrange}2 &\\
                    \color{IMSGreen}4 & \color{IMSBlue}0 & \color{IMSBlue}1 & \color{IMSBlue}0 & \color{IMSBlue}0 & \color{IMSBlue}0 & \color{IMSBlue}0 & \color{IMSOrange}\setminus &\\
                    \color{IMSGreen}5 & \color{IMSRed}0 & \color{IMSBlue}0 & \color{IMSBlue}0 & \color{IMSBlue}0 & \color{IMSRed}1 & \color{IMSRed}0 & \color{IMSOrange}1 & \color{IMSRed}\checkmark\\
                    \color{IMSGreen}6 & 0 & \color{IMSBlue}0 & \color{IMSBlue}0 & \color{IMSBlue}0 & 0 & 1 & \color{IMSOrange}\setminus &\\
                \end{block}
                \end{blockarray} \Rightarrow \begin{blockarray}{ccccccc}
                & \color{IMSGreen}1' & \color{IMSGreen}2' & \color{IMSGreen}3' & \color{IMSGreen}4' & \color{IMSGreen}5' & \color{IMSGreen}6'\\
                \begin{block}{c(cccccc)}
                    \color{IMSGreen}1 & \color{IMSBlue}0 & \color{IMSBlue}0 & \color{IMSBlue}0 & \color{IMSBlue}1 & \color{IMSBlue}0 & \color{IMSBlue}0 \\
                    \color{IMSGreen}2 & \color{IMSBlue}0 & \color{IMSBlue}0 & \color{IMSBlue}1 & \color{IMSBlue}0 & \color{IMSBlue}0 & \color{IMSBlue}0 \\
                    \color{IMSGreen}3 & 1 & \color{IMSBlue}0 & \color{IMSBlue}0 & \color{IMSBlue}0 & \color{IMSRed}1 & 0 \\
                    \color{IMSGreen}4 & \color{IMSBlue}0 & \color{IMSBlue}1 & \color{IMSBlue}0 & \color{IMSBlue}0 & \color{IMSBlue}0 & \color{IMSBlue}0 \\
                    \color{IMSGreen}5 & \color{IMSRed}0 & \color{IMSBlue}0 & \color{IMSBlue}0 & \color{IMSBlue}0 & \color{IMSRed}\mathbf{1} & \color{IMSRed}0 \\
                    \color{IMSGreen}6 & 0 & \color{IMSBlue}0 & \color{IMSBlue}0 & \color{IMSBlue}0 & \color{IMSRed}0 & 1 \\
                \end{block}
                \color{IMSOrange}\text{Sum} & \color{IMSOrange}\setminus & \color{IMSOrange}\setminus & \color{IMSOrange}\setminus & \color{IMSOrange}\setminus & \color{IMSOrange}2 & \color{IMSOrange}\setminus \\
                \color{IMSRed}\text{Column} & & & & & \color{IMSRed}\checkmark &
            \end{blockarray}
        \]
        
        \item[Eliminate the chosen row and column:] We start by eliminating column $5'$ to $e_5^{\intercal}$, then $S = \{3,5\}$. The Steiner tree $\mathcal{T}_{5,S}$ has root $5$ and a set of terminals $S$:
        \begin{center}        
            \begin{tikzpicture}
                \node at (0,2) {$\mathcal{T}_{5,S}=$};
                \draw (1.2,1)--(2.8,1); 
                \draw [ultra thick] (3,3) circle [radius=0.2];
                \node  at (3,3) {$3$};
                \draw [fill=black] (1,1) circle [radius=0.2];
                \node [white] at (1,1) {$5$};
                \draw [ultra thick] (3,1) circle [radius=0.2];
                \node at (3,1) {$6$};
                \node at (1,0) {$ $};
                \draw (3,2.8)--(3,1.2); 
            \end{tikzpicture}
        \end{center}
        Thus, algorithm \textit{PermRowCol} assigns $\CNOT(6,3)\CNOT(3,6)\CNOT(6,5)$. It follows that
        \begin{align}
            \vect{A}''' = \begin{blockarray}{ccccccc}
                & \color{IMSGreen}1' & \color{IMSGreen}2' & \color{IMSGreen}3' & \color{IMSGreen}4' & \color{IMSGreen}5' & \color{IMSGreen}6'\\
                \begin{block}{c(cccccc)}
                    \color{IMSGreen}1 & \color{IMSBlue}0 & \color{IMSBlue}0 & \color{IMSBlue}0 & \color{IMSBlue}1 & \color{IMSBlue}0 & \color{IMSBlue}0 \\
                    \color{IMSGreen}2 & \color{IMSBlue}0 & \color{IMSBlue}0 & \color{IMSBlue}1 & \color{IMSBlue}0 & \color{IMSBlue}0 & \color{IMSBlue}0 \\
                    \color{IMSGreen}3 & 1 & \color{IMSBlue}0 & \color{IMSBlue}0 & \color{IMSBlue}0 & \color{IMSRed}1 & 0 \\
                    \color{IMSGreen}4 & \color{IMSBlue}0 & \color{IMSBlue}1 & \color{IMSBlue}0 & \color{IMSBlue}0 & \color{IMSBlue}0 & \color{IMSBlue}0 \\
                    \color{IMSGreen}5 & \color{IMSRed}0 & \color{IMSBlue}0 & \color{IMSBlue}0 & \color{IMSBlue}0 & \color{IMSRed}1 & \color{IMSRed}0 \\
                    \color{IMSGreen}6 & 0 & \color{IMSBlue}0 & \color{IMSBlue}0 & \color{IMSBlue}0 & \color{IMSRed}0 & 1 \\
                \end{block}
            \end{blockarray} &\xrightarrow{R(3,6)} \begin{blockarray}{ccccccc}
                & \color{IMSGreen}1' & \color{IMSGreen}2' & \color{IMSGreen}3' & \color{IMSGreen}4' & \color{IMSGreen}5' & \color{IMSGreen}6'\\
                \begin{block}{c(cccccc)}
                    \color{IMSGreen}1 & \color{IMSBlue}0 & \color{IMSBlue}0 & \color{IMSBlue}0 & \color{IMSBlue}1 & \color{IMSBlue}0 & \color{IMSBlue}0 \\
                    \color{IMSGreen}2 & \color{IMSBlue}0 & \color{IMSBlue}0 & \color{IMSBlue}1 & \color{IMSBlue}0 & \color{IMSBlue}0 & \color{IMSBlue}0 \\
                    \color{IMSGreen}3 & 1 & \color{IMSBlue}0 & \color{IMSBlue}0 & \color{IMSBlue}0 & \color{IMSRed}1 & 0 \\
                    \color{IMSGreen}4 & \color{IMSBlue}0 & \color{IMSBlue}1 & \color{IMSBlue}0 & \color{IMSBlue}0 & \color{IMSBlue}0 & \color{IMSBlue}0 \\
                    \color{IMSGreen}5 & \color{IMSRed}0 & \color{IMSBlue}0 & \color{IMSBlue}0 & \color{IMSBlue}0 & \color{IMSRed}1 & \color{IMSRed}0 \\
                    \color{IMSGreen}6 & 1 & \color{IMSBlue}0 & \color{IMSBlue}0 & \color{IMSBlue}0 & \color{IMSRed}1 & 1 \\
                \end{block}
            \end{blockarray}\\
            &\xrightarrow{R(6,3)} \begin{blockarray}{ccccccc}
                & \color{IMSGreen}1' & \color{IMSGreen}2' & \color{IMSGreen}3' & \color{IMSGreen}4' & \color{IMSGreen}5' & \color{IMSGreen}6'\\
                \begin{block}{c(cccccc)}
                    \color{IMSGreen}1 & \color{IMSBlue}0 & \color{IMSBlue}0 & \color{IMSBlue}0 & \color{IMSBlue}1 & \color{IMSBlue}0 & \color{IMSBlue}0 \\
                    \color{IMSGreen}2 & \color{IMSBlue}0 & \color{IMSBlue}0 & \color{IMSBlue}1 & \color{IMSBlue}0 & \color{IMSBlue}0 & \color{IMSBlue}0 \\
                    \color{IMSGreen}3 & 0 & \color{IMSBlue}0 & \color{IMSBlue}0 & \color{IMSBlue}0 & \color{IMSRed}0 & 1 \\
                    \color{IMSGreen}4 & \color{IMSBlue}0 & \color{IMSBlue}1 & \color{IMSBlue}0 & \color{IMSBlue}0 & \color{IMSBlue}0 & \color{IMSBlue}0 \\
                    \color{IMSGreen}5 & \color{IMSRed}0 & \color{IMSBlue}0 & \color{IMSBlue}0 & \color{IMSBlue}0 & \color{IMSRed}1 & \color{IMSRed}0 \\
                    \color{IMSGreen}6 & 1 & \color{IMSBlue}0 & \color{IMSBlue}0 & \color{IMSBlue}0 & \color{IMSRed}1 & 1 \\
                \end{block}
            \end{blockarray}\\
            &\xrightarrow{R(5,6)} \begin{blockarray}{ccccccc}
                & \color{IMSGreen}1' & \color{IMSGreen}2' & \color{IMSGreen}3' & \color{IMSGreen}4' & \color{IMSGreen}5' & \color{IMSGreen}6'\\
                \begin{block}{c(cccccc)}
                    \color{IMSGreen}1 & \color{IMSBlue}0 & \color{IMSBlue}0 & \color{IMSBlue}0 & \color{IMSBlue}1 & \color{IMSBlue}0 & \color{IMSBlue}0 \\
                    \color{IMSGreen}2 & \color{IMSBlue}0 & \color{IMSBlue}0 & \color{IMSBlue}1 & \color{IMSBlue}0 & \color{IMSBlue}0 & \color{IMSBlue}0 \\
                    \color{IMSGreen}3 & 0 & \color{IMSBlue}0 & \color{IMSBlue}0 & \color{IMSBlue}0 & \color{IMSBlue}0 & 1 \\
                    \color{IMSGreen}4 & \color{IMSBlue}0 & \color{IMSBlue}1 & \color{IMSBlue}0 & \color{IMSBlue}0 & \color{IMSBlue}0 & \color{IMSBlue}0 \\
                    \color{IMSGreen}5 & \color{IMSBlue}0 & \color{IMSBlue}0 & \color{IMSBlue}0 & \color{IMSBlue}0 & \color{IMSBlue}1 & \color{IMSBlue}0 \\
                    \color{IMSGreen}6 & 1 & \color{IMSBlue}0 & \color{IMSBlue}0 & \color{IMSBlue}0 & \color{IMSBlue}0 & 1 \\
                \end{block}
            \end{blockarray}.
        \end{align}
        Since row $5$ is in fact $e_5$, this step is complete.
 
        \item[Update the output qubit allocation:]The output qubit allocation after eliminating row $5$ and column $5'$ is updated in \cref{table:rcallocationstep4}.
        
        \begin{table}[!ht]
        \centering
            \begin{tabular}{|l|l|l|l|l|l|l|}
            \hline
            Logical qubit/r     & 1 & 2 & 3 & 4 & \color{IMSRed}5 & 6 \\ \hline
            Physical register/c & 4 & 3  &   & 2 & \color{IMSRed}5  &   \\ \hline
            \end{tabular}
            \caption{Logical qubit $5$ is stored in the physical register $5$.}
            \label{table:rcallocationstep4}
        \end{table}
    \end{description}
    \item[Elimination step 5] After the fourth elimination step, the parity matrix $\vect{A}''''$ and the constrained topology $G''''$ are shown in \cref{fig:afterstep4}.\\
    \begin{figure}[!ht]
        \renewcommand{\thefigure}{(a)}
        \begin{subfigure}{.5\textwidth}
        \centering
        \[
            \vect{A}'''' = \begin{blockarray}{ccccccc}
                & \color{IMSGreen}1' & \color{IMSGreen}2' & \color{IMSGreen}3' & \color{IMSGreen}4' & \color{IMSGreen}5' & \color{IMSGreen}6'\\
                \begin{block}{c(cccccc)}
                    \color{IMSGreen}1 & \color{IMSBlue}0 & \color{IMSBlue}0 & \color{IMSBlue}0 & \color{IMSBlue}1 & \color{IMSBlue}0 & \color{IMSBlue}0 \\
                    \color{IMSGreen}2 & \color{IMSBlue}0 & \color{IMSBlue}0 & \color{IMSBlue}1 & \color{IMSBlue}0 & \color{IMSBlue}0 & \color{IMSBlue}0 \\
                    \color{IMSGreen}3 & 0 & \color{IMSBlue}0 & \color{IMSBlue}0 & \color{IMSBlue}0 & \color{IMSBlue}0 & 1 \\
                    \color{IMSGreen}4 & \color{IMSBlue}0 & \color{IMSBlue}1 & \color{IMSBlue}0 & \color{IMSBlue}0 & \color{IMSBlue}0 & \color{IMSBlue}0 \\
                    \color{IMSGreen}5 & \color{IMSBlue}0 & \color{IMSBlue}0 & \color{IMSBlue}0 & \color{IMSBlue}0 & \color{IMSBlue}1 & \color{IMSBlue}0 \\
                    \color{IMSGreen}6 & 1 & \color{IMSBlue}0 & \color{IMSBlue}0 & \color{IMSBlue}0 & \color{IMSBlue}0 & 1 \\
                \end{block}
            \end{blockarray}
        \]
        \caption{The updated parity matrix $\vect{A}''''$.}
        \end{subfigure}
        \renewcommand{\thefigure}{(b)}
        \begin{subfigure}{.5\textwidth}
        \centering
        \begin{tikzpicture}
            \node at (2,2) {$G''''=$};
            \draw (3,3) circle [radius=0.2];
            \node at (3,3) {$3$};
            \draw (3,1) circle [radius=0.2];
            \node at (3,1) {$6$};
            \node at (1,4.2) {$ $};
            \node at (1,0) {$ $};
            \draw (3,2.8)--(3,1.2);    
        \end{tikzpicture}
        \caption{The $2$-qubit line $G''''$.}
        \end{subfigure}
        \renewcommand{\thefigure}{16}\caption{After elimination step 4, \textit{PermRowCol} eliminates the parity matrix $\vect{A}'''$ to $\vect{A}''''$ in (a) under the constrained topology $G'''$. Accordingly, $G'''$ is reduced to $G''''$ in (b).}
        \label{fig:afterstep4} 
    \end{figure}
    
    \FloatBarrier
    
    \begin{description}
        \item[Choose the row and column to eliminate:] The set of non-cutting vertices of $G''''$ is $V_s = \{3,6\}.$ Then $r = 3$ and $c = 6$ since
        
        \[
            \vect{A}'''' = \begin{blockarray}{ccccccccc}
                & \color{IMSGreen}1' & \color{IMSGreen}2' & \color{IMSGreen}3' & \color{IMSGreen}4' & \color{IMSGreen}5' & \color{IMSGreen}6' & \color{IMSOrange}\text{Sum} & \color{IMSRed}\text{Row}\\
                \begin{block}{c(cccccc)cc}
                    \color{IMSGreen}1 & \color{IMSBlue}0 & \color{IMSBlue}0 & \color{IMSBlue}0 & \color{IMSBlue}1 & \color{IMSBlue}0 & \color{IMSBlue}0 & \color{IMSOrange}\setminus& \\
                    \color{IMSGreen}2 & \color{IMSBlue}0 & \color{IMSBlue}0 & \color{IMSBlue}1 & \color{IMSBlue}0 & \color{IMSBlue}0 & \color{IMSBlue}0 & \color{IMSOrange}\setminus& \\
                    \color{IMSGreen}3 & \color{IMSRed}0 & \color{IMSBlue}0 & \color{IMSBlue}0 & \color{IMSBlue}0 & \color{IMSBlue}0 & \color{IMSRed}1 & \color{IMSOrange}1& \color{IMSRed}\checkmark\\
                    \color{IMSGreen}4 & \color{IMSBlue}0 & \color{IMSBlue}1 & \color{IMSBlue}0 & \color{IMSBlue}0 & \color{IMSBlue}0 & \color{IMSBlue}0 & \color{IMSOrange}\setminus& \\
                    \color{IMSGreen}5 & \color{IMSBlue}0 & \color{IMSBlue}0 & \color{IMSBlue}0 & \color{IMSBlue}0 & \color{IMSBlue}1 & \color{IMSBlue}0 & \color{IMSOrange}\setminus& \\
                    \color{IMSGreen}6 & 1 & \color{IMSBlue}0 & \color{IMSBlue}0 & \color{IMSBlue}0 & \color{IMSBlue}0 & 1 & \color{IMSOrange}2& \\
                \end{block}
                \end{blockarray} \Rightarrow \begin{blockarray}{ccccccc}
                & \color{IMSGreen}1' & \color{IMSGreen}2' & \color{IMSGreen}3' & \color{IMSGreen}4' & \color{IMSGreen}5' & \color{IMSGreen}6'\\
                \begin{block}{c(cccccc)}
                    \color{IMSGreen}1 & \color{IMSBlue}0 & \color{IMSBlue}0 & \color{IMSBlue}0 & \color{IMSBlue}1 & \color{IMSBlue}0 & \color{IMSBlue}0 \\
                    \color{IMSGreen}2 & \color{IMSBlue}0 & \color{IMSBlue}0 & \color{IMSBlue}1 & \color{IMSBlue}0 & \color{IMSBlue}0 & \color{IMSBlue}0 \\
                    \color{IMSGreen}3 & \color{IMSRed}0 & \color{IMSBlue}0 & \color{IMSBlue}0 & \color{IMSBlue}0 & \color{IMSBlue}0 & \color{IMSRed}\mathbf{1} \\
                    \color{IMSGreen}4 & \color{IMSBlue}0 & \color{IMSBlue}1 & \color{IMSBlue}0 & \color{IMSBlue}0 & \color{IMSBlue}0 & \color{IMSBlue}0 \\
                    \color{IMSGreen}5 & \color{IMSBlue}0 & \color{IMSBlue}0 & \color{IMSBlue}0 & \color{IMSBlue}0 & \color{IMSBlue}1 & \color{IMSBlue}0 \\
                    \color{IMSGreen}6 & 1 & \color{IMSBlue}0 & \color{IMSBlue}0 & \color{IMSBlue}0 & \color{IMSBlue}0 & \color{IMSRed}1 \\
                \end{block}
                \color{IMSOrange}\text{Sum} & \color{IMSOrange}\setminus & \color{IMSOrange}\setminus & \color{IMSOrange}\setminus & \color{IMSOrange}\setminus & \color{IMSOrange}\setminus & \color{IMSOrange}1 \\
                \color{IMSRed}\text{Column} & & & & & & \color{IMSRed}\checkmark
            \end{blockarray}
        \]
        
        \item[Eliminate the chosen row and column:] We start by eliminating column $6'$ to $e_3^{\intercal}$, then $S = \{3,6\}$. The Steiner tree $\mathcal{T}_{3,S}$ has root $3$ and a set of terminals $S$:
        \begin{center}        
            \begin{tikzpicture}
            \node at (2,2) {$\mathcal{T}_{3,S}=$};
            \draw [fill=black] (3,3) circle [radius=0.2];
            \node [white] at (3,3) {$3$};
            \draw (3,1) circle [radius=0.2];
            \node [ultra thick] at (3,1) {$6$};
            \node at (1,0) {$ $};
            \draw (3,2.8)--(3,1.2);    
        \end{tikzpicture}
        \end{center}
        Thus, algorithm \textit{PermRowCol} assigns $\CNOT(6,3)$. It follows that
        \begin{align}
            \vect{A}'''' = \begin{blockarray}{ccccccc}
                & \color{IMSGreen}1' & \color{IMSGreen}2' & \color{IMSGreen}3' & \color{IMSGreen}4' & \color{IMSGreen}5' & \color{IMSGreen}6'\\
                \begin{block}{c(cccccc)}
                    \color{IMSGreen}1 & \color{IMSBlue}0 & \color{IMSBlue}0 & \color{IMSBlue}0 & \color{IMSBlue}1 & \color{IMSBlue}0 & \color{IMSBlue}0 \\
                    \color{IMSGreen}2 & \color{IMSBlue}0 & \color{IMSBlue}0 & \color{IMSBlue}1 & \color{IMSBlue}0 & \color{IMSBlue}0 & \color{IMSBlue}0 \\
                    \color{IMSGreen}3 & \color{IMSRed}0 & \color{IMSBlue}0 & \color{IMSBlue}0 & \color{IMSBlue}0 & \color{IMSBlue}0 & \color{IMSRed}1 \\
                    \color{IMSGreen}4 & \color{IMSBlue}0 & \color{IMSBlue}1 & \color{IMSBlue}0 & \color{IMSBlue}0 & \color{IMSBlue}0 & \color{IMSBlue}0 \\
                    \color{IMSGreen}5 & \color{IMSBlue}0 & \color{IMSBlue}0 & \color{IMSBlue}0 & \color{IMSBlue}0 & \color{IMSBlue}1 & \color{IMSBlue}0 \\
                    \color{IMSGreen}6 & 1 & \color{IMSBlue}0 & \color{IMSBlue}0 & \color{IMSBlue}0 & \color{IMSBlue}0 & \color{IMSRed}1 \\
                \end{block}
            \end{blockarray} &\xrightarrow{R(3,6)} \begin{blockarray}{ccccccc}
                & \color{IMSGreen}1' & \color{IMSGreen}2' & \color{IMSGreen}3' & \color{IMSGreen}4' & \color{IMSGreen}5' & \color{IMSGreen}6'\\
                \begin{block}{c(cccccc)}
                    \color{IMSGreen}1 & \color{IMSBlue}0 & \color{IMSBlue}0 & \color{IMSBlue}0 & \color{IMSBlue}1 & \color{IMSBlue}0 & \color{IMSBlue}0 \\
                    \color{IMSGreen}2 & \color{IMSBlue}0 & \color{IMSBlue}0 & \color{IMSBlue}1 & \color{IMSBlue}0 & \color{IMSBlue}0 & \color{IMSBlue}0 \\
                    \color{IMSGreen}3 & \color{IMSBlue}0 & \color{IMSBlue}0 & \color{IMSBlue}0 & \color{IMSBlue}0 & \color{IMSBlue}0 & \color{IMSBlue}1 \\
                    \color{IMSGreen}4 & \color{IMSBlue}0 & \color{IMSBlue}1 & \color{IMSBlue}0 & \color{IMSBlue}0 & \color{IMSBlue}0 & \color{IMSBlue}0 \\
                    \color{IMSGreen}5 & \color{IMSBlue}0 & \color{IMSBlue}0 & \color{IMSBlue}0 & \color{IMSBlue}0 & \color{IMSBlue}1 & \color{IMSBlue}0 \\
                    \color{IMSGreen}6 & 1 & \color{IMSBlue}0 & \color{IMSBlue}0 & \color{IMSBlue}0 & \color{IMSBlue}0 & \color{IMSBlue}0 \\
                \end{block}
            \end{blockarray}.
        \end{align}
        Since row $3$ is in fact $e_6$, this step is complete.
 
        \item[Update the output qubit allocation:]The output qubit allocation after eliminating row $3$ and column $6'$ is updated in \cref{table:rcallocationstep5}.
        
        \begin{table}[!ht]
        \centering
            \begin{tabular}{|l|l|l|l|l|l|l|}
            \hline
            Logical qubit/r     & 1 & 2 & \color{IMSRed}3 & 4 & 5 & 6 \\ \hline
            Physical register/c & 4 & 3  & \color{IMSRed}6 & 2 & 5 &   \\ \hline
            \end{tabular}
            \caption{Logical qubit $3$ is stored in the physical register $6$.}
            \label{table:rcallocationstep5}
        \end{table}
    \end{description}
\end{description}

\FloatBarrier
\subsection{Output from \textit{PermRowCol}}
    
After the fifth elimination step, \textit{PermRowCol} terminates as there is precisely one vertex left in the constrained topology. The parity matrix $\vect{A}$ is reduced to a permutation $P$. Accordingly, the final output qubit allocation is shown below.
    
\[
    P = \begin{blockarray}{ccccccc}
        & \color{IMSGreen}1' & \color{IMSGreen}2' & \color{IMSGreen}3' & \color{IMSGreen}4' & \color{IMSGreen}5' & \color{IMSGreen}6'\\
        \begin{block}{c(cccccc)}
        \color{IMSGreen}1 & 0 & 0 & 0 & 1 & 0 & 0 \\
        \color{IMSGreen}2 & 0 & 0 & 1 & 0 & 0 & 0 \\
        \color{IMSGreen}3 & 0 & 0 & 0 & 0 & 0 & 1 \\
        \color{IMSGreen}4 & 0 & 1 & 0 & 0 & 0 & 0 \\
        \color{IMSGreen}5 & 0 & 0 & 0 & 0 & 1 & 0 \\
        \color{IMSGreen}6 & 1 & 0 & 0 & 0 & 0 & 0 \\
    \end{block}
    \end{blockarray}
\]

\begin{table}[!ht]
    \centering
    \begin{tabular}{|l|l|l|l|l|l|l|}
        \hline
        Logical qubit/r     & 1 & 2 & 3 & 4 & 5 & 6 \\ \hline
        Physical register/c & 4 & 3  & 6 & 2 & 5 &  1 \\ \hline
    \end{tabular}
    \caption{Qubit allocation after resynthesizing the CNOT circuit $\vect{C}$ under the constrained topology $G$.}
    \label{table:rcallocationstep6}
\end{table}

\subsection{Examination of the output validity}
By concatenating CNOTs produced from each elimination step, the re-synthesized circuit $\vect{C}'$ and the corresponding parity matrix $\vect{M}$ is shown in \cref{fig:output}. With the input parity matrix $\vect{A}$ and the permutation matrix $P$ output by \textit{PermRowCol}, we have $\vect{M}P = \vect{A}$. This corresponds to the qubit allocation after re-synthesizing the CNOT circuit $\vect{C}$ over $G$, as described by \cref{table:rcallocationstep6}. Hence, our re-synthesized circuit $\vect{C}'$ is equivalent to circuit $\vect{C}$ up to column permutation specified by $P$.

\begin{figure}[!ht]
  \renewcommand{\thefigure}{(a)}
  \begin{subfigure}{1\textwidth}
    \[
\Qcircuit @C=.9em @R=.9em @!R {
          \lstick{\ket{1}} & \targ & \ctrl{1} & \ctrl{3} & \qw & \qw & \qw & \qw & \qw & \qw & \qw & \qw & \qw & \qw & \qw & \rstick{\ket{1 \oplus 2}}\\
          \lstick{\ket{2}} & \ctrl{-1} & \targ & \qw & \qw & \qw & \targ & \qw & \ctrl{1} & \ctrl{3} & \qw & \qw & \qw & \qw & \qw & \rstick{\ket{2 \oplus 4 \oplus 5 \oplus 6}}\\
          \lstick{\ket{3}} & \qw & \qw & \qw & \qw & \qw & \qw & \ctrl{3} & \targ & \qw & \targ & \ctrl{3} &  \qw & \targ & \qw & \rstick{\ket{3 \oplus 6}}\\
          \lstick{\vect{C}' \: =\:\ket{4}} & \qw & \qw & \targ & \ctrl{1} & \qw & \qw & \qw & \qw & \qw & \qw & \qw & \qw & \qw & \qw & \rstick{\ket{1 \oplus 2 \oplus 4}}\\
          \lstick{\ket{5}} & \qw & \qw & \qw & \targ & \targ & \ctrl{-3} & \qw & \qw & \targ & \qw & \qw & \targ & \qw & \qw & \rstick{\ket{1 \oplus 2 \oplus 3 \oplus 4 \oplus 5 \oplus 6}}\\
          \lstick{\ket{6}} & \qw & \qw & \qw & \qw & \ctrl{-1} & \qw & \targ & \qw & \qw & \ctrl{-3} & \targ & \ctrl{-1} & \ctrl{-3} & \qw & \rstick{\ket{2 \oplus 3 \oplus 4 \oplus 5 \oplus 6}}
  }
\]
\label{appL-a}
\caption{The re-synthesized CNOT circuit $\vect{C}'$ after running \textit{PermRowCol} with input parity matrix and constrained topology defined in \cref{fig:initial}.}
  \end{subfigure}
  \renewcommand{\thefigure}{(b)}
\begin{subfigure}{1\textwidth}
      \centering
 \vect{M} = \begin{blockarray}{ccccccc}
        & \color{IMSGreen}1' & \color{IMSGreen}2' & \color{IMSGreen}3' & \color{IMSGreen}4' & \color{IMSGreen}5' & \color{IMSGreen}6'\\
        \begin{block}{c(cccccc)}
        \color{IMSGreen}1 & 1 & 0 & 0 & 1 & 1 & 0 \\
        \color{IMSGreen}2 & 1 & 1 & 0 & 1 & 1 & 1 \\
        \color{IMSGreen}3 & 0 & 0 & 1 & 0 & 1 & 1 \\
        \color{IMSGreen}4 & 0 & 1 & 0 & 1 & 1 & 1 \\
        \color{IMSGreen}5 & 0 & 1 & 0 & 0 & 1 & 1 \\
        \color{IMSGreen}6 & 0 & 1 & 1 & 0 & 1 & 1 \\
    \end{block}
    \end{blockarray}
\label{appL-b}
\caption{$\vect{M}$ is the parity matrix of $\vect{C}'$.}
\end{subfigure}
  \renewcommand{\thefigure}{17}\caption{The circuit $\vect{C}'$ in (a) can be exactly represented by the $6 \times 6$ parity matrix $\vect{M}$ in (b).}
\label{fig:output} 
\end{figure}

\end{document}